\documentclass[11pt,a4paper]{article} 

\pdfoutput=1

\newcommand{\Path}{.}

\usepackage[a4paper,left=2cm, right=2cm, top=2.5cm, bottom=2.5cm]{geometry}

\usepackage[utf8]{inputenc} 
\usepackage{lmodern}
\usepackage[T1]{fontenc}

\usepackage[ngerman,english]{babel} 
\usepackage{csquotes}

\usepackage{graphicx}
\usepackage{rotating} 
\usepackage{capt-of}

\usepackage{amsmath, amsthm}
\usepackage{amssymb}

\usepackage{booktabs}

\allowdisplaybreaks[2] 

\usepackage{xcolor}
\definecolor{myblue}{RGB}{25,25,112}

\usepackage{natbib}
\usepackage[
    colorlinks,
    linkcolor=myblue!80!blue,
    citecolor=myblue!80!blue,
    urlcolor=myblue!60!blue
    ]{hyperref}

\usepackage{placeins} 

\usepackage{dsfont} 

\newcommand{\IR}{{\mathbb R}}
\newcommand{\IN}{{\mathbb N}}
\newcommand{\IZ}{{\mathbb Z}}

\newcommand{\calT}{\mathcal{T}}

\newcommand{\N}{\operatorname{N}} 
\newcommand{\IG}{\operatorname{IG}} 
\newcommand{\Ising}{\operatorname{Ising}} 

\newcommand{\h}[1]{^{(#1)}}

\newcommand{\Ind}{\mathds{1}}
\newcommand{\usw}{, \ldots ,}
\newcommand{\drm}{\mathrm{d}}
\newcommand{\eps}{\varepsilon}
\newcommand{\inv}{^{-1}}

\newcommand{\rank}{\operatorname{rank}} 
\newcommand{\trans}{^\top}
\newcommand{\orth}{^\perp}

\newcommand{\norm}[1]{{\left\vert\kern-0.25ex\left\vert #1      \right\vert\kern-0.25ex\right\vert_2}}   
\newcommand{\scal}[2]{{\langle #1, #2\rangle_2}}

\newcommand{\SNR}{\operatorname{SNR}}

\usepackage{csvsimple}
\usepackage{multirow}
\usepackage{authblk}

\title{The Impact of Model Assumptions in Scalar-on-Image Regression}

\author{Clara Happ}
\author{Sonja Greven}
\author{Volker J. Schmid\\for the Alzheimer's Disease Neuroimaging Initiative
\thanks{Data used  in  preparation  of  this  article  were  obtained  from  the  Alzheimer’s  Disease Neuroimaging  Initiative  (ADNI)  database  (\url{http://adni.loni.usc.edu}). As  such,  the  investigators  within the ADNI contributed to the design and implementation of ADNI and/or provided data but  did  not  participate  in  analysis  or  writing  of  this  paper.  A  complete  listing  of  ADNI  investigators can be found at: \url{http://adni.loni.usc.edu/wp-content/uploads/how_to_apply/ADNI_Acknowledgement_List.pdf}. 
}}

\affil{Department of Statistics, LMU Munich, Munich, Germany}

\date{}

\begin{document}

\maketitle

\abstract{Complex statistical models such as scalar-on-image regression often require strong assumptions to overcome the issue of non-identifiability. While in theory it is well understood that model assumptions can strongly influence the results, this seems to be underappreciated, or played down, in practice. 

The article gives a systematic overview of the main approaches for scalar-on-image regression with a special focus on their assumptions. We categorize the assumptions and develop measures to quantify the degree to which they are met. The impact of model assumptions  and the practical usage of the proposed measures are illustrated in a simulation study and in an application to neuroimaging data. The results show that different assumptions indeed lead to quite different estimates with similar predictive ability, raising the question of their interpretability.
We give recommendations for making modeling and interpretation decisions in practice, based on the new measures and simulations  using hypothetic coefficient images and the observed data.
}

\newpage

\setlength{\parindent}{0pt}
\setlength{\parskip}{0pt}

\section{Introduction}

Medical images allow us to look into the human body and therefore provide a rich source of information in statistical analyses. Scalar-on-image regression aims at finding a relationship between a scalar response and an image covariate. The results can be used to make subject-specific predictions, but also for finding interpretable associations and for generating hypotheses in neuroscientific research. In contrast to the widely used and very popular pixelwise or voxelwise methods, as for example statistical parametric mapping (SPM \cite{SPMbook}), which fit a separate model in each pixel with the image information as the response, scalar-on-image regression models include the whole image as a predictor in a single model. Consequently, the number of variables in principle equals at least the number of pixels in the image, which is typically much larger than the sample size. 
The model is hence inherently unidentifiable and requires strong structural assumptions on the coefficients to overcome the issue of non-identifiability. 
While this is less problematic for prediction (different coefficient images may give similar predictions), it remains an issue for the estimation and particularly for the interpretability of the coefficient image. Although all this is well understood from a theoretical point of view, we consider it an underappreciated or underplayed problem in practice, which entails the risk of over-interpreting effects that are mainly driven by the model assumptions. 

The aim of this paper is three-fold. First, we provide a review of different conceptual approaches for scalar-on-image regression, including their assumptions and currently available implementations. Overall, we discuss eight models that represent the principal approaches for scalar-on-image regression.  Some reduce the complexity by means of basis function representations of the coefficient image and can therefore be related to the broad field of scalar-on-function regression methods \cite{ReissEtAl:2016, MuellerStadtmueller:2005, CardotEtAl:1999}. Others apply dimension reduction methods, partly combined with a basis function expansion \cite{ReissOgden:2010, ReissEtAl:2015}. Finally, we also consider methods that formulate the model assumptions in terms of spatial Gaussian Markov random field priors \cite{Besag:1974, GoldsmithEtAl:2014}.
We systematically compare the models from a theoretical point of view as well as in simulations and in a case study based on neuroimaging data from a study on Alzheimer's disease \cite{WeinerEtAl:2015}.
Second, due to the inherent non-identifiability of scalar-on-image regression models, we investigate the influence of the model assumptions on the coefficient estimates and examine the extent of the problem in practice.  We argue that the structural assumptions made in the different models can come in different levels of abstraction that we characterize as underlying and parametric model assumptions. We show that different assumptions can indeed lead to quite different estimates, raising the question of interpretability of the resulting estimates.
Third, we give recommendations and develop measures that can help to  make modeling and interpretation decisions in practice.

The article is structured as follows: Section~\ref{sec:methods} introduces the scalar-on-image regression model and the different estimation methods. Particular emphasis is put on the model assumptions. In Section~\ref{sec:MeasureAssumptions}, they are compared and categorized and measures are developed that allow to assess to what extent the assumptions of a certain model are met. Section~\ref{sec:simStudy} contains the simulation study and Section~\ref{sec:app} presents the neuroimaging application. 
The paper concludes with a discussion and an outlook to potential future research in Section~\ref{sec:out}.

\section{Overview of Methods for Scalar-on-Image Regression}
\label{sec:methods}

This section introduces the scalar-on-image regression model and provides a systematic overview of the approaches considered in this paper. The presented models have been selected to represent the most important assumptions and all relevant model classes in scalar-on-image regression. In addition, we have focused on easily accessible methods, for which software solutions are already available or which we could implement without much effort. 
An overview of the implementations for the studied methods is given at the end of this section and in the code supplement of this article (available on GitHub: \url{https://github.com/ClaraHapp/SOIR}).

\subsection{The Scalar-on-Image Regression Model}
\label{sec:SOIRmodel}

The scalar-on-image regression model studied in this paper is assumed to be of the following form:
\begin{equation}
y_i = \sum_{j = 1}^p w_{i,j} \alpha_j + \sum_{l = 1}^L x_{i,l} \beta_l + \eps_i, \quad i = 1 \usw N.
\label{eq:ModelEquation}
\end{equation}
The observed data for each of the $N \in \IN$ observation units, as for example subjects in a medical study, consist of a scalar response $y_i$, an image covariate $x_i$ with $L \in \IN$ pixels and scalar covariates $w_i \in \IR^p$, including an intercept term. The images are assumed to be demeaned over all observations in the following. As in the standard linear model, the vector $\alpha \in \IR^p$ contains the coefficients for $w_i$ and the error term $\eps_i$ is assumed to be i.i.d.\ Gaussian with variance $\sigma^2_\eps > 0$. The coefficient image $\beta$ relates the observed images $x_i$ to the response in terms of a discrete inner product and therefore has the same size as $x_i$. Note that this inner product can be seen as an approximation to an integral $\int_\calT x_i(t) \beta(t) \drm t$, by interpreting the images as functions $x_i(\cdot)$ and $\beta(\cdot)$, mapping from a typically two- or three-dimensional domain $\calT$ to $\IR$. This representation is particularly useful in basis function approaches (Section~\ref{sec:BasisFunMethods}). 
The approximation in general must include integration weights to be valid. In most cases, however, the pixels are all equidistant and the weights can be set to one, at most changing the scale of $\beta(\cdot)$.
Alternatively, the model can be written in matrix-form
\begin{equation}
y = W \alpha + X\beta + \eps
\label{eq:ModelEquationMatrix}
\end{equation}
with $y = (y_1 \usw y_N)$, $W \in \IR^{N \times p}$ the matrix of scalar covariates,  $X \in \IR^{N \times L}$ the matrix of vectorized image covariates, $\beta \in \IR^L$ the vectorized coefficient image and $\eps \sim \N(0, \sigma^2_\eps I_N)$ with $I_N \in \IR^{N \times N}$ the identity matrix. Note that theoretically, the images $x_i$ and therefore also the coefficient image $\beta$ can be three- or even higher dimensional. In practice, increasing the dimensionality of the images is frequently associated with a considerable computational burden and is not supported by all implementations. For reasons of simplicity and comparability, only 2D images are considered in the following analysis. 

Model~\eqref{eq:ModelEquation} is effectively a standard linear model with coefficients $\alpha$ and $\beta$. In most cases, however, the total number of coefficients $p + L$ will exceed the number of observation units $N$ by far, i.e. the  model will in general be unidentifiable. On the other hand, the coefficients $\beta_l$ are known to form an image and thus will show dependencies between neighbouring pixels. It is therefore natural to make structural assumptions about $\beta$. These assumptions imply restrictions on the coefficients $\beta_l$ and can thus help to overcome the issue of non-identifiability. 
As the true $\beta$ coefficient is unknown, the structural assumptions on $\beta$ have to be made prior to the analysis. They reflect prior beliefs about the unknown image and can be expected to have an influence on the result. 
In the following, we present the most common approaches for scalar-on-image regression. They can broadly be categorized into basis function approaches (Section~\ref{sec:BasisFunMethods}) and random field methods (Section~\ref{sec:RFmethods}).

\subsection{Basis Function Approaches}
\label{sec:BasisFunMethods}

Basis function approaches start from the idea that the unknown coefficient image is generated by a function $\beta(\cdot) \colon \calT \to \IR$. The function is evaluated at a rectangular grid of observation points $t_l \in \calT$ (the pixels), such that $\beta_l = \beta(t_l)$, and assumed to lie in the span of $K$ known basis functions $B_1 \usw B_K$ on $\calT$, which is a $K$-dimensional space.  Then~\eqref{eq:ModelEquation} translates to
\begin{equation}
y_i = \sum_{j = 1}^p w_{i,j} \alpha_j + \sum_{l = 1}^L x_{i,l} \beta_l + \eps_i
=\sum_{j = 1}^p w_{i,j} \alpha_j + \sum_{l = 1}^L x_{i,l} \sum_{k = 1}^K b_k B_k(t_l) + \eps_i. 
\label{eq:ModelBasisFuns}
\end{equation}
This assumption reduces the estimation of $\beta$ from $L$ coefficients $\beta_l$ to $K$ coefficients $b_k$, as usually the number of basis functions $K$ is chosen much smaller than the number of pixels $L$. If further  $p + K < N$, this solves the identifiability issue. Otherwise, one can make additional assumptions on the coefficients $b_k$, depending on the basis functions used.

If the basis functions $B_k$ are orthonormal with respect to the standard inner product in $L^2(\calT)$, it can be useful to interpret the observed images $x_i$ as functions, too, and to expand them in the same basis functions as $\beta(\cdot)$ with coefficients $ \theta_{i,m}$, as then
\begin{align}
y_i &= \sum_{j = 1}^p w_{i,j} \alpha_j + \sum_{l = 1}^L x_{i,l} \beta_l + \eps_i
= \sum_{j = 1}^p w_{i,j} \alpha_j + \sum_{l = 1}^L   \sum_{m = 1}^\infty \theta_{i,m}   B_{m}(t_l)  \sum_{k = 1}^K  b_k  B_{k}(t_l)   + \eps_i \notag \\
& \approx  \sum_{j = 1}^p w_{i,j} \alpha_j +  \sum_{m = 1}^K \sum_{k = 1}^K  \theta_{i,m}  b_k  \int_\calT B_{m}(t)    B_{k}(t) \drm t  + \eps_i
= w_i \trans \alpha + \theta_i \trans b + \eps_i,
\label{eq:ONBexpansion}
\end{align}
which is a standard linear regression with the covariate vectors $w_i = (w_{i,1} \usw w_{i,p})$ and $\theta_i = (\theta_{i,1} \usw \theta_{i,K})$ and the coefficient vectors $\alpha$ and $b = (b_1 \usw b_K)$. Given an estimate $\hat b$, a simple plug-in estimate for $\beta$ then is $\hat \beta_l = \sum_{k = 1}^K \hat b_k B_k(t_l)$.

The choice of the basis functions has a considerable influence on the estimate $\hat \beta$. We divide the methods into three classes  with fixed basis functions, data-driven basis functions or a combination of the two.

\subsubsection{Fixed basis functions}

\textbf{(Penalized) B-Splines}, in the following referred to as \textit{Splines}:\\
B-Splines \cite{EilersMarx:1996, MarxEilers:1999, DeBoor:1972} are a popular class of bases for representing smooth functions. In the case of a two-dimensional function evaluated on a grid of pixels, one can use tensor product splines, giving 
$
\beta(t) =  \sum_{k_x = 1}^{K_x} \sum_{k_y = 1}^{K_y} b_{k_x, k_y} B_{k_x}(t_x) B_{k_y}(t_y)
$
for $t = (t_x, t_y)$. In the B-spline based scalar-on-image regression model \cite{MarxEilers:2005} the unknown coefficients $b$ and $\alpha$ are found by minimizing a penalized least squares criterion
including a quadratic penalty for penalizing differences in $b$ along the $x$- and $y$- axes. This yields a smooth function and -- in most cases -- an identifiable model \cite{Happ:2013,ScheiplGreven:2016}. The penalty parameters can be found e.g.\  by (generalized) cross-validation \cite{MarxEilers:2005} or using a restricted maximum likelihood (REML) approach \cite{Wood:2011}.

The main assumption here is that the unknown coefficient function $\beta(\cdot)$ can be represented well by the $K_x \cdot K_y$ tensor product spline basis functions and that it has smooth variation. In the context of neuroimaging this would mean that brain areas that are close to each other have a similar association with the response without abrupt changes in $\beta(\cdot)$.

\bigskip
\textbf{Wavelets}, \textit{WNET}:\\
Given a so-called pair of mother and father wavelet functions $\psi$ and $\phi$, an arbitrary function $f$ on a real interval can  be expressed as
$f(t) = \sum_{n \in \IZ} c_{M_0,n} \phi_{M_0,n}(t) + \sum_{m = -\infty}^{M_0} \sum_{n \in \IZ} d_{m,n} \psi_{m,n}(t)$
with coefficients $c_{M_0,n} = \scal{f}{\phi_{M_0,n}}$ and $  d_{m,n} = \scal{f}{\psi_{m,n}}$. The basis functions $\phi_{M_0,n}$ and $\psi_{m,n}$ are orthonormal for a given resolution level $M_0$ (and $m \leq M_0$) \cite{Daubechies:1988,ReissEtAl:2015} and derive from the original mother and father wavelets via dilatation and translation: $\psi_{m,n}(t) = 2^{-m/2} \psi(2^{-m}t - n)$  and $\phi_{m,n}(t) = 2^{-m/2} \phi(2^{-m}t -n)$ with $m,n \in \IZ$ \cite{Daubechies:1988}. In practical applications, $f$ will be observed on a finite grid $\{t_1 \usw t_L\}$, and thus the infinite sums will be truncated. 
For the two-dimensional case, one can again use a tensor-type approach, defining basis functions for the $x$-, $y$- and $xy$-directions. The basis coefficients can be obtained efficiently if the side length of the image is a power of $2$ \cite{Mallat:1989}.

In practice, one observes that only a few basis functions are needed to describe most functions well, even those with sharp, highly localized features, due to the different resolutions of the basis functions. The majority of the coefficients $c_{M_0,n},~d_{m,n}$ can therefore be set to $0$ without affecting the important characteristics of the function. This is the basic idea of the WNET model \cite{ReissEtAl:2015} for scalar-on-image regression, where the expansion of the unknown coefficient function $\beta(\cdot)$ in wavelet basis functions is combined with a variable selection step. The authors propose to add a (na\"ive) elastic net penalty \cite{ZouHastie:2005} which combines the smoothing property of Ridge regression with the variable selection obtained by LASSO. The algorithm can be extended by an additional screening step, retaining only the $K^\ast < K$ coefficients with the highest variance \cite{JohnstoneLu:2009}.

As a full wavelet basis could perfectly reproduce the observed image, the main assumption on $\beta(\cdot)$ is sparsity of the coefficients $b_k$, i.e.\  that the signal concentrates on a few basis functions which may lead to local variations of smoothness. From a neuroimaging point of view, this means that the model should be able to produce estimates $\hat \beta(\cdot)$ with local adaptivity, which can capture sharp features in some areas and have a high degree of smoothness elsewhere \cite{ReissEtAl:2015}. The preselection step further assumes that the non-zero coefficients are those corresponding to the highest variation in the observed images $x_i$ and thus variation patterns in the data translate to the coefficient image.

\subsubsection{Data-driven basis functions}

\bigskip
\textbf{Principal component regression}, \textit{PCR2D}:\\
Functional principal component regression \cite{MuellerStadtmueller:2005} expands the unknown function $\beta(\cdot)$ in principal components (functions or images), that are obtained from the data. They represent orthogonal modes of variation in the data and thus provide the most parsimonious representation of the data in terms of the number of basis functions needed to explain a given degree of variation in the data. Expanding $x_i$ and $\beta(\cdot)$ in the same $K$ leading principal component functions and making use of their orthonormality yields~\eqref{eq:ONBexpansion} with covariates  $\theta_{i,k}$, the individual principal component scores for each observation and each principal component, and coefficients $b_k$ to be estimated.
In most cases the total number of unknown variables $p + K$ will be much smaller than the number of observations $N$, thus the resulting model is identifiable.
Smooth principal component images can be calculated using a regularized tensor product decomposition \cite{Allen:2013} based on a rank-one approximation (CANDECOMP/PARAFRAC) \cite{CarrollChang:1970}.  In principle, the MFPCA approach for (multivariate) functional data \cite{HappGreven:2016} can also be used to calculate eigenimages, interpreting the images as multivariate functional data with a single element on a two- or three-dimensional domain.

The crucial assumption on the coefficient function $\beta(\cdot)$ is that $\beta(\cdot)$ is a linear combination of the first $K$ principal components, i.e.\ that  $\beta(\cdot)$ shares the same modes of variation as the observed images $x_i$. On the one hand, this seems plausible, as areas with high variation in the observed images are likely to explain differences in the response values. On the other hand, this corresponds to a strong restriction of $\beta(\cdot)$ to a $K$-dimensional subspace of $L^2(\calT)$. The critical number $K$ can be found e.g.\ by cross-validation.

\subsubsection{Combined Methods}

The following methods combine a basis function expansion of $\beta(\cdot)$ with a subsequent data-dependent dimension reduction based on principal component analysis or partial least squares.

\bigskip
\textbf{Principal component regression based on splines}, \textit{FPCR}:\\
The FPCR method \cite{ReissOgden:2010} proposes to expand $\beta(\cdot)$ in a spline basis and add a smoothness penalty on the coefficients $b$ in order to impose smoothness on $\beta(\cdot)$, as in \textit{Splines}. The least squares criterion to minimize thus becomes
\begin{equation}
\norm{y - W\alpha - X B b}^2 + \lambda b \trans P b \to \min_{\alpha, b}
\label{eq:penModel}
\end{equation}
with $B \in \IR^{L \times K}$ the matrix of the basis functions $B_1 \usw B_K$ evaluated on the observation grid $\{t_1 \usw t_L\}$, $\lambda > 0$ a regularization parameter and $P \in \IR^{K \times K}$ an appropriate penalty matrix, e.g.\  for penalizing first differences. This corresponds to a penalized linear model with design matrix $XB$ for the coefficients $b$.
In a next step, the singular value decomposition of $XB$ is calculated: $XB = U D V\trans$ with $V \in \IR^{K\times K}$ containing the principal components of $XB$. Then $b$ is assumed to lie in the span of the  leading $K_0 < K$ principal components of $XB$ \cite{ReissOgden:2010}, i.e.\  $b = V_0 \tilde b$ with $V_0 \in \IR^{K \times K_0}$ the matrix containing the first $K_0$ columns of $V$. Then~\eqref{eq:penModel} can be written as a model in $\tilde b$
\[
\norm{y - W\alpha - X B V_0 \tilde b}^2 + \lambda \tilde b V_0 \trans P V_0 \tilde b \to \min_{\alpha, \tilde b}.
\]
$K_0$ can be chosen by cross-validation and is usually much smaller than $K$, which makes the model identifiable in $\tilde b$ if $K_0 < n$. Once an estimate for $\tilde b$ is found, the  estimated coefficient image is given by $\hat \beta = B V_0 \tilde b$.

In this approach, $\beta(\cdot)$ is assumed to lie in the span of  a given spline basis with coefficients $b$ and to be a smooth function, which is induced by a smoothness penalty, as in \textit{Splines}. Moreover, the coefficient vector is assumed to lie in the span of the leading principal components of the matrix $XB$, which corresponds to a considerable dimension reduction, similar to PCR2D.

\bigskip
\textbf{Principal component regression in wavelet space}, \textit{WCR}:\\
The WCR method \cite{ReissEtAl:2015} proposes to transform the unknown coefficient function $\beta(\cdot)$ to the wavelet space with coefficients $b = (b_1 \usw b_K)$. 
 In a subsequent screening step, only the $K^\ast$ coefficients $\theta_{i,k}$ with the highest sample-variance across the images are retained, giving a matrix $X^\ast \in \IR^{N \times K^\ast}$ and the corresponding vector of unknown coefficients $b^\ast \in \IR^{K^\ast}$ (cf.\ \textit{WNET}). Next, the singular value decomposition of $X^\ast = U^\ast D ^\ast {V^\ast} \trans$ is calculated with $V^\ast \in \IR^{K^\ast \times K^\ast}$ containing the principal components of $X^\ast$. It is then assumed that $b^\ast$ lies in the span of the first $K_0$ principal components of $X^\ast$, i.e.\  $b^\ast = V^\ast_0 \tilde b^\ast$ with $V^\ast_0$ the matrix containing the first $K_0$ columns of $V^\ast$ as in the spline-based approach. 
Given the estimated values $\tilde b^\ast$, the estimated coefficient function $\hat \beta(\cdot)$ can be obtained by calculating $b^\ast = V^\ast_0 \tilde b^\ast$, setting all other coefficients in $b$ to zero and retransforming $b$ to the original space. 

The coefficient function $\beta(\cdot)$ here is assumed to be representable by given wavelet basis functions, where only a small number $K^\ast$ of wavelet coefficients are assumed to be non-zero, notably those coefficients which have the highest variation in the images. Moreover, the coefficient vector is assumed to lie in the span of the leading principal components of the non-zero wavelet coefficients of the images. Overall, this means that $\beta(\cdot)$ is assumed to be similar to the observed images concerning patterns of variation, including local variation of smoothness \cite{ReissEtAl:2015}.

\bigskip
\textbf{Partial least squares in wavelet space}, \textit{WPLS}:\\
A variant of the last method is WPLS \cite{ReissEtAl:2015}, where principal component analysis is replaced by partial least squares. While principal component analysis focuses on the most important modes of variation in the covariate images $x_i$ or their wavelet coefficients $\theta_{i,k}$, partial least squares finds the components in $x_i$ that are most relevant for predicting the outcome $y_i$. 

Similarly to the previous approach, $\beta(\cdot)$ is assumed to lie in the span of  wavelets with a sparse coefficient vector $b$, having non-zero values only for those entries where the corresponding wavelet coefficients of the images have the highest covariation with the response. Moreover, the non-zero coefficients $b^\ast$ are assumed to lie in the span of the leading principal least squares components derived from the wavelet coefficients $\theta_{i,k}$ of the observed images $x_i$, but also from the response values $y_i$. 

\subsection{Random Field Methods}
\label{sec:RFmethods}

Random fields are frequently used to model the coefficient image $\beta$ in a Bayesian framework. In contrast to basis function approaches,  $\beta$ is modeled directly on the pixel level, i.e.\  the unknown coefficient is $\beta = (\beta_1 \usw \beta_L)$. Following the Bayesian paradigm, one assumes a prior distribution for all variables in model~\eqref{eq:ModelEquationMatrix}, assuming that $\alpha, \beta$ and $\sigma^2_\eps$ are independent:
$
\left. y \middle| \alpha, \beta, \sigma^2_\eps \right.  \sim \N(W \alpha + X\beta, \sigma^2_\eps I_N) 
$
with a constant prior for $\alpha$ and $\sigma^2_\eps \sim \IG(\delta_\eps\h{1}, \delta_\eps\h{2})$ for some $\delta_\eps \h{1},\delta_\eps\h{2} > 0$. The full conditionals for $\alpha$ and $\sigma^2_\eps$ are then given by
$
 \left. \alpha \middle| \cdot \right.  \sim \N \left( (W \trans W)\inv W\trans(y - X\beta), \sigma^2_\eps (W \trans W)\inv\right) $ and $ \left. \sigma^2_\eps \middle| \cdot \right. \sim \IG \Big(\delta_\eps\h{1} + \frac{N}{2}, \delta_\eps\h{2} + \frac{1}{2} (y- W\alpha - X\beta)\trans (y - W\alpha - X\beta) \Big)$,
i.e.\  they are known distributions. Samples from the posterior distribution can thus be obtained by a simple Gibbs sampler.
Random field priors for $\beta$ model the spatial dependence between pixels. In order to facilitate sampling from the posterior, one often chooses priors that yield simple full conditionals.

\bigskip

\textbf{Gaussian Markov Random Fields}, \textit{GMRF}:\\
A commonly used class of priors for $\beta$ are (intrinsic) Gaussian Markov Random Fields (GMRF), which can induce smoothness and constitute a conjugate prior for $\beta$. The value of $\beta$ for a pixel $l$ is assumed to depend only on the values of $\beta$ in the neighbourhood of $l$ (Markov property), which can be modeled as
$
\left. \beta_l \middle| \beta_{\delta(l)}, \sigma^2_\beta \right.  \sim \N \Big(\frac{1}{d_l} \sum_{j \sim l} \beta_j, \frac{\sigma^2_\beta}{d_l} \Big)
$. 
Here $d_l = \#\{j = 1 \usw L \colon j \sim l\}$ denotes the number of neighbours of $l$ and $\beta_{\delta(l)}= \{ \beta_j \colon j \sim l\}$ is the set of all neighbouring coefficients of the $l$-th location, where $j \sim l$ means that the pixels $j$ and $l$ are neighbours \cite{Besag:1974, RueHeld:2005}. The choice of the neighbourhood thus models the dependence structure in $\beta$. 
The common variance parameter $\sigma^2_\beta$ is again assumed to have an $\IG(\delta_\beta\h{1}, \delta_\beta\h{2})$ distribution with $\delta_\beta \h{1},\delta_\beta\h{2} > 0$, which can be shown to be conjugate in this case. 
The prior assumption for $\beta$ can be rewritten in an unconditional form: $p\left(\beta \middle|\sigma^2_\beta \right) \propto (\sigma^2_\beta)^{- \rank(P)/2} \exp\Big(-\frac{1}{2\sigma^2_\beta}\beta\trans P \beta \Big)$ 
with $P \in \IR^{L \times L}$ the neighbourhood matrix with
$p_{j,l} = d_l$ for $j = l$, $p_{j,l} = -1$ for $j \sim l$ and $p_{j,l} = 0$ otherwise.
This is not a proper distribution, as $P$ does not have full rank ($\rank(P) = L-1$). However, this prior assumption yields a proper Gaussian full conditional for $\beta$ if the data contains enough information, and hence samples from the posterior can be drawn by simple Gibbs sampling. 
The Bayesian approach with Gaussian Markov random field priors has an interesting correspondence to penalized  basis function methods with constant local basis functions $ \Ind_l$ for each pixel, where the Gaussian prior corresponds to the quadratic penalty. The smoothing parameter in this penalized formulation is given by the fraction of the prior variances $\lambda =  \frac{\sigma_\eps^2}{\sigma_\beta^2}$. 

The assumptions for the Bayesian GMRF models are given in terms of the priors. For the coefficient image $\beta$ the GMRF prior induces smoothness. As in the \textit{Splines} model, this means that e.g.\ in neuroimaging adjacent brain areas have a similar association with the response without abrupt changes in $\beta$. However, the coefficient images can be expected to have more small scale structure, as smoothness is induced on the level of pixels rather than on the level of basis functions.

\bigskip

\textbf{Sparse Gaussian Markov Random Field}, \textit{SparseGMRF}:\\
The sparse GMRF method \cite{GoldsmithEtAl:2014} adds a variable selection aspect to the \textit{GMRF} model to combine smoothness with sparsity.
The basic idea here is that in general, not the full image $x_i$ will show a relevant association with the response and thus major parts of the coefficient image $\beta$ can be assumed to be zero. At the same time, the non-zero pixels of interest ideally should form smooth coherent clusters.

The authors propose to combine the GMRF prior for $\beta$ with a latent binary Ising prior $\gamma \sim \Ising(a,b)$. The corresponding prior for $\beta$ is given as
\begin{align*}
\left. \beta_l \middle| \beta_{\delta(l)}, \gamma_l, \sigma^2_\beta \right. & \sim 
\begin{cases}
\delta(0) & \gamma_l = 0\\
\N \left(\frac{1}{d_l} \sum_{j \sim l} \beta_j, \frac{\sigma^2_\beta}{d_l} \right)
& \gamma_l = 1
\end{cases}
\end{align*}
with $\delta(0)$ the Dirac measure at $0$. This model has an additional level $\gamma$ in the hierarchical Bayesian model structure. Depending on the value of the Ising field $\gamma$ in a pixel $l$, the corresponding $\beta$ coefficient is either set to $0$ (if $\gamma_l = 0$, pixel is not selected) or follows the GMRF prior distribution (if $\gamma_l = 1$, i.e.\  pixel is selected). Samples from the joint full conditional of $\beta$ and $\gamma$ can be obtained by single-site Gibbs sampling \cite{GoldsmithEtAl:2014}. The authors propose to choose the  hyperparameters $\sigma_\eps^2, \sigma_\beta^2, a$ and $b$ via cross-validation with extremely short MCMC chains (e.g.\  $250$ iterations). 

This model assumes the true $\beta$ image to be sparse with a few coherent smooth areas of non-zero pixels, which is modelled by a combination of a GMRF and a latent Ising field. It is particularly designed for neuroimaging data, where the assumptions translate to most image locations not being predictive for the response, while in the relevant parts of the brain neighbouring pixels  have similar effects \cite{GoldsmithEtAl:2014}.

\subsection{Implementations}
\label{sec:modelImplement}

For most of the considered approaches, software implementations are available in existing \texttt{R}-packages or easily made available. This was one of the inclusion criteria. 

The spline regression model (\textit{Splines}) can be fit using the \texttt{gam} function for generalized additive models in the \texttt{R}-package \texttt{mgcv} \cite{Wood:2011, mgcv}. The implementation is very flexible and can handle 2D, 3D or even higher dimensional data and many different basis functions. 

All wavelet-based approaches (\textit{WCR}, \textit{WPLS} and \textit{WNET}) are implemented in the \texttt{refund.wave} package \cite{refund.wave} (\textit{WCR} and \textit{WPLS} in \texttt{wcr} and \textit{WNET} in \texttt{wnet}), heavily building on the \texttt{wavethresh} package \cite{wavethresh} for calculating the transformations to the wavelet space. They all can handle 2D and 3D images with the restriction that the sidelength of the images must be the same power of $2$ for all dimensions.
 For \textit{WNET}, the \texttt{glmnet} package \cite{glmnet} is used for the elastic net part. 
 
The principal component regression approach based on splines (\textit{FPCR}) is available  in the function \texttt{fpcr} in the related package \texttt{refund} \cite{refund}. The implementation currently accepts only 2D images, but without restrictions on the sidelengths of the images.

For the calculation of the eigenimages in \textit{PCR2D} we use the implementation of the rank-one based approach \cite{Allen:2013} in the \texttt{MFPCA} package \cite{MFPCA} which at present works only for 2D images.  The reconstruction of the coefficient image $\hat \beta$ using the estimated eigenimages and the regression coefficients can easily be done using the \texttt{expandBasisFunction} method in \texttt{MFPCA}.

For \textit{SparseGMRF}, an \texttt{R} implementation of the Gibbs sampler in the supplementary files or the original article \cite{GoldsmithEtAl:2014}. For reasons of performance we (re-)implemented the Gibbs samplers in \textit{SparseGMRF} and \textit{GMRF} in \texttt{C}. The code is provided in the code supplement including an \texttt{R} interface.  Both models are currently implemented only for 2D images, but can easily be extended to the 3D case by properly defining the neighbourhood structure.
Usage examples for all methods are given in the code supplement of this article (\url{https://github.com/ClaraHapp/SOIR}).

\section{Discussion and Measures for Model Assumptions}
\label{sec:MeasureAssumptions}

As discussed in Section~\ref{sec:SOIRmodel}, the scalar-on-image regression model~\eqref{eq:ModelEquation} in general is not identifiable, as the total number of model coefficients in most applications exceeds the number of observation units. All proposed models therefore need to make structural assumptions on $\beta$ to overcome the issue of non-identifiability and make  estimation possible. However, ``we buy information with assumptions'' \cite{Coombs:1964}. This comes at the price that the estimate found by a certain model contains not only information from the data, but also from the model assumptions. It is hence important to be aware of the assumptions made and to understand how they influence the estimate.

\subsection{Underlying and Parametric Model Assumptions}

In the following, we distinguish between underlying and parametric model assumptions. Conceptually, this reflects two different steps of the modelling process: The underlying assumptions describe the fundamental model assumptions, such as smoothness or sparsity, and the general class of coefficient images that a model can handle, e.g.\  linear combinations of splines or wavelets.
The parametric model assumptions reflect model-specific restrictions of the parameters in the estimation process, in terms of penalties or variable selection, in order to achieve the goals of the underlying assumptions. Although there may be some overlap between the two concepts, we believe that this distinction can be helpful for comparing and categorizing the models.

For the discussed models, the underlying model assumptions can be broadly  divided into three categories (cf. Table~\ref{tab:impAss}). They are 1. smoothness, which we interpret as neighbouring pixels having similar values, 2. sparsity, meaning that a few coefficients dominate all others and 3. projection, which reflects the assumption that the coefficient image  can be expanded in  given basis functions.
While the latter is often motivated by technical considerations, sparsity and smoothness are very plausible assumptions e.g.\ in neuroimaging: If a certain number of pixels in a particular brain region shows a strong relationship with the response, one can expect that this effect carries over to the whole region rather than affecting only single, disconnected pixels. On the other hand, it is also likely that not all parts and structures of the brain or all coefficients of basis functions are equally associated with the response and some may even be completely omitted from the model by setting their coefficients to zero. 

The underlying smoothness assumption translates to parametric assumptions in terms of penalties on the coefficients (\textit{Splines, FPCR}) or priors  (\textit{GMRF, SparseGMRF}). Both enforce similarity among neighbours. 
In Bayesian models (\textit{GMRF, SparseGMRF}) smoothness is further affected by assumptions on the prior variance  $\sigma^2_\beta$, which controls the variability in $\beta$.
Sparsity is achieved by variable selection methods (\textit{WNET, SparseGMRF}) or restrictions on e.g.\ the number of principal components to include in the model (\textit{FPCR, PCR2D, WCR, WPLS}) on the level of voxels (\textit{SparseGMRF}), basis functions (\textit{FPCR, PCR2D}) or wavelet coefficients (\textit{WCR, WPLS, WNET}).
This shows that a global, underlying assumption can be translated into different parametric assumptions, which might result in quite different coefficient images $\hat \beta$.
In other words, different parametric assumptions can aim at the same underlying assumption, but they achieve it in very different ways and not necessarily equally well.
If the true coefficient image does not fulfill the assumptions, e.g.\ does not fall into the space spanned by the basis functions, then the estimate will be a (biased) approximation to $\beta$ in the given space. This bias cannot be detected from in-sample prediction error due to the non-identifiability of $\beta$, as different estimates $\hat \beta$ can give equally good predictions.

In order to understand how strongly the assumptions affect the estimate, we develop measures that quantify how well the underlying and parametric model assumptions are met.

\begin{table}
\center
\caption{Underlying model assumptions for the considered models.
The order of the models has been slightly rearranged with respect to the presentation in Section~\ref{sec:methods} according to their assumptions.}
\label{tab:impAss}
\footnotesize
\begin{tabular}{lccc}
\toprule
Method &  Smoothness & Sparsity & Projection  \\
\cmidrule{1-1} \cmidrule{2-4} 
\textit{Splines} &  image & - & spline basis  \\
\textit{FPCR}& image & PCs of $XB$ & spline basis \\
\textit{PCR2D} & - & PCs of images & PCs of images   \\
\textit{WCR} & - & wavelet coefficients & wavelet space \\
\textit{WPLS} & - & wavelet coefficients & wavelet space  \\
\textit{WNET} & - & wavelet coefficients & wavelet space   \\
\textit{SparseGMRF}& image & pixels  & -  \\
\textit{GMRF} & image & - & -  \\
\bottomrule
\end{tabular}
\end{table}


\subsection{Measures for Quantifying the Impact of Model Assumptions}
\label{sec:measureDefs}

For better comparability, all following measures are constructed such that they take values between $0$ and $1$ with $0$ meaning that the model assumptions are perfectly met and $1$ meaning that the assumptions are not met at all.

\bigskip

\textbf{Smoothness:} Smoothness is interpreted as neighbouring pixels having similar values. The sum of squared differences between neighbours can thus be used as a measure of smoothness. For $\beta \in \IR^L$ with a given neighbourhood structure, a natural smoothness measure is $\sum_{i \sim j} (\beta_i - \beta_j)^2 = \beta \trans P \beta$ 
for the symmetric and positive semidefinite neighbourhood matrix $P \in \IR^{L \times L}$ (see GMRF in Section~\ref{sec:RFmethods}).
By the theorem of Rayleigh-Ritz \cite{HornJohnson:1985} and as the smallest eigenvalue of $P$ is $0$, we have that the smoothness measure
\[m_\text{Smoothness}(\beta) = \frac{\beta \trans P \beta}{\lambda_\text{max}(P) \beta \trans \beta},\]
with $\lambda_\text{max}(P)$ the maximal eigenvalue of $P$ lies between $0$ (constant, i.e.\  extremely smooth image) and 1 (extremely nonsmooth images).

This measure can be used to assess the smoothness of an image as an underlying model assumption and therefore can be compared over all models assuming smoothness. It can also measure the degree to which the parametric smoothness assumptions made in the GMRF based models are met. For the approaches using splines, $\beta$ can be replaced by the vector of spline coefficients $(b_1 \usw b_K)$ and $P \in \IR^{K \times K}$ by the associated penalty matrix to measure the parametric model assumptions.
 
 \bigskip
\textbf{Sparsity:} The Gini index $  G(\beta) = 1 - 2 \sum_{l= 1}^L \frac{\beta_{(l)}}{\Vert \beta \Vert_1} \left( \frac{	L - l+ \frac{1}{2}}{L} \right )$ is  a reasonable measure for sparsity of an image $\beta \in \IR^L \setminus \{0\}$ \cite{HurleyRickard:2009}. Here $\beta_{(1)} \leq \beta_{(2)} \leq \ldots \leq \beta_{(L)}$ denotes the ordered values of $\left | \beta_l \right|,~ l = 1\usw L$ and $\Vert \beta \Vert_1 =\sum_{i = l}^L \left | \beta_l \right|$. We define
\[m_\text{Sparsity}(\beta) = 1 - G(\beta)\]
with 
$m_\text{Sparsity}(\beta) = 0$ for complete inequality of $\beta$ across all pixels (very sparse case) and $m_\text{Sparsity}(\beta)  = 1$ indicating complete equality of $ \beta$ across all entries (non-sparse case).
This measure can also be applied to a coefficient vector $b = (b_1 \usw b_K)$, e.g.\  of wavelet coefficients to measure the underlying sparsity assumption in \textit{WCR, WPLS} and \textit{WNET} and therefore allows a comparison between these methods.

Parametric sparsity assumptions are in general implemented by variable selection methods. A sparsity measure for a  coefficient vector $b \in \IR^K$ is hence given by the proportion of non-zero coefficients in $b$:
\[m_\text{Selection}(b) = \frac{\#\{k = 1 \usw K \colon b_k \neq 0\}}{K}.\]
Values close to $0$ indicate extreme sparsity ($b \equiv 0$), a value of 1 means no sparsity. The sparse GMRF approach \cite{GoldsmithEtAl:2014} assumes sparsity on the pixel level, i.e.\  here one can apply $m_\text{Selection}$ to the vectorized posterior mean of the Ising field $\gamma$, thresholded at $0.5$. As $m_\text{Selection}(b)$ is based on the model-specific coefficients, it should not be compared across models.

\bigskip
\textbf{Projection:}
 Basis function approaches assume that the function $\beta(\cdot)$ generating the coefficient image lies in the span of some predefined basis functions $B_1 \usw B_K$, which can be splines, wavelets or principal component functions.
A suitable measure for this assumption is
\[m_\text{Projection}(\beta) = \frac{\Vert P^\perp \beta \Vert^2}{\Vert \beta \Vert^2} =  1 - \frac{\Vert P \beta \Vert^2}{\Vert \beta \Vert^2} \]
with $P \beta$ the orthogonal projection of $\beta$ onto the space spanned by $B_1 \usw B_K$, $P \orth \beta$ the projection onto the orthogonal complement of that space and $\Vert \beta \Vert^2 = \sum_{l = 1}^L \beta_l^2$. A value of $1$ means that $\beta$ lies completely in the orthogonal complement of the basis functions and $m_\text{Projection}(\beta) = 0$, if $\beta$ is indeed a linear combination of the basis functions. 

For a given model, the estimate $\hat \beta$ will always lie in the given model class. However, we will use this measure for underlying assumptions for the true $\beta$ in the simulation in Section~\ref{sec:sim}.

\bigskip

\textbf{Prior variability:}
We use the Kullback-Leibler divergence between the (conditional) prior of a parameter and its (full conditional) posterior as a measure for the prior impact. A similar approach using the full prior and posterior densities has been proposed in the literature \cite{IttiBaldi:2005}.

For the \textit{GMRF} model, we choose the full conditional $\IG(a_\text{post}, b_\text{post})$ for $\sigma^2_\beta$ as the reference and calculate the Kullback-Leibler divergence $D$ to the prior $\IG(a_\text{pri}, b_\text{pri})$.
For \textit{SparseGMRF}, the authors propose to choose $\sigma_\beta^2$ via cross-validation \cite{GoldsmithEtAl:2014}. This can be interpreted as a discrete uniform prior on the set of possible values $\{\sigma^2_1 \usw \sigma^2_K\}$ for $\sigma^2_\beta$ and the full conditional  is a point measure on the optimal value $\sigma^2_\ast$ found by cross-validation. The Kullback-Leibler divergence is $D =  \log(K)$, hence grows logarithmically with the number of possible values for $\sigma^2_\beta$. 
We divide $D$ by $10$ for numerical reasons and transform the result to $[0,1]$, giving as measure for the impact of the prior variability
\[m_\text{Prior}(\beta) = 1 - \exp(-D/10).\]
Values close to $1$ correspond to $D \to \infty$, i.e.\  situations where the information from the prior has little influence on the full conditional. In this case, model assumptions will in general not be met for the full conditional.
By contrast, values close to $0$ correspond to $D \approx 0$. Here prior and full conditional are very similar, meaning that the full conditional is close to the prior assumptions. Prior assumptions are clearly model-specific. Therefore, this measure should only be compared within a model class.

\bigskip

Interpreting the measures for an estimated coefficient image might be nontrivial in practical applications. We therefore propose to create artificial, hypothetic coefficient images, that can for example be motivated by the question of interest. The values for the estimated coefficient image can then be compared to the corresponding values for the hypothetic images to reveal e.g.\ differences in smoothness or sparsity.
In addition, the hypothetic images can be used to generate new response values by combining them with the real data, as done in the following simulation study. Smoothness or sparsity measures for estimates obtained from this simulated data can also serve as reference values for interpreting the measures, as illustrated in Section~\ref{sec:app}.

\section{Simulation Study}
\label{sec:simStudy}

In this section, the performance of different scalar-on-image regression approaches is analyzed for various coefficient images $\beta$, reflecting the assumptions in the different models, and using real data from the Alzheimer's Disease Neuroimaging Initiative study (ADNI \cite{WeinerEtAl:2015}) that are also considered in the application in Section~\ref{sec:app}. On the one hand, this ensures that the image covariates have a realistic degree of complexity \cite{ReissOgden:2007, ReissOgden:2010}. On the other hand, this allows to study systematically which kind of features in the coefficient images can be found with the data at hand and how this translates to the measures proposed in the previous section.

\subsection{Simulation Settings}
\begin{figure}[ht]
\centering
\includegraphics[width = 0.65\textwidth]{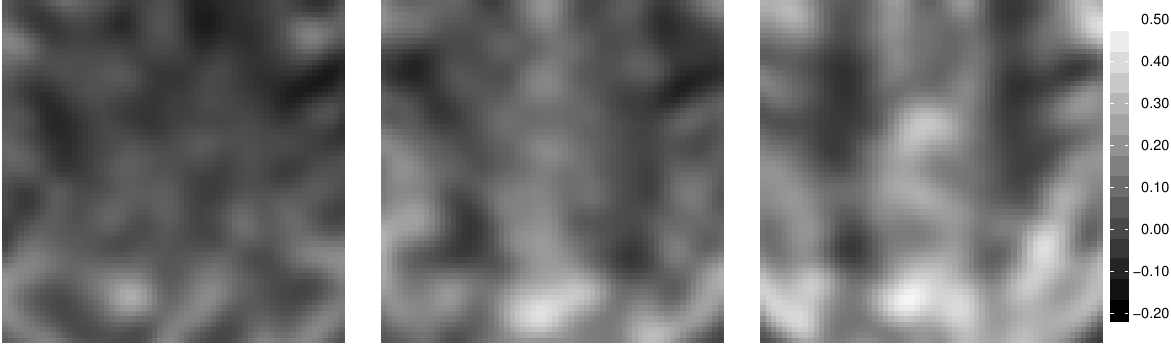}
\caption{The first three image covariates $x_1, x_2, x_3$ after demeaning.}
\label{fig:imageCovariates}
\end{figure}

The image covariates stem from FDG-PET scans, which measure the glucose uptake in the brain. The original scans were co-registered to simultaneously measured MRI scans in order to reduce registration effects \cite{AraqueCaballeroEtAl:2015}.
We use $64 \times 64$  subimages of the first  $N = 250$ or $N = 500$ images in the original data as covariates $x_1 \usw x_N$. Three example images are shown in  Fig.~\ref{fig:imageCovariates}. The image size is determined by the wavelet-based methods, which require the sidelength of the images to be a power of 2.
The demeaned images take values between $-1$ and $1.24$.

\begin{figure}
\centering
\includegraphics[width = .85\textwidth]{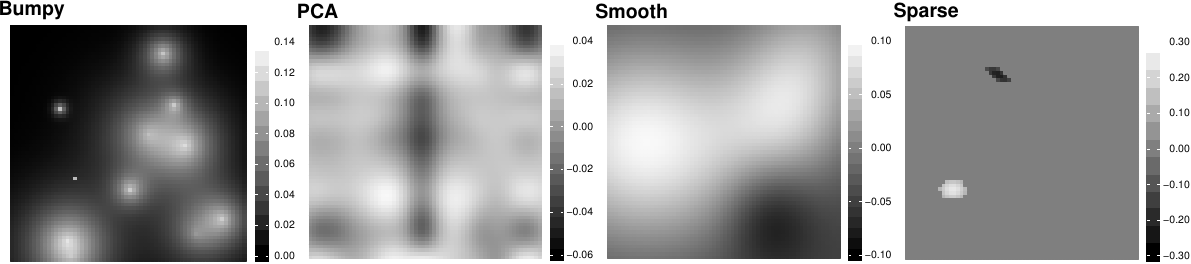}
\caption{Coefficient images $\beta$ used for the simulation. From left to right: 
\textit{bumpy}, \textit{pca} (based on the first $N = 250$ images in the dataset), \textit{smooth} and \textit{sparse}. Note the individual scale for each image.}
\label{fig:betaFuns}
\end{figure}

We consider four different coefficient images that reflect the main assumptions in the models (see Fig.~\ref{fig:betaFuns}):
\begin{itemize}
\item \textit{bumpy} \cite{ReissEtAl:2015}, an image with some high-peaked, clearly defined ``bumps''. It is a two-dimensional version of the \textit{bump} function \cite{DonohoJohnstone:1994} which has become a common benchmark for one-dimensional wavelet models. It is thus expected that the wavelet-based methods should be the most suitable ones for estimation.
\item \textit{pca}, an image constructed as a linear combination of the first $K=5$ principal components of the image covariates $x_1 \usw x_N$ found by the rank-one based method \cite{Allen:2013} with coefficients $b_k = (-1)^k \exp(-\frac{k}{5}),~ k = 1 \usw 5$. Obviously, the principal component based method should work very well in this case.
\item \textit{smooth}, a smooth image which corresponds to the smoothness assumption made in the spline-based models and for the Bayesian models using Gaussian Markov random fields. It is constructed as a mixture of three 2D normal densities.
\item \textit{sparse}, an image that is mostly zero with two small, smooth spikes \cite{GoldsmithEtAl:2014}. This image corresponds to the assumption made for the \textit{SparseGMRF} model. 
\end{itemize}

The ADNI roster IDs (RID) of the subjects whose data was used for this simulation together with code for reproducing the coefficient images can be found in the code supplement (\url{https://github.com/ClaraHapp/SOIR}).

The response is constructed as
$y_i = \alpha + \sum_{l = 1}^L x_{i,l} \beta_{l} + \eps_i$ for each $i = 1\usw N$
with $\alpha = -1$ as intercept, a total number of $L = 64^2 = 4096$ pixels and $\eps_i$ chosen such that the signal-to-noise ratio
\[
\text{SNR} = \frac{\widehat{\text{sd}}(\sum_{l =1}^L x_{i,l} \beta_l)}{\text{sd}(\eps_i)}
\]
is either equal to $4$ or to $1$ \cite{GoldsmithEtAl:2014} which corresponds to $R^2 = 0.94$ and $0.5$ \cite{ReissOgden:2007}.

All eight models presented in Section~\ref{sec:methods} are considered in the simulation study. As the \textit{GMRF} model with a commonly used $\IG(1,1)$ prior for the variance parameters $\sigma_\eps^2, \sigma_\beta^2$, which is considered to be rather uninformative, performed poorly, we added another model, \textit{GMRF2}, with a highly informative $\IG(10,10^{-3})$ prior  (prior mean: $10^{-3}$, prior variance: $10^{-9}$) for the variance parameters. The detailed settings for each of the models are given in Section~7.1 in the appendix. 
In total, the simulation study comprises nine different models, four coefficient images and two sample sizes and signal-to-noise ratios, each. For each setting, the simulation and analysis is repeated 100 times.
A sensitivity study with varying coefficient images gave similar results, showing that the spatial distribution of features in the coefficient images has only a marginal impact on the results (see appendix, Section~7.3).

The resulting estimates $\hat \beta$ and the fitted values
$\hat y_i = \hat \alpha + \sum_{l = 1}^L x_{i,l} \hat \beta_{l}$ for each $i =1 \usw N$
are evaluated with respect to the relative estimation accuracy and the relative (in-sample) prediction error
\[
\frac{\sum_{i = l}^L (\beta_l - \hat \beta_l)^2}{\sum_{i = l}^L (\beta_l - \bar \beta)^2}
\qquad \text{and}
\qquad
\frac{\sum_{i = 1}^n (y_i - \hat y_i)^2}{\sum_{i = 1}^n (y_i - \bar y)^2 }
\]
with $\bar \beta = \frac{1}{L} \sum_{l = 1}^L \beta_l$ and $\bar y = \frac{1}{N}\sum_{i = 1}^N y_i$.
Taking the relative errors allows to compare the results across coefficient images $\beta$ and datasets $\{(x_i, y_i),~ i = 1 \usw N\}$ generated in different iterations of the study. A relative estimation error of $1$ means that the estimated coefficient image gives equally good results as a constant image, taking the average value of the true $\beta$ in each pixel. This corresponds to a simpler model in the mean of the image covariate over pixels. Analogously, a relative prediction error of $1$ means that the prediction is comparable to a simple intercept model, not taking the image information into account. Relative errors above $1$ therefore are indicators for poor performance. In addition, the measures for underlying  and parametric model assumptions from Section~\ref{sec:MeasureAssumptions} are calculated for each estimate $\hat \beta$ and -- for the underlying assumptions -- compared with those of the true images. As computation time plays an important role for the practical usability of the models, it is also recorded.

\FloatBarrier
\subsection{Results}
\label{sec:sim}

\begin{figure}
\centering
\includegraphics[width = 0.9\textwidth]{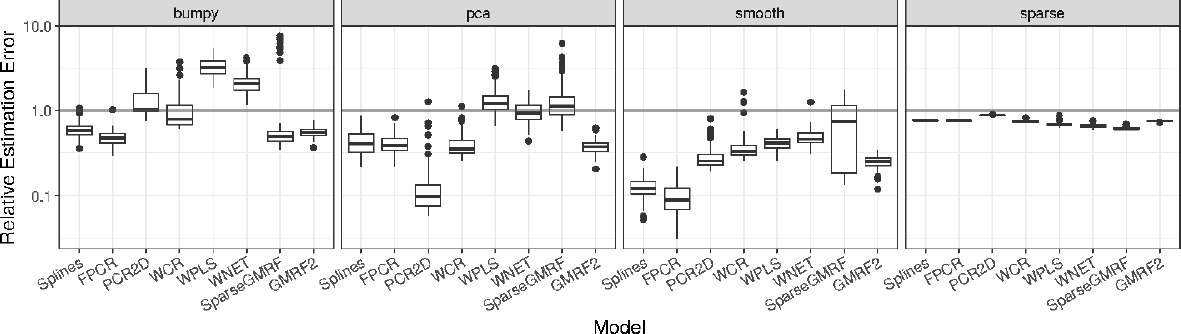}

\caption{Relative estimation errors for $N = 250$ observations and $\SNR = 4$ over all $100$ simulation runs.  
Boxplots show the errors for all models except \textit{GMRF} depending on the true coefficient image.
Gray horizontal lines mark $1$, which corresponds to a constant coefficient image, having the average value of the true $\beta$ image.}
\label{fig:simRes250}
\end{figure}

\begin{figure}[ht]
\centering
\includegraphics[width = 0.9\textwidth]{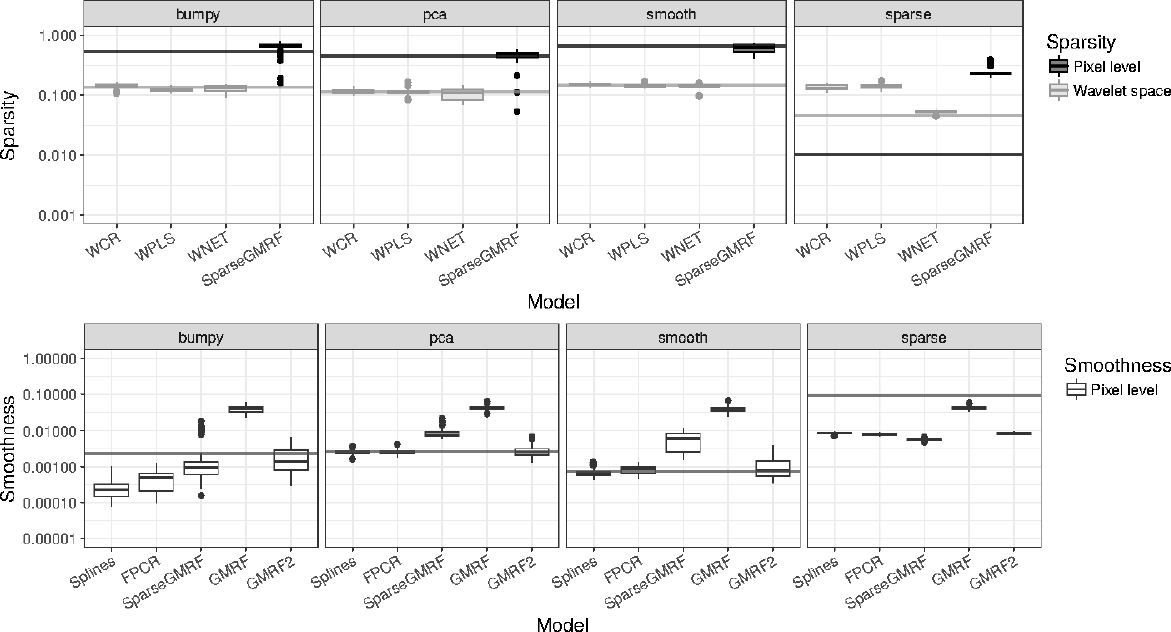}

\caption{Measures for underlying model assumptions in the simulation for $N = 250$  observations and $\SNR = 4$ over all $100$ simulation runs. Boxplots show the measures for the different models depending on the true coefficient image. All values on log-scale. Gray horizontal lines correspond to the values for the true coefficient images.}
\label{fig:simImplicitAss250}
\end{figure}

\begin{figure}[ht]
\centering
\includegraphics[width = \textwidth]{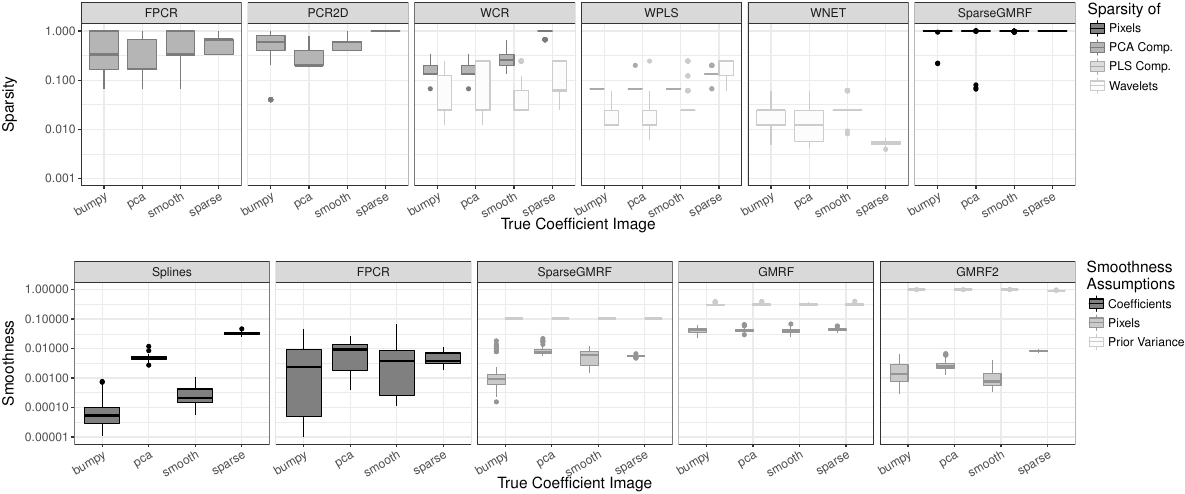}

\caption{Measures for parametric model assumptions in the simulation for $N = 250$ observations and $\SNR = 4$ over all $100$ simulation runs. Boxplots show the measures for the different coefficient images depending on the model. All values on log-scale.}
\label{fig:simExplicitAss250}
\end{figure}

\begin{figure}[t]
\centering
\includegraphics[width = \textwidth]{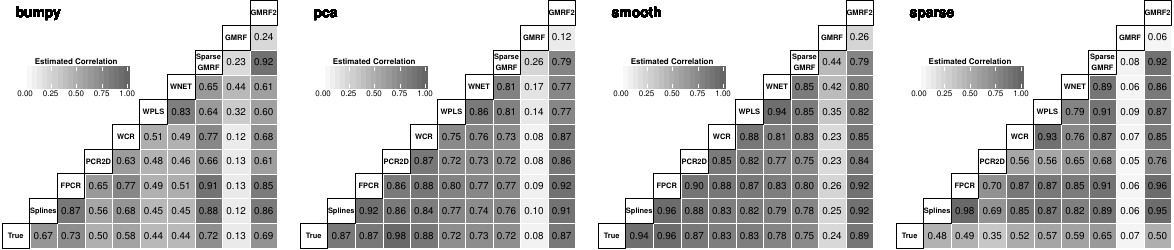}
\caption{Median correlation between the true coefficient images and the estimates for $N = 250$ observations and $\SNR = 4$ over all $100$ simulation runs. The figures show the median  correlation of the vectorized images depending on the true images and the models.}
\label{fig:simBetaCorr250}
\end{figure}

Figures~\ref{fig:simRes250} to~\ref{fig:simBetaCorr250} show the results of the simulation study for $N = 250$ and $\SNR = 4$. The settings for $N = 500$ and/or $\SNR = 1$ gave similar results, which are shown in the online appendix together with example plots and predictions from all models in the $N = 250$/$\SNR = 4$ setting.

Overall, \textit{GMRF} gives very poor results with extremely high estimation errors (median: $63.59$, sd: $68.58$ for $N = 250$/$\SNR = 4$) and above average prediction errors (median: $0.71$, sd: $64.03$ for $N = 250$/$\SNR = 4)$. As \textit{GMRF2} performs reasonably well, this indicates that the choice of the prior for the variance parameters in Bayesian models matters and highly informative priors are required in this case. The \textit{GMRF} model is therefore not considered in the following analysis.

The predictive accuracy for the different coefficient images is rather constant over all models with values close to $0.05$ (cf.\ Fig.~8 in the appendix), i.e. the models clearly perform better than the intercept model. For $\SNR = 1$, the errors increase to values around $0.5$ for all models.  If the focus is only on prediction, the different models and their assumptions hence lead to equally good results. The scalar-on-image regression model, however, also aims  at an interpretable coefficient image $\hat\beta$, showing how the observed image covariates $x_i$ influence the response $y_i$. The relative estimation error is thus of greater importance for assessing a model's ability of giving interpretable results.

As seen in Fig.~\ref{fig:simRes250}, relative estimation errors can take values close to or considerably above $1$, indicating poor results, even in the idealistic case of $\SNR = 4$. In total, the error rates are lowest for \textit{smooth}, meaning that this coefficient image is captured best. Except for \textit{sparse}, Fig.~\ref{fig:simRes250} shows a lot of variation between the models, i.e.\ the different model assumptions indeed lead to quite different estimates with different quality.
The best results are found for \textit{pca} with \textit{PCR2D} estimation and \textit{smooth} when estimated by \textit{Splines} or \textit{FPCR}, hence settings in which the true coefficient image meets the models assumptions very well. This is also seen in the low corresponding measures for the underlying model assumptions, c.f.\  $m_\text{Projection}$ in Table~\ref{tab:projBeta} (relevant for \textit{pca/PCR2D}) and $m_\text{Smoothness}$ in Fig.~\ref{fig:simImplicitAss250} (for the \textit{smooth} image when estimated by \textit{Splines} or \textit{FPCR}). Notably, the measures for \textit{SparseGMRF} and \textit{GMRF2}, which assume smoothness on a pixel level, are somewhat higher, which leads to somewhat worse estimation results (cf.\ Fig.~\ref{fig:simRes250}).

\begin{table}[ht]
\caption{Values of $m_\text{Projection}$ for the different coefficient images depending on the basis functions.}
\label{tab:projBeta}
\center
\footnotesize
\begin{tabular}{lrrrr}
\toprule
& \multicolumn{4}{c}{Coefficient Image}  \\
  \cmidrule(lr){2-5} 
Projection on &  \multicolumn{1}{c}{\textit{bumpy}} & \multicolumn{1}{c}{\textit{pca}} & \multicolumn{1}{c}{\textit{smooth}} & \multicolumn{1}{c}{\textit{sparse}} \\
  \cmidrule(lr){1-1} \cmidrule(lr){2-5}
PCs & $ 1.22 \cdot 10^{-01} $ & $ 3.65 \cdot 10^{-30} $ & $ 7.05 \cdot 10^{-02} $ &$ 8.54 \cdot 10^{-01} $ \\
splines & $ 4.16 \cdot 10^{-03} $ & $ 3.79 \cdot 10^{-04} $ & $ 3.22 \cdot 10^{-08} $ & $ 5.03 \cdot 10^{-01} $ \\
wavelets & $ 1.73 \cdot 10^{-18} $ & $ 2.43 \cdot 10^{-18} $ & $ 2.36 \cdot 10^{-18} $ & $ 4.39 \cdot 10^{-18} $ \\
\bottomrule
\end{tabular}
\end{table}

While for \textit{pca} and \textit{smooth} the outcomes are mostly as expected in terms of best-performing methods,  the result  for \textit{bumpy} is more surprising. One would expect the wavelet-based methods to perform best \cite{ReissEtAl:2015}. By contrast, all wavelet methods are outperformed by \textit{Splines}, \textit{FPCR} and \textit{GMRF2}, hence methods that assume smoothness. This, however, is in line with previous results who found that wavelet methods did not clearly outperform non-wavelet methods for the \textit{bumpy} coefficient image \cite{ReissEtAl:2015}. The measures for the underlying model assumptions give an explanation for this result: The \textit{bumpy} image can be perfectly projected into the wavelet space, just as all coefficient images can (see Table~\ref{tab:projBeta}), but has a similar sparsity in the wavelet space as \textit{pca} and \textit{smooth} (cf.\ Fig.~\ref{fig:simImplicitAss250}), meaning that  the sparsity assumption in \textit{WCR},  \textit{WPLS} and \textit{WNET} has no advantage for the estimation. The underlying smoothness measures for \textit{Splines}, \textit{FPCR} and \textit{GMRF2} in Fig.~\ref{fig:simImplicitAss250} show that the resulting estimates are a bit too smooth, but they still yield better results than the wavelet-based methods. 

The \textit{sparse} coefficient image is the most difficult to estimate, as it has two rather spiky features and the rest of the image is equal to zero.
Indeed, all models have relative estimation errors close to $1$, which means that the methods perform similarly as a simpler model, taking the average value of $\beta$ as a constant coefficient. In the case of \textit{sparse}, the average is close to $0$, hence the simpler model corresponds to a pure intercept model, without the image. 
Contrary to expectation, the \textit{SparseGMRF} model does not clearly outperform the other models, although it involves a pixelwise variable selection step and hence the possibility to set entire areas of the image to zero.
The model measures for parametric assumptions in Fig.~\ref{fig:simExplicitAss250} show that it produces estimates that are quite smooth, but completely non-sparse. This means that the sparsity assumption is more or less ignored in the estimation process. \textit{SparseGMRF} hence behaves just as a non-sparse GMRF with the variance parameters chosen via cross-validation. This corresponds to a discrete prior with three values for each parameter, hence a highly informative prior.

In order to check the agreement of the estimated coefficient images among each other and with the true $\beta$, correlations of the vectorized images were calculated, which are given in Fig.~\ref{fig:simBetaCorr250}.
Notably, for \textit{sparse}, all estimated coefficient images show medium to high correlation among themselves, but only moderate correlation with the true coefficient image. Moreover, the correlation between \textit{GMRF} and all other models as well as the true coefficient image is rather weak, which reflects the extremely poor results for this model.

In summary, the simulation shows that the assumptions made in the different models can lead to quite different results in terms of estimation accuracy, depending on the structure of the true  coefficient image. At the same time, the predictive performance is quite similar over all models. This is a clear sign of non-identifiability, putting the interpretability of the estimates into question. 
For a higher SNR, the relative errors for estimation and prediction generally decrease. However, the methods still result in relatively high error rates for \textit{bumpy} and \textit{sparse}, coefficient images with highly localized features.
In an overall comparison of the models  in this simulation \textit{FPCR} seems to give the best results, as it is always among the best two models in terms of estimation accuracy. Moreover, it is also by far the model with the shortest computation time (see Fig.~29 in the appendix).  In particular, the combination of a spline basis representation and a principal component analysis for $XB$ appears to be advantageous compared to the pure \textit{Splines} models. Similarly, \textit{WCR} performs better than the other wavelet basis methods in all settings considered in this study. As expected, \textit{PCR2D} clearly outperforms all other methods for the \textit{pca} coefficient image, which perfectly meets the assumptions made in this model. For all other coefficient images, the method gives intermediate results. Finally, for the GMRF based models, the highly informative \textit{GMRF2} model performs best, followed by \textit{SparseGMRF}. The latter, however, does not make use of the integrated variable selection. In addition, it is computationally very demanding, as it required around $85\%$ of the total computation time of the study, although a relatively simple setting was chosen with only three possible values for each hyperparameter.

\FloatBarrier

\section{Application}
\label{sec:app}

In this section, the scalar-on-image regression models are applied to data from the ADNI study \cite{WeinerEtAl:2015} to illustrate the impact that model assumptions can have in practice. Moreover, we show how the simulation results from Section~\ref{sec:simStudy} can be used as reference for interpreting the measures introduced in Section~\ref{sec:MeasureAssumptions}.
We use data from $N = 754$ subjects in the study having an FDG-PET scan and an MRI scan at baseline. Their roster IDs are given in the code supplement of this article. The aim is to find a relation between a neuropsychological test (Alzheimer's disease assessment scale - cognitive subscale, ADAS-Cog \cite{ADASCog}) at baseline and the FDG-PET scans, which were co-registered to the MRI scans in order to reduce registration effects.
ADAS-Cog is a current standard for diagnosing Alzheimer's disease (AD). It takes values between $0$ and $70$, where higher values indicate worse global cognition and thus a higher risk of AD. In order to obtain approximate residual normality, the ADAS-Cog values were square root transformed before the analysis (see Fig.~36 in the appendix). 
For FDG-PET, which reflects the neural integrity of the brain, we use the same $64 \times 64$ subimages as in the simulation study for all $N = 754$ subjects. As additional covariates, we consider  age at baseline, gender and years of education, which are known risk factors for AD.

\begin{figure}[ht!]
\centering
\includegraphics[width = \textwidth]{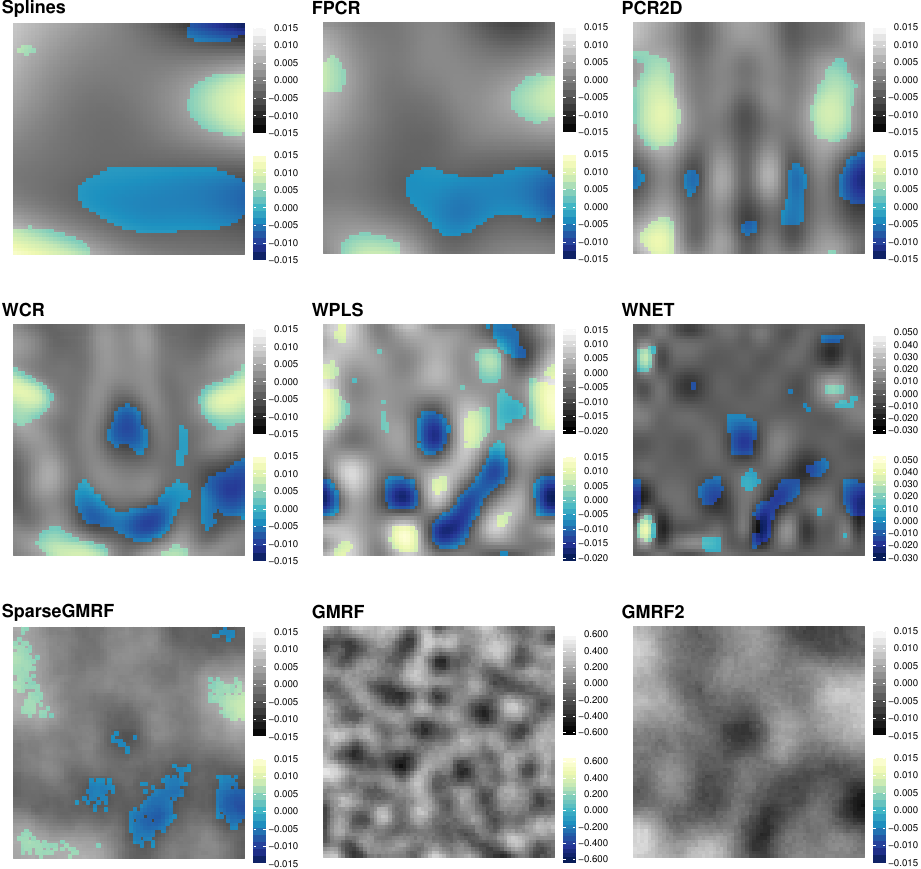}
\caption{Coefficient image estimates for the application with pointwise $95\%$ confidence bands/credible intervals. Coloured pixels correspond to ``significant'' pixels, i.e.\ the confidence band/credible interval in this pixel does not contain zero. Note that some pixels too many might be flagged, as the confidence bands do not account for e.g.\  variable selection via cross-validation and are mostly calculated on a pointwise basis.  Note also the different scales for \textit{WPLS}, \textit{WNET} and \textit{GMRF}.}
\label{fig:appBetaCI}
\end{figure}

For each model, we calculate estimates $\hat \alpha$ for the scalar covariates and $\hat \beta$ for the image covariates together with pointwise $95\%$ credibility/confidence bands to illustrate the variability in the estimates. The results are shown in Fig.~\ref{fig:appBetaCI} for $\hat \beta$ and in Fig.~37 for $\hat \alpha$. Details on the calculation of the credibility/confidence bands for the different models are given in Section~8.1 in the appendix.
All measures for underlying and parameteric assumptions for each model are found in Table~3 in the appendix together with an illustration of the goodness of fit (Fig.~36). 

All models have a very similar predictive performance, except for \textit{GMRF}, which is in line with the results from our simulation study. The prediction error of around $0.6$ is slightly higher than in the simulation setting with $\SNR = 1$.

The estimated coefficient images in Fig.~\ref{fig:appBetaCI} have some common features in the ``significant'' parts, i.e.\ pixels where the confidence bands do not contain zero, which we can interpret as data-driven:
Regions with negative values are found especially in the center, in the area of the cingulate gyrus, which is related to emotion processing, learning and memory, \cite{EncyclopediaNeuroscience} and in the lower right part of the images, that can be identified with the supramarginal gyrus and thus language processing \cite{Petrides:2014}. Below-average FDG-PET values in these areas would be associated with a higher risk of AD, which is in line with the literature on early AD \cite{MinoshimaEtAl:1997}.
Most estimated coefficient images further have positive ``bumps'' in the upper left and right part of the images, which correspond approximately to the precentral gyri and thus the primary motor cortex \cite{EncyclopediaNeuroscience}. Positive values are also found in the lower left corner in the area of the right angular gyrus, which links visual impressions to the corresponding notions \cite{EncyclopediaNeuroscience}. Above-average FDG-PET values in these areas would be associated with higher ADAS-Cog values, and thus a higher risk of Alzheimer's disease, which is more difficult to interpret.
However, the influence of the model assumptions is clearly seen and the percentage of the flagged pixels, their location and the shape of the ``significant'' regions differ considerably among the methods.

The estimated coefficient images for \textit{FPCR} and  \textit{Splines} are quite smooth, which is reflected in the rather  low values for the underlying smoothness assumption (0.002, comparable to the true \textit{pca} and \textit{bumpy} images in the simulation with $N = 500$ and $\SNR =1$). The pixels flagged as ``significant'' by both models form mostly large, round shaped areas.
\textit{GMRF2} and \textit{SparseGMRF} induce smoothness on the pixel level rather than for basis function coefficients. The estimates resulting from these methods show more fine-scale structure, which manifests in somewhat higher measures for the underlying smoothness assumption (0.007 for \textit{SparseGMRF}, 0.010 for \textit{GMRF2}) and is in line with the results from the simulation. The parametric sparsity measure for \textit{SparseGMRF} is equal to $1$, meaning that sparsity is not achieved, as it was the case in the simulation study. However, when comparing the estimates, the image produced by \textit{SparseGMRF} seems to be a bit blurred compared to \textit{GMRF2}. This might be caused by setting the $\beta$ coefficients to zero in some MCMC iterations, shrinking the posterior mean towards zero. The fine scale structure of the estimate is also seen in the ``significant'' regions for \textit{SparseGMRF}, which also contain ``non-significant'' pixels. \textit{GRMF2} does not flag any pixel at all. 
The wavelet-based methods show an even more pronounced small-scale structure with more abrupt changes between positive and negative values. A very characteristic feature here are the negative values in the center of the images (cingulate cortex), which might have been oversmoothed by the other models. The main assumption for all three models is  sparsity in the wavelet space. The corresponding measures for the underlying model assumptions are lowest for \textit{WNET}, followed by \textit{WPLS} (cf. Table~3 in the appendix, best comparable to the \textit{smooth} function in the simulation). This translates to a more ``spiky'' estimate in Fig.~\ref{fig:appBetaCI}.
The last model, \textit{PCR2D}, assumes sparsity in the principal component space. As seen for the parametric sparsity measure, the leading $20$ of $25$ possible eigenimages are selected by cross-validation to construct the estimate, which is relatively high compared to our simulation. What is striking here is the rectangular, ``streaky'' nature of the structures seen in the coefficient image. This is clearly caused by constructing the eigenimages based on rank-one tensor-product decompositions \cite{Allen:2013}.
Overall, comparing the measures for underlying and parametric model assumptions of the estimates with reference patterns from the simulation indicates that the estimates are less smooth than the true \textit{smooth} image in the simulation study and the majority of the corresponding estimates. Moreover, all methods that use principal component approaches for dimension reduction in different spaces yield estimates that require a rather high number of principal components, meaning that they exhibit a relatively complex structure. Overall, the lack of resemblance to patterns of the measures for the simulated coefficient images might indicate that none of the model assumptions for the used methods perfectly capture the structure of the true coefficient image. 

As in the simulation correlations are calculated between all estimates to measure similarities among the estimated coefficient images (see Fig.~38 in the appendix). The highest correlation is found between models that assume smoothness (\textit{SparseGMRF} and \textit{FPCR}: $92\%$, \textit{SparseGMRF} and \textit{GMRF2}: $91\%$, \textit{FPCR} and  \textit{Splines}: $90\%$). The \textit{GMRF} estimate shows no correlation with any other method, which is also seen in Fig.~\ref{fig:appBetaCI}. There are no clear similarities in the correlation structure to any of the simulation settings. 

For the scalar covariates, all models except for \textit{GMRF} find ``significant'' effects, as the confidence/credible intervals do not include zero (see Fig.~37 in the appendix).
There is agreement between methods that the estimated intercepts and the effects for age at baseline are both positive, which makes sense as ADAS-Cog takes positive values and age is known to be a main risk factor for AD. For gender and years of education, the estimated coefficients are negative, i.e.\ on average, being female and a longer period of education are associated with lower ADAS-Cog values and a lower risk of AD.
However, there is also notable variation between the methods, as the confidence bands do not necessarily overlap or contain the point estimates of all other methods.
Note that some of the differences in $\hat \alpha$ might be related to the different coefficient image estimates $\hat \beta$.

In total, the results of the application show that while some methods used here show some common patterns in their results, they differ substantially in their details, as model assumptions have a strong influence on the results. In practical applications such as this one, this can entail the risk of over-interpreting effects that are mainly driven by the model assumptions. 
\section{Discussion}
\label{sec:out}

Scalar-on-image regression is an inherently non-identifiable statistical problem due to the fact that the number of pixels in the coefficient image -- and therefore the number of coefficients -- exceeds the number of observations, in many cases by far. In order to overcome the issue of non-identifiability, different approaches have been proposed in the literature, making different structural assumptions on the coefficient image and including all forms and combinations of smoothness, sparsity or projection onto a subspace. Model assumptions mostly result from both methodological and applied considerations. In neuroimaging, one would for example assume that the coefficient image, which represents the association between the observed images and the responses, has no abrupt changes within contiguous brain regions, following human physiology. In practice, this can be achieved by minimizing differences between neighbouring coefficients. At the same time, this reduces the degrees of freedom in the estimation and mostly leads to an identifiable model under the assumptions made.

Whereas the beneficial aspects of making assumptions are well known and understood, their impact on the estimates seems underappreciated or played down in practice. From a theoretical point of view it is obvious that models  with different assumptions may lead to different estimates due to non-identifiability. In practical applications, however, it is not always clear which model is appropriate and to what extent the model assumptions influence the results. While this is less crucial for predictions, it strongly affects the interpretability of the coefficient image estimates, as one cannot directly see whether features in the estimate are dominated by the model assumptions, driven by the data or supported by both, as one would ideally assume.  

In this paper, we have provided a systematic overview of the principal approaches to scalar-on-image regression and the assumptions made in the different models. The assumptions have been characterized as underlying ones, that describe the fundamental assumptions of a model, and parametric assumptions, that are expressed in terms of model-specific penalties or priors on the parameters and translate the underlying assumptions into an estimation approach. 
The methods discussed in this paper do not completely represent all published scalar-on-image models, but largely cover all main classes and their assumptions and focus on methods with available implementations. Variations include e.g.\ the LASSO-variant of \textit{WNET} \cite{ZhaoEtAl:2015} (implemented in the \texttt{R} package \texttt{refund.wave} \cite{refund.wave}), all types of models that combine smoothness of the coefficient image with a sparsity assumption as in \textit{SparseGMRF} \cite{HuangEtAl:2013,ShiKang:2015,KangEtAl:2016,LiEtAl:2015}, 
tensor based methods as \textit{PCR2D} \cite{ZhouEtAl:2013, MirandaEtAl:2015}
 or methods for scalar-on-function regression that can easily be extended to the scalar-on-image case \cite{ReissEtAl:2016}.
All of these methods have in common that they build on a (linear) regression approach, which is obviously a strong (meta) assumption in itself. While linearity facilitates the estimation and interpretation of these high-dimensional models, it is of course questionable whether a simple linear association between image predictors and scalar responses adequately reflects the complexity in the data.
Particularly in neuroimaging, the advent of highly nonlinear machine learning methods such as support vector machines \cite{CortesVapnik:1995} and, more recently, deep neural networks \cite{Schmidhuber:2015}, have given rise to a vast body of literature that aims at relating imaging data (mainly structural and functional MRI) to scalar responses, with a main focus on classification \cite{ArbabshiraniEtAl:2017}. Making use of nonlinearity, these approaches can have a superior performance than the classical (generalized) linear regression model in terms of prediction accuracy. At the same time, a main criticism of these prediction based approaches is that they are mostly \textit{black box} algorithms and therefore hardly interpretable, although first steps in this direction have been taken \cite{RibeiroEtAl:2016}. Interpretability, which is one key selling point of regression models compared to machine learning approaches, \cite{Dunson:2018} may however be strongly influenced by model assumptions, as our results show.

In this case, one would ideally wish to have a diagnostic criterion that identifies problematic settings, i.e.\ settings in which the model assumptions dominate the estimate, in advance. This, however, seems very challenging, if even feasible.  The measures proposed in this paper constitute a first step in this direction, as they quantify the degree to which the model assumptions are met by a given image or estimate. Their usage and interpretation has been illustrated in the simulation and in the case study in Section~\ref{sec:app}, showing for example that the sparsity assumption in \textit{SparseGMRF}, which is particularly designed to identify predictive brain regions and set the remaining parts in the coefficient image to zero, is mostly ignored in the estimation process. 
 In Bayesian approaches, where model assumptions are formulated in terms of priors, alternative measures have been proposed, e.g.\ for prior-data conflict \cite{EvansMoshonov:2006}, prior informativeness \cite{Mueller:2012}, or prior data size with respect to the likelihood \cite{ReimherrEtAl:2014}. Together with our measures, they could serve as starting point for an overall measure for the appropriateness of model assumptions. However, most of these Bayesian measures are restricted to rather simple models and to proper priors. Further work is needed to be able to apply them to high-dimensional models such as scalar-on-image regression, improper priors such as the intrinsic GMRF priors or non-Bayesian models that include dimension reduction or variable selection steps.

For practical applications, we recommend to carefully check the assumptions in the models used. The measures proposed in this paper can help to interpret the results for real data, e.g.\ by relating values obtained for estimates from the observed data to the values for hypothetic coefficient images, as done in the application. Conducting simulations can also be indicative for the types of features that can be found with the chosen methods and the observed data. For the case of the FDG-PET images used in Sections~\ref{sec:simStudy} and~\ref{sec:app}, smooth coefficients and those lying in the span of the leading principal components were estimated quite well by methods with corresponding assumptions. At the same time, the coefficient images \textit{bumpy} and \textit{sparse}, which have highly localized features, are seen to be considerably more difficult to estimate with this data.
Further, it seems helpful to compare the results with those of other approaches, making different assumptions, in order to find common patterns. These may help understanding which features in the estimated coefficient image are mostly driven by the data, the model assumptions or seem to combine both sources of information. Empirical confidence bands as in the application can serve as a first indicator, which regions of the estimated coefficient images might be of interest. For the ADNI data studied in Section~\ref{sec:app}, the empirical confidence bands agree most in the right upper and lower part of the images, in the area of the precentral gyrus. Within these regions of interest, one could for example check the agreement of the estimated coefficient images as an indicator of data-driven effects.
The idea of combining different models is also adopted in ensemble methods, e.g.\ for scalar-on-function regression \cite{GoldsmithScheipl:2014}. A drawback of this approach, however, is that it is based on predictive performance in a cross-validation setting, which is not only associated with high computational costs, but also aims more at prediction and  not at interpretability. 

As an alternative, one could replace the single model that incorporates the full imaging information by multiple models in separate brain locations, as commented by one of the reviewers. This of course would reduce the number of predictors in the models and therefore lower the risk of non-identifiability.
Apart from the fact that appropriate regions of interest would need to be defined, e.g.\ data-driven or based on medical knowledge, such an approach makes two important assumptions on its own: First, the relevant information is assumed to stem from the selected ROIs only, ignoring potentially important data in the remaining brain regions. Second, treating different ROIs in separate models implies the assumption that there is no correlation between these regions, which is also a quite strong assumption given the complexity of the human brain and thus no confounding by left-out regions.

In summary, model assumptions are a necessary and helpful tool to overcome identifiability issues in complex and inherently non-identifiable models such as scalar-on-image regression. Our results show that in practical applications, it is very important to be aware of model assumptions and the impact that they can have on the coefficient estimates.

\section*{Acknowledgements}

The authors thank P.\ T.\ Reiss for providing the \texttt{R}-code for the \textit{bumpy} function   and M.\ Ewers and M.\ À.\ Araque Caballero for registering the FDG-PET scans. We would also like to thank two anonymous reviewers for their insightful and constructive comments.
C.\ Happ and S.\ Greven were supported by the German Research Foundation through Emmy Noether grant GR 3793/1-1.

Data collection and sharing for the neuroimaging data in Section~\ref{sec:app}  was funded by the Alzheimer's Disease Neuroimaging Initiative (ADNI, National Institutes of Health Grant U01 AG024904) and DOD ADNI (Department of Defense award number W81XWH-12-2-0012). A detailed list of ADNI funding is available at \url{http://adni.loni.usc.edu/about/funding/}. The grantee organization is the Northern California Institute for Research and Education, and the study is coordinated by the Alzheimer's Disease Cooperative Study at the University of California, San Diego. ADNI data are disseminated by the Laboratory for Neuro Imaging at the University of Southern California.


\newpage

\bibliographystyle{apalike}
{
\small
\bibliography{ScalarOnImage}

\begin{thebibliography}{}

\bibitem[Allen, 2013]{Allen:2013}
Allen, G.~I. (2013).
\newblock Multi-way functional principal components analysis.
\newblock In {\em IEEE 5th International Workshop on Computational Advances in
  Multi-Sensor Adaptive Processing (CAMSAP)}, pages 220--223.

\bibitem[{Araque Caballero} et~al., 2015]{AraqueCaballeroEtAl:2015}
{Araque Caballero}, M.~{\'{A}}., Brendel, M., Delker, A., Ren, J., Rominger,
  A., Bartenstein, P., Dichgans, M., Weiner, M.~W., and Ewers, M. (2015).
\newblock {Mapping 3{-}year changes in gray matter and metabolism in
  {A$\beta$-}positive nondemented subjects}.
\newblock {\em Neurobiol Aging}, 36(11):2913--2924.

\bibitem[Arbabshirani et~al., 2017]{ArbabshiraniEtAl:2017}
Arbabshirani, M.~R., Plis, S., Sui, J., and Calhoun, V.~D. (2017).
\newblock {Single subject prediction of brain disorders in neuroimaging:
  Promises and pitfalls}.
\newblock {\em Neuroimage}, 145:137--165.

\bibitem[Besag, 1974]{Besag:1974}
Besag, J. (1974).
\newblock {Spatial interaction and the statistical analysis of lattice
  systems}.
\newblock {\em J R Stat Soc Series B Stat Methodol}, 36(2):192--236.

\bibitem[Binder et~al., 2009]{EncyclopediaNeuroscience}
Binder, M.~D., Hirokawa, N., and Windhorst, U., editors (2009).
\newblock {\em Encyclopedia of Neuroscience}.
\newblock Springer, Berlin, Heidelberg.

\bibitem[Cardot et~al., 1999]{CardotEtAl:1999}
Cardot, H., Ferraty, F., and Sarda, P. (1999).
\newblock {Functional linear model}.
\newblock {\em Stat Probab Lett}, 45(1):11--22.

\bibitem[Carroll and Chang, 1970]{CarrollChang:1970}
Carroll, J.~D. and Chang, J.~J. (1970).
\newblock {Analysis of individual differences in multidimensional scaling via
  an n-way generalization of "Eckart-Young" decomposition}.
\newblock {\em Psychometrika}, 35(3):283--319.

\bibitem[Coombs, 1964]{Coombs:1964}
Coombs, C.~H. (1964).
\newblock {\em A theory of data}.
\newblock Wiley, New York.

\bibitem[Cortes and Vapnik, 1995]{CortesVapnik:1995}
Cortes, C. and Vapnik, V. (1995).
\newblock Support-vector networks.
\newblock {\em Mach Learn}, 20(3):273--297.

\bibitem[Daubechies, 1988]{Daubechies:1988}
Daubechies, I. (1988).
\newblock {Orthonormal bases of compactly supported bases}.
\newblock {\em Commun Pure Appl Math}, 41:909--996.

\bibitem[{De Boor}, 1972]{DeBoor:1972}
{De Boor}, C. (1972).
\newblock {On calculating with B-splines}.
\newblock {\em J Approx Theory}, 6(1):50--62.

\bibitem[Donoho and Johnstone, 1994]{DonohoJohnstone:1994}
Donoho, D.~L. and Johnstone, J.~M. (1994).
\newblock {Ideal spatial adaptation by wavelet shrinkage}.
\newblock {\em Biometrika}, 81(3):425--455.

\bibitem[Dunson, 2018]{Dunson:2018}
Dunson, D.~B. (2018).
\newblock {Statistics in the big data era: Failures of the machine}.
\newblock {\em Stat Probab Lett}.
\newblock In press.

\bibitem[Eilers and Marx, 1996]{EilersMarx:1996}
Eilers, P. H.~C. and Marx, B.~D. (1996).
\newblock {Flexible smoothing with B-splines and penalties}.
\newblock {\em Stat Sci}, 11(2):89--121.

\bibitem[Evans and Moshonov, 2006]{EvansMoshonov:2006}
Evans, M. and Moshonov, H. (2006).
\newblock {Checking for prior-data conflict}.
\newblock {\em Bayesian Anal}, 1(4):893--914.

\bibitem[Friedman et~al., 2010]{glmnet}
Friedman, J., Hastie, T., and Tibshirani, R. (2010).
\newblock Regularization paths for generalized linear models via coordinate
  descent.
\newblock {\em J Stat Softw}, 33(1):1--22.

\bibitem[Friston et~al., 2007]{SPMbook}
Friston, K.~J., Ashburner, J.~T., Kiebel, S.~J., Nichols, T.~E., and Penny,
  W.~D., editors (2007).
\newblock {\em Statistical parametric mapping: The analysis of functional brain
  images}.
\newblock Academic Press, London.

\bibitem[Gelman, 2006]{Gelman:2006}
Gelman, A. (2006).
\newblock {Prior distributions for variance parameters in hierarchical models
  (Comment on Article by Browne and Draper)}.
\newblock {\em Bayesian Analysis}, 1(3):515--534.

\bibitem[Goldsmith et~al., 2014]{GoldsmithEtAl:2014}
Goldsmith, J., Huang, L., and Crainiceanu, C.~M. (2014).
\newblock {Smooth Scalar-on-Image Regression via Spatial Bayesian Variable
  Selection}.
\newblock {\em J Comput Graph Stat}, 23(1):46--64.

\bibitem[Goldsmith and Scheipl, 2014]{GoldsmithScheipl:2014}
Goldsmith, J. and Scheipl, F. (2014).
\newblock {Estimator selection and combination in scalar-on-function
  regression}.
\newblock {\em Comput Stat Data Anal}, 70:362--372.

\bibitem[Goldsmith et~al., 2016]{refund}
Goldsmith, J., Scheipl, F., Huang, L., Wrobel, J., Gellar, J., Harezlak, J.,
  McLean, M.~W., Swihart, B., Xiao, L., Crainiceanu, C., and Reiss, P.~T.
  (2016).
\newblock {\em refund: Regression with Functional Data}.
\newblock R package version 0.1-15.

\bibitem[Happ, 2013]{Happ:2013}
Happ, C. (2013).
\newblock Identifiability in scalar-on-functions regression.
\newblock Master's thesis, LMU Munich.

\bibitem[Happ, 2016]{MFPCA}
Happ, C. (2016).
\newblock {\em MFPCA: Multivariate Functional Principal Component Analysis for
  Data Observed on Different Dimensional Domains}.
\newblock R package version 1.0-1.

\bibitem[Happ and Greven, 2018]{HappGreven:2016}
Happ, C. and Greven, S. (2018).
\newblock Multivariate functional principal component analysis for data
  observed on different (dimensional) domains.
\newblock {\em J Am Stat Assoc}.
\newblock To appear.

\bibitem[Horn and Johnson, 1985]{HornJohnson:1985}
Horn, R.~A. and Johnson, C.~R. (1985).
\newblock {\em Matrix Analysis}.
\newblock Cambridge University Press, Cambridge.

\bibitem[Huang et~al., 2013]{HuangEtAl:2013}
Huang, L., Goldsmith, J., Reiss, P.~T., Reich, D.~S., and Crainiceanu, C.~M.
  (2013).
\newblock {Bayesian scalar-on-image regression with application to association
  between intracranial DTI and cognitive outcomes.}
\newblock {\em Neuroimage}, 83:210--23.

\bibitem[Huo et~al., 2014]{refund.wave}
Huo, L., Reiss, P., and Zhao, Y. (2014).
\newblock {\em refund.wave: Wavelet-Domain Regression with Functional Data}.
\newblock R package version 0.1.

\bibitem[Hurley and Rickard, 2009]{HurleyRickard:2009}
Hurley, N. and Rickard, S. (2009).
\newblock Comparing measures of sparsity.
\newblock {\em IEEE Trans Inf Theory}, 55(10):4723--4741.

\bibitem[Itti and Baldi, 2005]{IttiBaldi:2005}
Itti, L. and Baldi, P.~F. (2005).
\newblock A principled approach to detecting surprising events in video.
\newblock In {\em Proc. IEEE Conference on Computer Vision and Pattern
  Recognition (CVPR)}, pages 631--637, San Diego, CA.

\bibitem[Johnstone and Lu, 2009]{JohnstoneLu:2009}
Johnstone, I.~M. and Lu, A.~Y. (2009).
\newblock {On Consistency and Sparsity for Principal Components Analysis in
  High Dimensions}.
\newblock {\em J Am Stat Assoc}, 104(486):682--693.

\bibitem[Kang et~al., 2018]{KangEtAl:2016}
Kang, J., Reich, B.~J., and Staicu, A.-M. (2018).
\newblock {Scalar-on-image regression via the soft-thresholded Gaussian
  process}.
\newblock {\em Biometrika}, 105(1):165--184.

\bibitem[Li et~al., 2015]{LiEtAl:2015}
Li, F., Zhang, T., Wang, Q., Gonzalez, M.~Z., Maresh, E.~L., and Coan, J.~A.
  (2015).
\newblock {Spatial Bayesian variable selection and grouping for
  high-dimensional scalar-on-image regression}.
\newblock {\em Ann Appl Stat}, 9(2):687--713.

\bibitem[Mallat, 1989]{Mallat:1989}
Mallat, S.~G. (1989).
\newblock {A Theory for Multiresolution Signal Decomposition: The Wavelet
  Representation}.
\newblock {\em IEEE Trans Pattern Anal Mach Intell}, 11(7):674--693.

\bibitem[Marx and Eilers, 1999]{MarxEilers:1999}
Marx, B.~D. and Eilers, P. H.~C. (1999).
\newblock {Generalized Linear Regression on Sampled Signals and Curves: A
  P-Spline Approach}.
\newblock {\em Technometrics}, 41(1):1--13.

\bibitem[Marx and Eilers, 2005]{MarxEilers:2005}
Marx, B.~D. and Eilers, P. H.~C. (2005).
\newblock {Multidimensional Penalized Signal Regression}.
\newblock {\em Technometrics}, 47(1):13--22.

\bibitem[Minoshima et~al., 1997]{MinoshimaEtAl:1997}
Minoshima, S., Giordani, B., Berent, S., Frey, K.~A., Foster, N.~L., and Kuhl,
  D.~E. (1997).
\newblock Metabolic reduction in the posterior cingulate cortex in very early
  alzheimer's disease.
\newblock {\em Ann Neurol}, 42(1):85--94.

\bibitem[Miranda et~al., 2018]{MirandaEtAl:2015}
Miranda, M.~F., Zhu, H., and Ibrahim, J.~G. (2018).
\newblock {TPRM: Tensor partition regression models with applications in
  imaging biomarker detection}.
\newblock {\em Ann Appl Stat}.
\newblock To appear. arXiv: 1505.05482.

\bibitem[M{\"{u}}ller and Stadtm{\"{u}}ller, 2005]{MuellerStadtmueller:2005}
M{\"{u}}ller, H.~G. and Stadtm{\"{u}}ller, U. (2005).
\newblock {Generalized functional linear models}.
\newblock {\em Ann Stat}, 33(2):774--805.

\bibitem[M{\"{u}}ller, 2012]{Mueller:2012}
M{\"{u}}ller, U.~K. (2012).
\newblock {Measuring prior sensitivity and prior informativeness in large
  Bayesian models}.
\newblock {\em J Monet Econ}, 59(6):581--597.

\bibitem[Nason, 2016]{wavethresh}
Nason, G. (2016).
\newblock {\em wavethresh: Wavelets Statistics and Transforms}.
\newblock R package version 4.6.8.

\bibitem[Petrides, 2014]{Petrides:2014}
Petrides, M. (2014).
\newblock {\em Neuroanatomy of Language Regions of the Human Brain}.
\newblock Academic Press, San Diego.

\bibitem[Reimherr et~al., 2014]{ReimherrEtAl:2014}
Reimherr, M., Meng, X.-L., and Nicolae, D.~L. (2014).
\newblock {Being an informed Bayesian: Assessing prior informativeness and
  prior – likelihood conflict}.
\newblock arXiv: 1406.5958.

\bibitem[Reiss et~al., 2017]{ReissEtAl:2016}
Reiss, P.~T., Goldsmith, J., Shang, H.~L., and Ogden, R.~T. (2017).
\newblock Methods for scalar-on-function regression.
\newblock {\em Int Stat Rev}, 85(2):228--249.

\bibitem[Reiss et~al., 2015]{ReissEtAl:2015}
Reiss, P.~T., Huo, L., Zhao, Y., Kelly, C., and Ogden, R.~T. (2015).
\newblock {Wavelet-domain regression and predictive inference in psychiatric
  neuroimaging}.
\newblock {\em Ann Appl Stat}, 9(2):1076--1101.

\bibitem[Reiss and Ogden, 2007]{ReissOgden:2007}
Reiss, P.~T. and Ogden, R.~T. (2007).
\newblock {Functional Principal Component Regression and Functional Partial
  Least Squares}.
\newblock {\em J Am Stat Assoc}, 102(479):984--996.

\bibitem[Reiss and Ogden, 2010]{ReissOgden:2010}
Reiss, P.~T. and Ogden, R.~T. (2010).
\newblock {Functional Generalized Linear Models with Images as Predictors}.
\newblock {\em Biometrics}, 66:61--69.

\bibitem[Ribeiro et~al., 2016]{RibeiroEtAl:2016}
Ribeiro, M.~T., Singh, S., and Guestrin, C. (2016).
\newblock "why should i trust you?": Explaining the predictions of any
  classifier.
\newblock In {\em KDD}, pages 1135--1144, New York, NY, USA. ACM.

\bibitem[Rosen et~al., 1984]{ADASCog}
Rosen, W.~G., Mohs, R.~C., and Davis, K.~L. (1984).
\newblock A new rating scale for alzheimer's disease.
\newblock {\em Am J Psychiatry}, 141(11):1356--1364.

\bibitem[Rue and Held, 2005]{RueHeld:2005}
Rue, H. and Held, L. (2005).
\newblock {\em {Gaussian Markov Random Fields: Theory and Applications}}.
\newblock Chapman {\&} Hall/CRC, Boca Raton.

\bibitem[Scheipl and Greven, 2016]{ScheiplGreven:2016}
Scheipl, F. and Greven, S. (2016).
\newblock {Identifiability in penalized function-on-function regression
  models}.
\newblock {\em Electron J Stat}, 10(1):495--526.

\bibitem[Schmidhuber, 2015]{Schmidhuber:2015}
Schmidhuber, J. (2015).
\newblock {Deep Learning in neural networks: An overview}.
\newblock {\em Neural Netw}, 61:85--117.

\bibitem[Shi and Kang, 2015]{ShiKang:2015}
Shi, R. and Kang, J. (2015).
\newblock {Thresholded Multiscale Gaussian Processes with Application to
  Bayesian Feature Selection for Massive Neuroimaging Data}.
\newblock arXiv: 1504.06074.

\bibitem[Weiner et~al., 2015]{WeinerEtAl:2015}
Weiner, M.~W., Veitch, D.~P., Aisen, P.~S., Beckett, L.~A., Cairns, N.~J.,
  Cedarbaum, J., Donohue, M.~C., Green, R.~C., Harvey, D., Jack, C.~R., Jagust,
  W., Morris, J.~C., Petersen, R.~C., Saykin, A.~J., Shaw, L., Thompson, P.~M.,
  Toga, A.~W., and Trojanowski, J.~Q. (2015).
\newblock {Impact of the Alzheimer's Disease Neuroimaging Initiative, 2004 to
  2014}.
\newblock {\em Alzheimers Dement}, 11(7):865--884.

\bibitem[Wood, 2011]{Wood:2011}
Wood, S.~N. (2011).
\newblock {Fast stable restricted maximum likelihood and marginal likelihood
  estimation of semiparametric generalized linear models}.
\newblock {\em J R Stat Soc Series B Stat Methodol}, 73:3--36.

\bibitem[Wood, 2016]{mgcv}
Wood, S.~N. (2016).
\newblock {\em mgcv: Mixed GAM Computation Vehicle with GCV/AIC/REML Smoothness
  Estimation}.
\newblock R package version 1.8.17.

\bibitem[Zhao et~al., 2015]{ZhaoEtAl:2015}
Zhao, Y., Chen, H., and Ogden, R.~T. (2015).
\newblock Wavelet-based weighted lasso and screening approaches in functional
  linear regression.
\newblock {\em J Comput Graph Stat}, 24(3):655--675.

\bibitem[Zhou et~al., 2013]{ZhouEtAl:2013}
Zhou, H., Li, L., and Zhu, H. (2013).
\newblock {Tensor regression with applications in neuroimaging data analysis}.
\newblock {\em J Am Stat Assoc}, 108(502):540--552.

\bibitem[Zou and Hastie, 2005]{ZouHastie:2005}
Zou, H. and Hastie, T. (2005).
\newblock {Regularization and variable selection via the elastic-net}.
\newblock {\em J R Stat Soc Series B Stat Methodol}, 67(2):301--320.

\end{thebibliography}
}

\FloatBarrier

\newpage 

\section{Appendix -- Simulation}
\subsection{Model Settings}
\label{sec:modelSettings}

\textit{Splines}: The unknown $\beta$ image is expanded in $K_x = K_y = 15$ cubic B-spline basis functions in each direction, penalizing the second squared differences of the corresponding coefficients. The smoothing parameter $\lambda$ is found via REML. The calculations can be done using the \texttt{gam} function in the \texttt{R}-package \texttt{mgcv} \citep{Wood:2011, mgcv}.

\textit{FPCR}: As for the pure spline approach we use $K_x = K_y = 15$ basis functions for each marginal and choose the smoothing parameter via REML. The number $K_0$ of principal components retained for regression is chosen via five-fold cross-validation from $\{5,10,25,50,100,150\}$. The model is fit using the function \texttt{fpcr} in the \texttt{R}-package \texttt{refund} \citep{refund}. 

\textit{PCR2D}: We calculate $25$ two-dimensional principal components of the observed images using the approach of \citet{Allen:2013} as implemented in the \texttt{MFPCA} package \citep{MFPCA} with second difference penalty for smoothing in each direction. The smoothing parameters $\lambda_v, \lambda_w$ are chosen via GCV within the boundaries $10^{-4}$ and $10^2$. The response $y$ is regressed on the first $K  \in \{1,5,10,15,20,25\}$ score vectors to find the coefficients for the unknown coefficient image. An optimal choice of $K$ is found via five-fold cross-validation.

\textit{WCR}: We use the function \texttt{wcr} in the package \texttt{refund.wave} \citep{refund.wave}. The observed images are transformed to the wavelet space using Daubechies least-asymmetric orthonormal compactly supported wavelets with 10 vanishing moments. The resolution level $M_0$ is fixed to $3$. Only the $K^\ast \in \{10,25,50,100,250, 500,1000\}$ coefficients having the highest variance are retained. The response $y$ is regressed on the leading $K_0\in \{5,10,15,25,50,75\}$ principal components of the remaining coefficients (restricting $K_0 \leq K^\ast$) and the result is transformed back to the original space. An optimal combination of $K^\ast$ and $K_0$ is found via five-fold cross-validation.

\textit{WPLS}: The wavelet-based principal least squares method is implemented in the same function \texttt{wcr} of the \texttt{refund.wave} package, using the option \texttt{method = "pls"}. For all parameters ($M_0, K^\ast, K_0$) we use the same specifications as for \textit{WCR}.

\textit{WNET}: Here also we use Daubechies least-asymmetric orthonormal compactly supported wavelets with 10 vanishing moments and a resolution level $M_0 = 3$ to obtain wavelet coefficients from the observed images. The model is estimated using the \texttt{wnet} function in the \texttt{R}-package \texttt{refund.wave} \citep{refund.wave}. As for the other two wavelet methods, the number of wavelet coefficients that are retained for the regression is chosen from $K^\ast \in \{10,25,50,100,250, 500,1000\}$. For the elastic net part, the mixing parameter $\eta$ can take values in $\{0,0.25,0.5,0.75,1\}$, with $0$ corresponding to the Ridge penalty and $1$ giving the LASSO approach.
 Candidate values for the penalty parameter $\lambda$ are automatically generated by the  \texttt{glmnet} function. An optimal combination of $K^\ast$ and $\eta$ is chosen via five-fold cross-validation.

\textit{SparseGMRF}:
A constant prior for $\alpha$ is used. The hyperparameters are chosen via five-fold cross-validation from $a \in \{-4,-2,-0.5\},~ b \in \{0.1, 0.5, 1.5\},~ \sigma^2_\eps, \sigma^2_\beta \in \{10^{-5},10^{-3}, 10^{-1}\}$.  For each parameter combination and each fold (in total $81 \cdot 5 = 405$ combinations), a short Gibbs sampling is run with $250$ iterations, of which $100$ are discarded as burnin (no thinning), following the settings in \citet{GoldsmithEtAl:2014}. For the starting values, $\gamma_l$ is sampled randomly from $\{0,1\}$ and if $\gamma_l = 1$, $\beta_l$ is sampled from $\N(0, \sigma^2_\beta)$, otherwise $\beta_l = 0$. The pixels are updated in random order.

\textit{GMRF:}
The prior for the unknown coefficient image $\beta$ is chosen as an intrinsic GMRF with four neighbours and for $\alpha$ a constant prior is used. The priors for the variance parameters $\sigma_\eps^2$ and $\sigma_\beta^2$ are chosen as conjugate inverse gamma distributions with $\sigma_\eps^2, \sigma_\beta^2 \sim \IG(1,1)$ (which is considered rather uninformative, but not entirely without controversy, see \citet{Gelman:2006}; model \textit{GMRF}) and $\sigma_\eps^2, \sigma_\beta^2 \sim \IG(10,10^{-3})$ (highly informative with a prior mean of $10^{-3}$ and a prior variance of $10^{-9}$; model \textit{GMRF2}). For both models, the Gibbs Sampler is run over $5000$ iterations, of which $500$ are discarded as burnin and saving each $20$th step (thinning). As a starting value, $\beta_l$ is initialized with $\N(0, \tilde \sigma^2_\beta)$ with $\tilde \sigma^2_\beta$ the prior mode. The pixels are updated in random order.

\FloatBarrier

\subsection{Supplementary Results for the Simulation}

\subsubsection{Results for N = 250 and SNR = 4}
\label{sec:examplePlots}

\begin{minipage}{\textwidth}
\centering
\includegraphics[width = 0.85\textwidth]{\Path/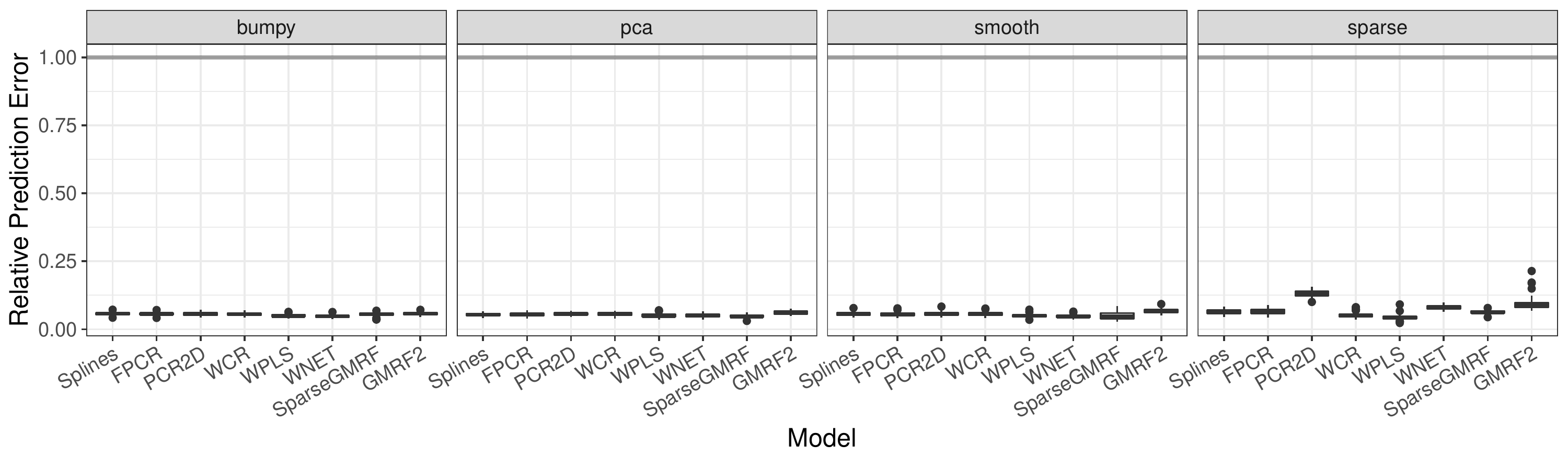}

\captionof{figure}{\protect Relative prediction errors for $N = 250$ observations and $\SNR = 4$ over all $100$ simulation runs.  
Boxplots show the errors for all models except \textit{GMRF} depending on the true coefficient image.
Gray horizontal lines mark $1$, which corresponds to a constant coefficient image, having the average value of the true $\beta$ image.}
\label{fig:predErr_250_4}
\end{minipage}


\begin{figure}[!h]
\centering
\includegraphics[page = 1, width = 0.3\textwidth]{\Path/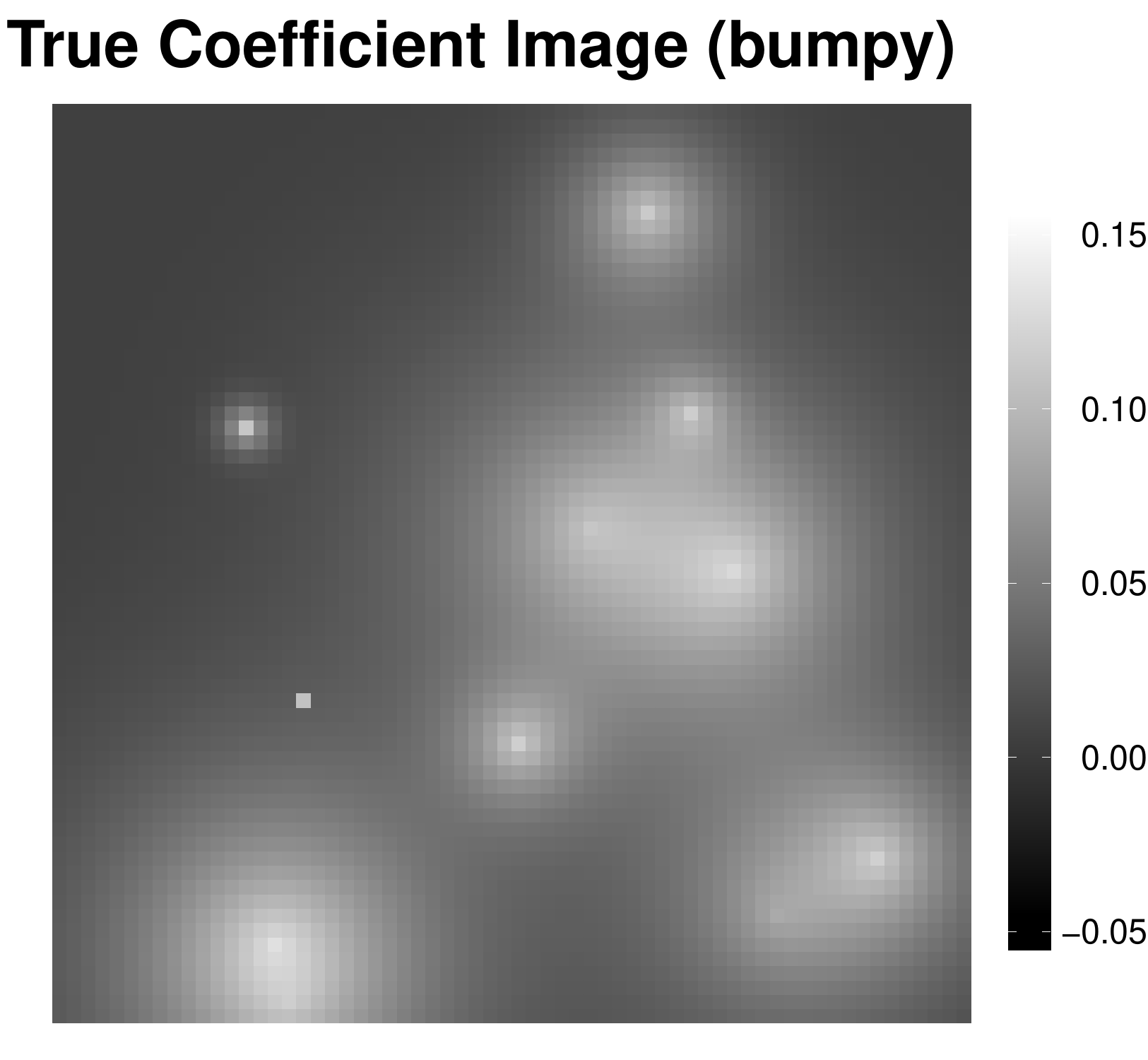}

\includegraphics[page = 2, width = 0.3\textwidth]{\Path/Code/Simulation/ExamplePlots/beta_bumpy_SNR_4.pdf}
\includegraphics[page = 3, width = 0.3\textwidth]{\Path/Code/Simulation/ExamplePlots/beta_bumpy_SNR_4.pdf}
\includegraphics[page = 4, width = 0.3\textwidth]{\Path/Code/Simulation/ExamplePlots/beta_bumpy_SNR_4.pdf}

\includegraphics[page = 5, width = 0.3\textwidth]{\Path/Code/Simulation/ExamplePlots/beta_bumpy_SNR_4.pdf}
\includegraphics[page = 6, width = 0.3\textwidth]{\Path/Code/Simulation/ExamplePlots/beta_bumpy_SNR_4.pdf}
\includegraphics[page = 7, width = 0.3\textwidth]{\Path/Code/Simulation/ExamplePlots/beta_bumpy_SNR_4.pdf}

\includegraphics[page = 8, width = 0.3\textwidth]{\Path/Code/Simulation/ExamplePlots/beta_bumpy_SNR_4.pdf}
\includegraphics[page = 9, width = 0.3\textwidth]{\Path/Code/Simulation/ExamplePlots/beta_bumpy_SNR_4.pdf}
\includegraphics[page = 10, width = 0.3\textwidth]{\Path/Code/Simulation/ExamplePlots/beta_bumpy_SNR_4.pdf}
\caption{\protect\input{\Path/captions/simBetaExample_bumpy}}
\label{fig:simBetaExample_bumpy}
\end{figure}


\begin{figure}
\centering
\includegraphics[page = 1, width = 0.3\textwidth]{\Path/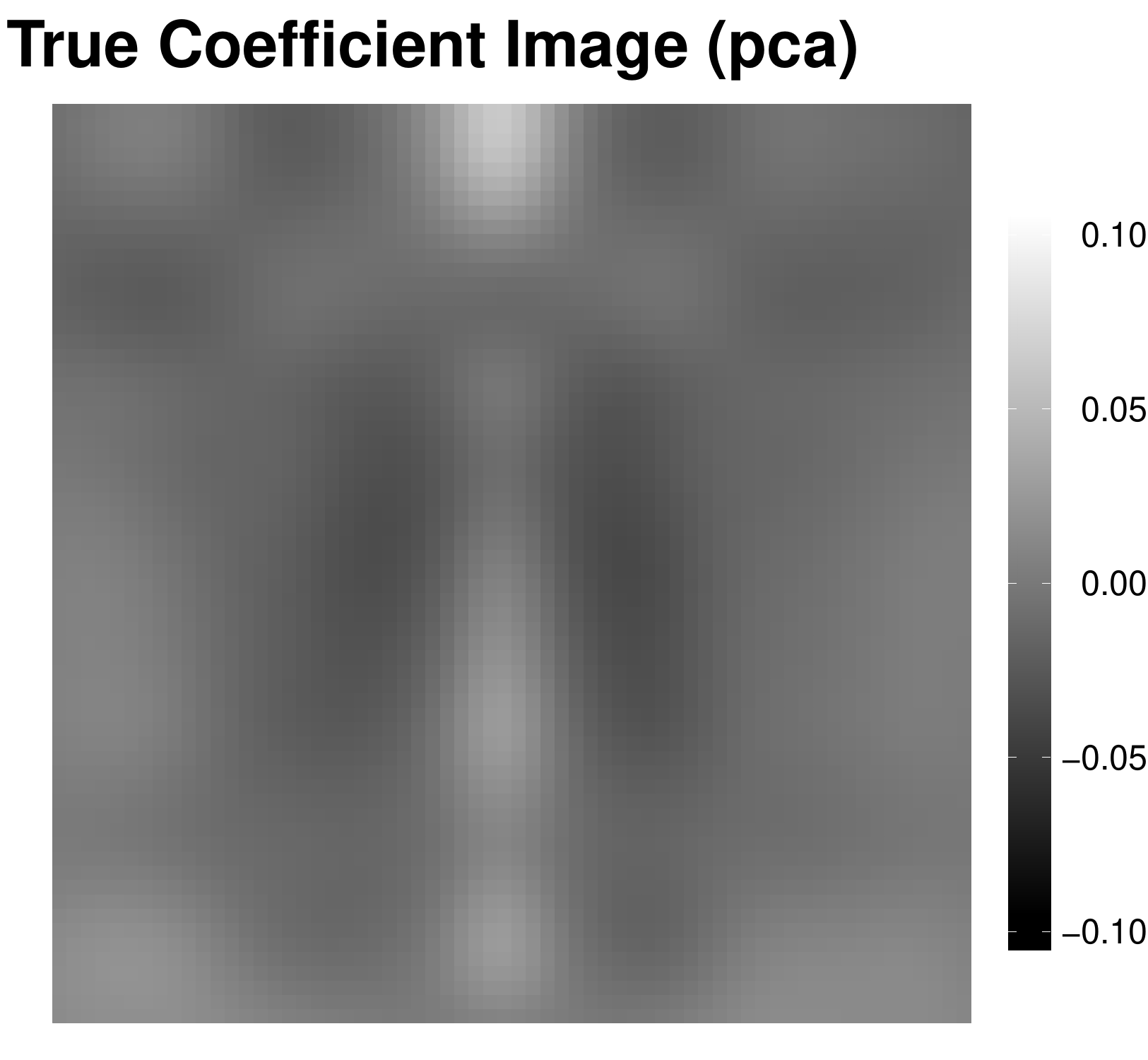}

\includegraphics[page = 2, width = 0.3\textwidth]{\Path/Code/Simulation/ExamplePlots/beta_pca_SNR_4.pdf}
\includegraphics[page = 3, width = 0.3\textwidth]{\Path/Code/Simulation/ExamplePlots/beta_pca_SNR_4.pdf}
\includegraphics[page = 4, width = 0.3\textwidth]{\Path/Code/Simulation/ExamplePlots/beta_pca_SNR_4.pdf}

\includegraphics[page = 5, width = 0.3\textwidth]{\Path/Code/Simulation/ExamplePlots/beta_pca_SNR_4.pdf}
\includegraphics[page = 6, width = 0.3\textwidth]{\Path/Code/Simulation/ExamplePlots/beta_pca_SNR_4.pdf}
\includegraphics[page = 7, width = 0.3\textwidth]{\Path/Code/Simulation/ExamplePlots/beta_pca_SNR_4.pdf}

\includegraphics[page = 8, width = 0.3\textwidth]{\Path/Code/Simulation/ExamplePlots/beta_pca_SNR_4.pdf}
\includegraphics[page = 9, width = 0.3\textwidth]{\Path/Code/Simulation/ExamplePlots/beta_pca_SNR_4.pdf}
\includegraphics[page = 10, width = 0.3\textwidth]{\Path/Code/Simulation/ExamplePlots/beta_pca_SNR_4.pdf}
\caption{\protect\input{\Path/captions/simBetaExample_pca}}
\label{fig:simBetaExample_pca}
\end{figure}


\begin{figure}
\centering
\includegraphics[page = 1, width = 0.3\textwidth]{\Path/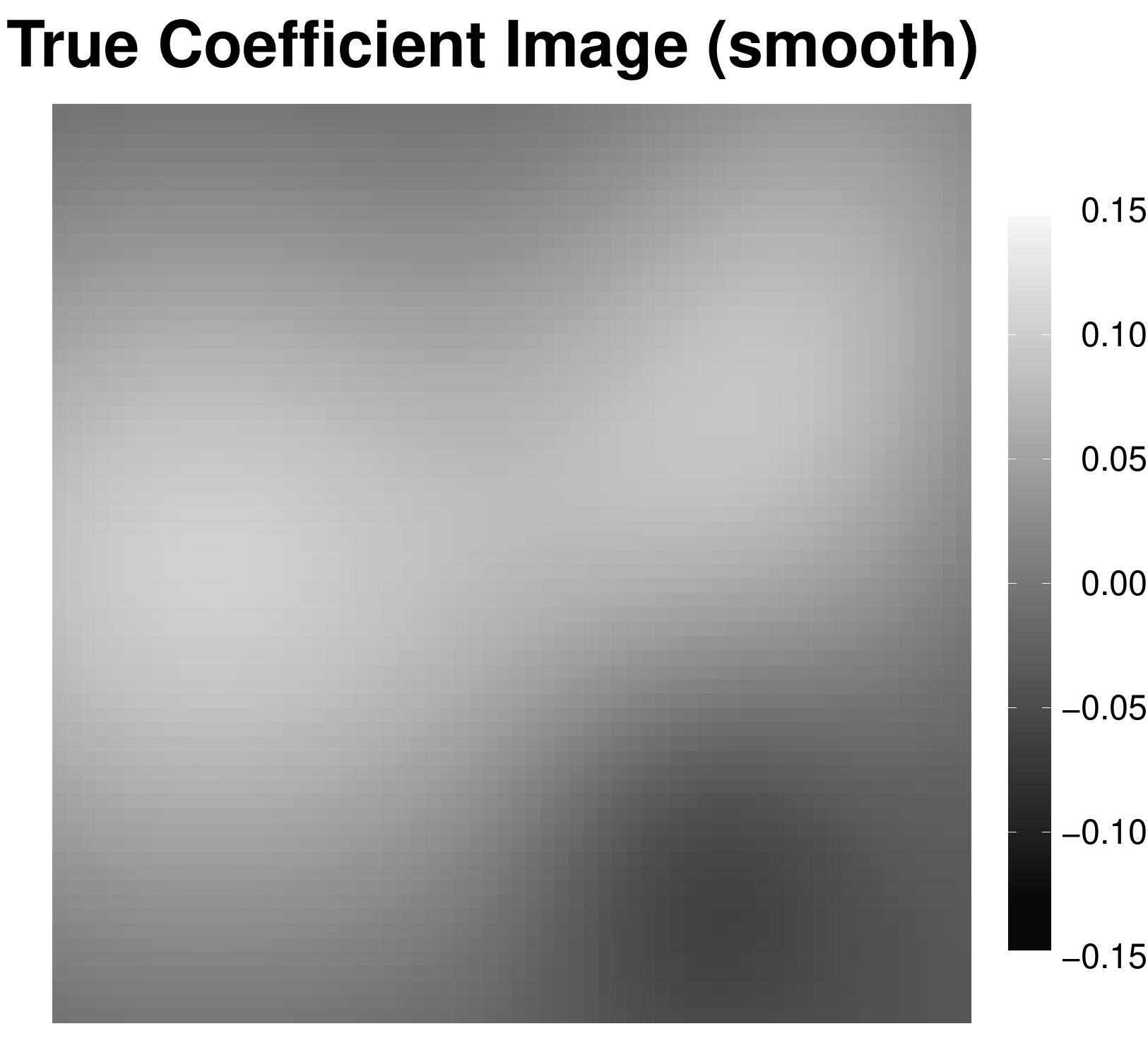}

\includegraphics[page = 2, width = 0.3\textwidth]{\Path/Code/Simulation/ExamplePlots/beta_smooth_SNR_4.pdf}
\includegraphics[page = 3, width = 0.3\textwidth]{\Path/Code/Simulation/ExamplePlots/beta_smooth_SNR_4.pdf}
\includegraphics[page = 4, width = 0.3\textwidth]{\Path/Code/Simulation/ExamplePlots/beta_smooth_SNR_4.pdf}

\includegraphics[page = 5, width = 0.3\textwidth]{\Path/Code/Simulation/ExamplePlots/beta_smooth_SNR_4.pdf}
\includegraphics[page = 6, width = 0.3\textwidth]{\Path/Code/Simulation/ExamplePlots/beta_smooth_SNR_4.pdf}
\includegraphics[page = 7, width = 0.3\textwidth]{\Path/Code/Simulation/ExamplePlots/beta_smooth_SNR_4.pdf}

\includegraphics[page = 8, width = 0.3\textwidth]{\Path/Code/Simulation/ExamplePlots/beta_smooth_SNR_4.pdf}
\includegraphics[page = 9, width = 0.3\textwidth]{\Path/Code/Simulation/ExamplePlots/beta_smooth_SNR_4.pdf}
\includegraphics[page = 10, width = 0.3\textwidth]{\Path/Code/Simulation/ExamplePlots/beta_smooth_SNR_4.pdf}
\caption{\protect\input{\Path/captions/simBetaExample_smooth}}
\label{fig:simBetaExample_smooth}
\end{figure}


\begin{figure}
\centering
\includegraphics[page = 1, width = 0.3\textwidth]{\Path/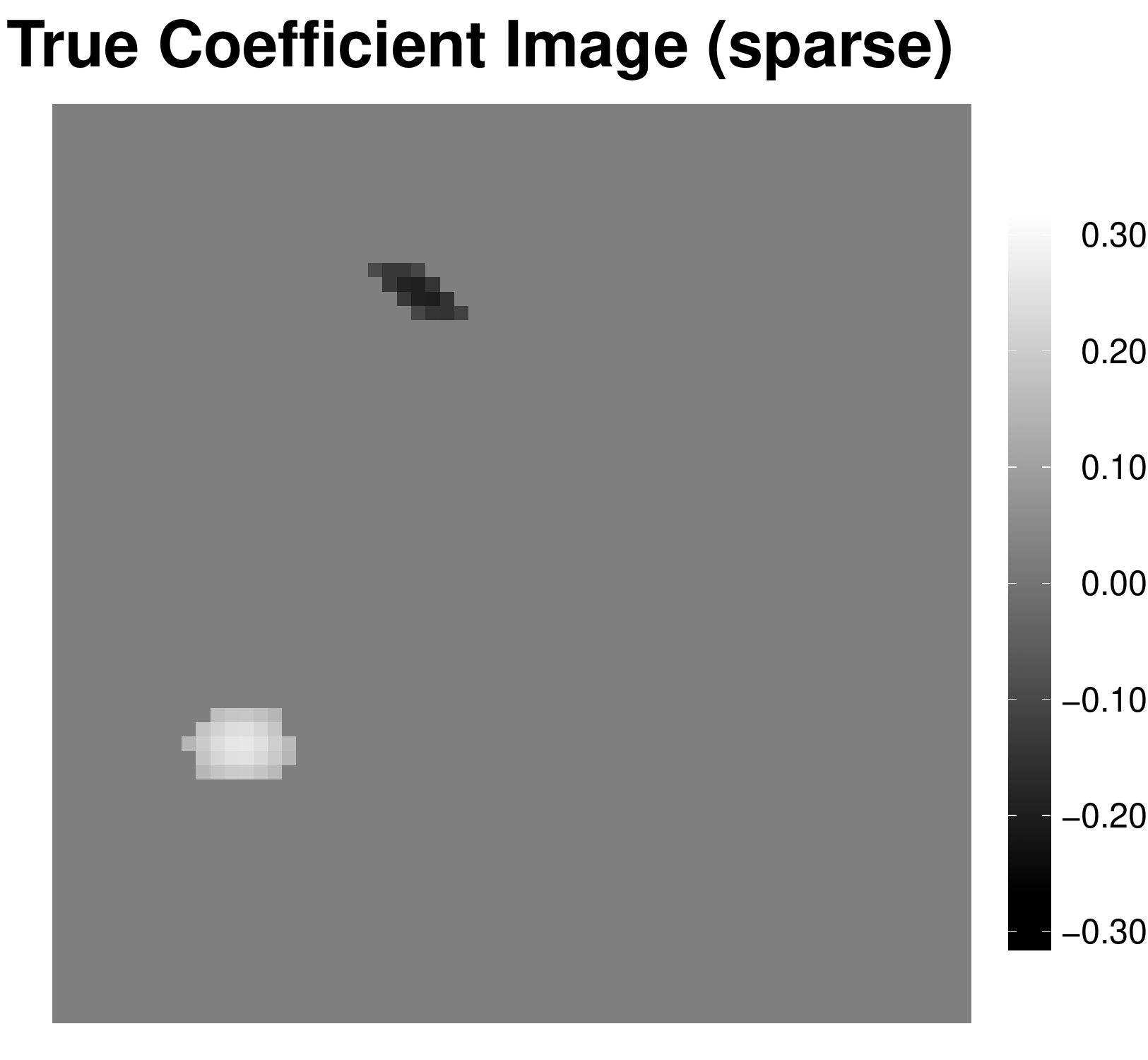}

\includegraphics[page = 2, width = 0.3\textwidth]{\Path/Code/Simulation/ExamplePlots/beta_sparse_SNR_4.pdf}
\includegraphics[page = 3, width = 0.3\textwidth]{\Path/Code/Simulation/ExamplePlots/beta_sparse_SNR_4.pdf}
\includegraphics[page = 4, width = 0.3\textwidth]{\Path/Code/Simulation/ExamplePlots/beta_sparse_SNR_4.pdf}

\includegraphics[page = 5, width = 0.3\textwidth]{\Path/Code/Simulation/ExamplePlots/beta_sparse_SNR_4.pdf}
\includegraphics[page = 6, width = 0.3\textwidth]{\Path/Code/Simulation/ExamplePlots/beta_sparse_SNR_4.pdf}
\includegraphics[page = 7, width = 0.3\textwidth]{\Path/Code/Simulation/ExamplePlots/beta_sparse_SNR_4.pdf}

\includegraphics[page = 8, width = 0.3\textwidth]{\Path/Code/Simulation/ExamplePlots/beta_sparse_SNR_4.pdf}
\includegraphics[page = 9, width = 0.3\textwidth]{\Path/Code/Simulation/ExamplePlots/beta_sparse_SNR_4.pdf}
\includegraphics[page = 10, width = 0.3\textwidth]{\Path/Code/Simulation/ExamplePlots/beta_sparse_SNR_4.pdf}
\caption{\protect\input{\Path/captions/simBetaExample_sparse}}
\label{fig:simBetaExample_sparse}
\end{figure}


\begin{figure}[ht!]
\centering

\includegraphics[width = 0.48\textwidth]{\Path/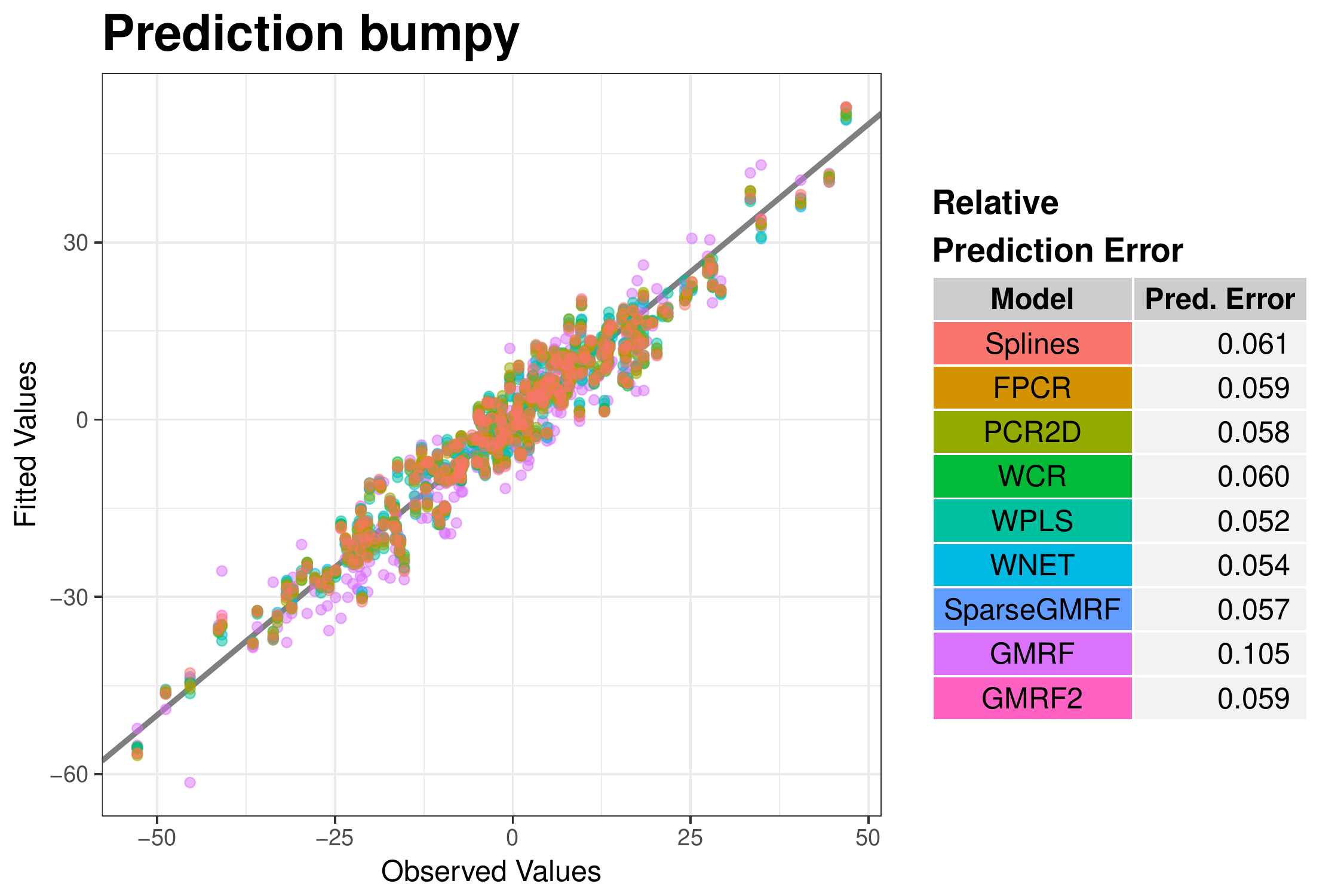}
\includegraphics[width = 0.48\textwidth]{\Path/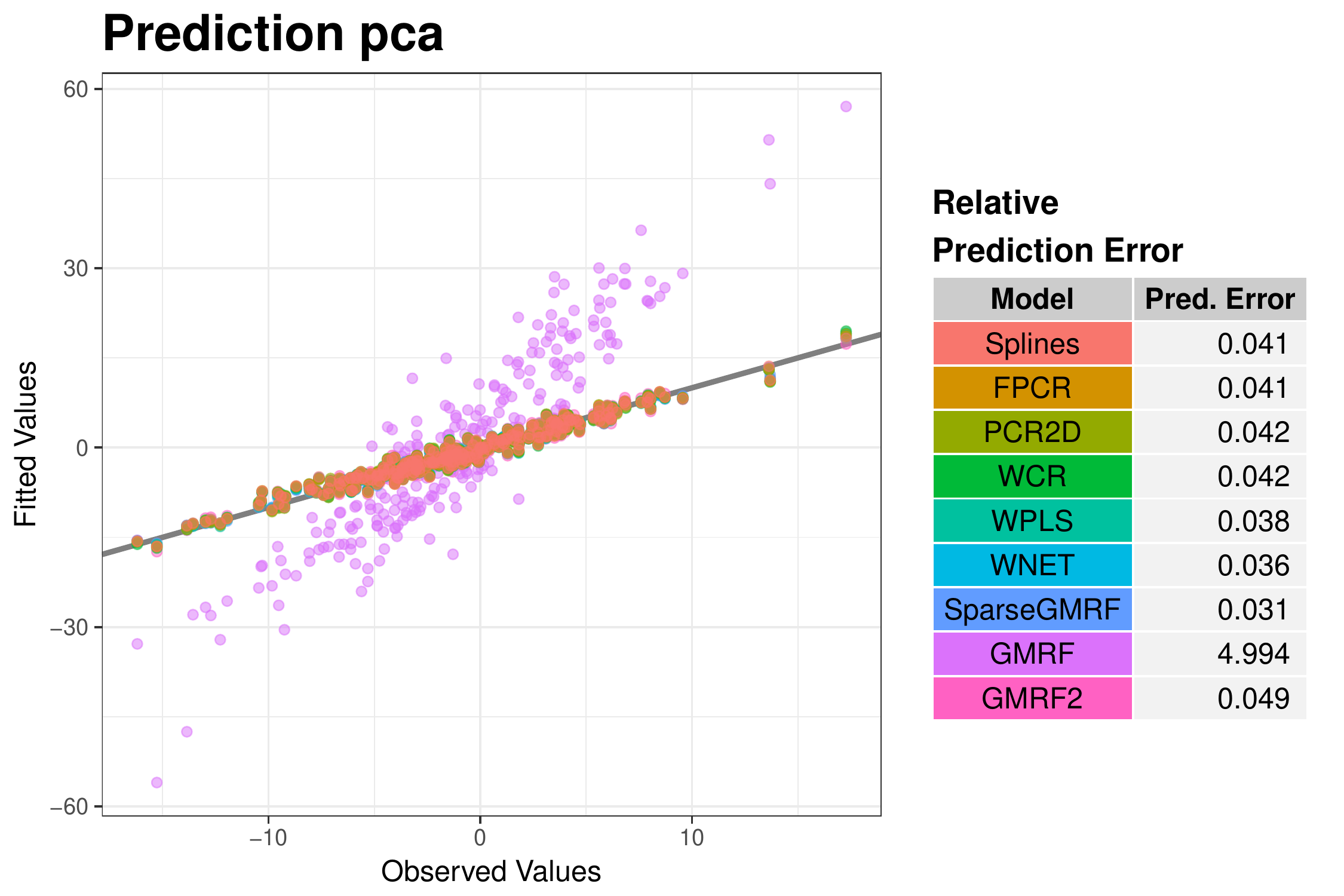}

\includegraphics[width = 0.48\textwidth]{\Path/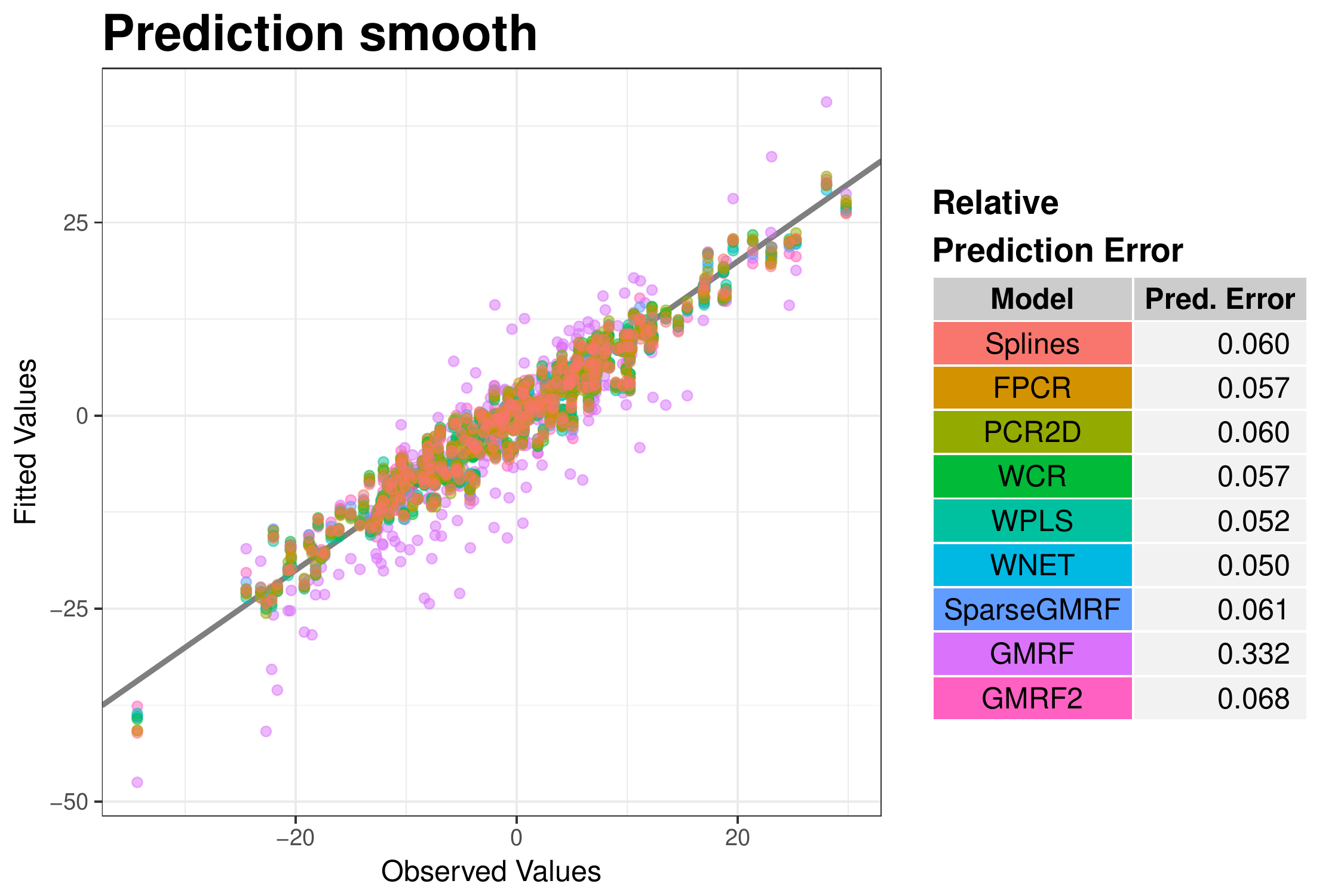}
\includegraphics[width = 0.48\textwidth]{\Path/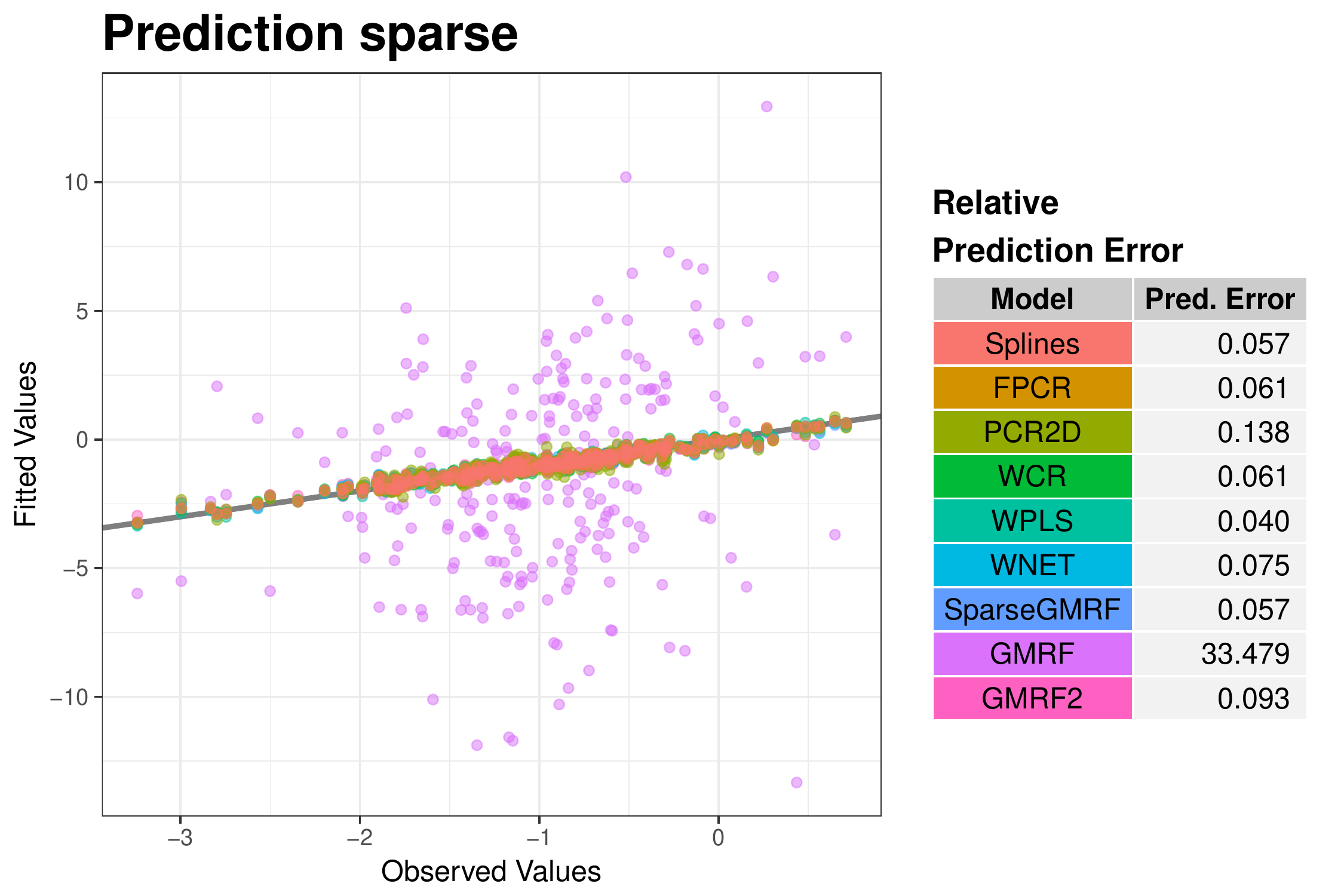}

\caption{\protect\input{\Path/captions/simPredExample}}
\label{fig:simPredExample}
\end{figure}

\FloatBarrier

\subsubsection{Results for N = 250 and SNR = 1}

\begin{minipage}{\textwidth}
\centering
\includegraphics[width = 0.85\textwidth]{\Path/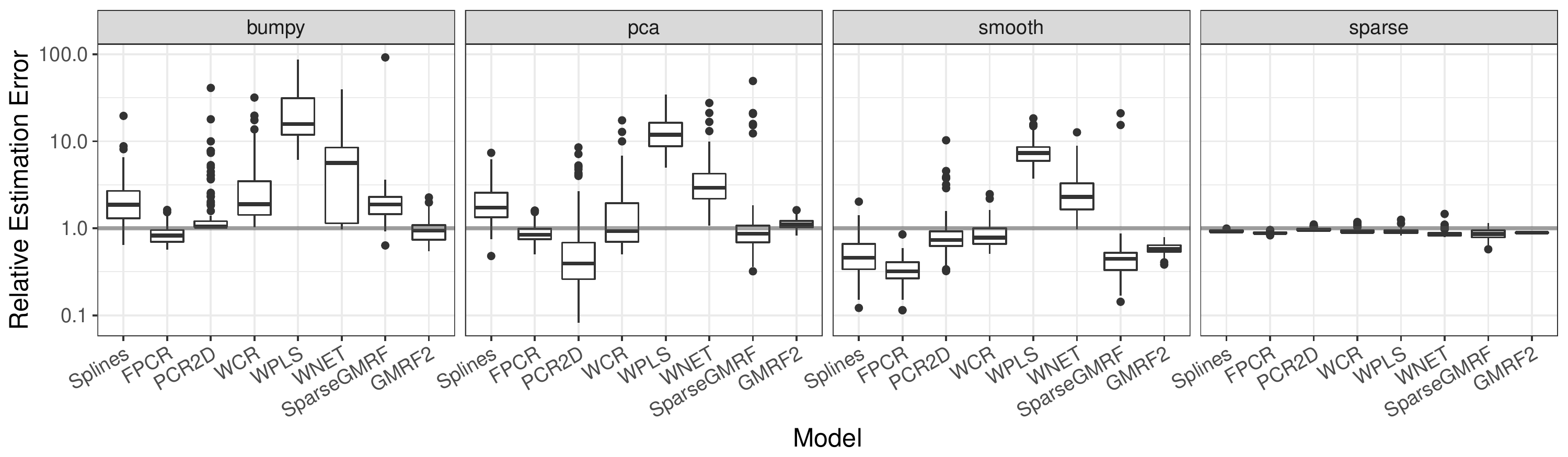}

\captionof{figure}{\protect Relative estimation errors for $N = 250$ observations and $\SNR = 1$ over all $100$ simulation runs.  
Boxplots show the errors for all models except \textit{GMRF} depending on the true coefficient image (\textit{GMRF}: median: $73.93$, sd: $142.78$).
Gray horizontal lines mark $1$, which corresponds to a constant coefficient image, having the average value of the true $\beta$ image.}
\label{fig:estErr_250_1}
\end{minipage}

\bigskip

\begin{minipage}{\textwidth}
\centering
\includegraphics[width = 0.85\textwidth]{\Path/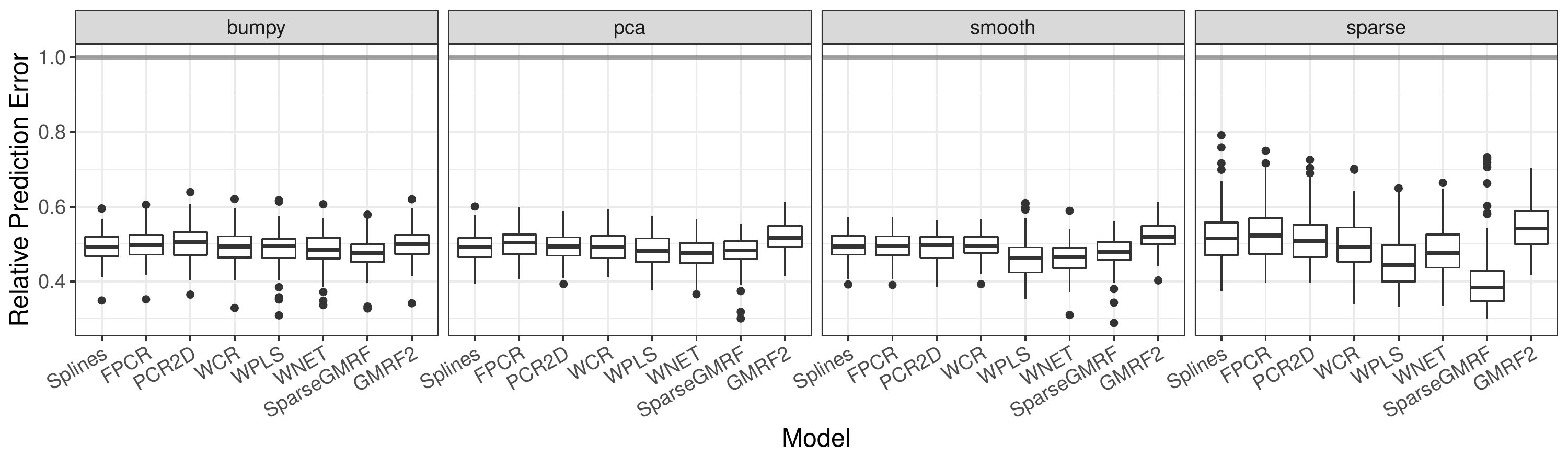}

\captionof{figure}{\protect Relative prediction errors for $N = 250$ observations and $\SNR = 1$ over all $100$ simulation runs.  
Boxplots show the errors for all models except \textit{GMRF} depending on the true coefficient image  (\textit{GMRF}: median: $0.68$, sd: $36.67$).
Gray horizontal lines mark $1$, which corresponds to the simple intercept model.}
\label{fig:predErr_250_1}
\end{minipage}

\bigskip

\begin{minipage}{\textwidth}
\centering
\includegraphics[width = 0.85\textwidth]{\Path/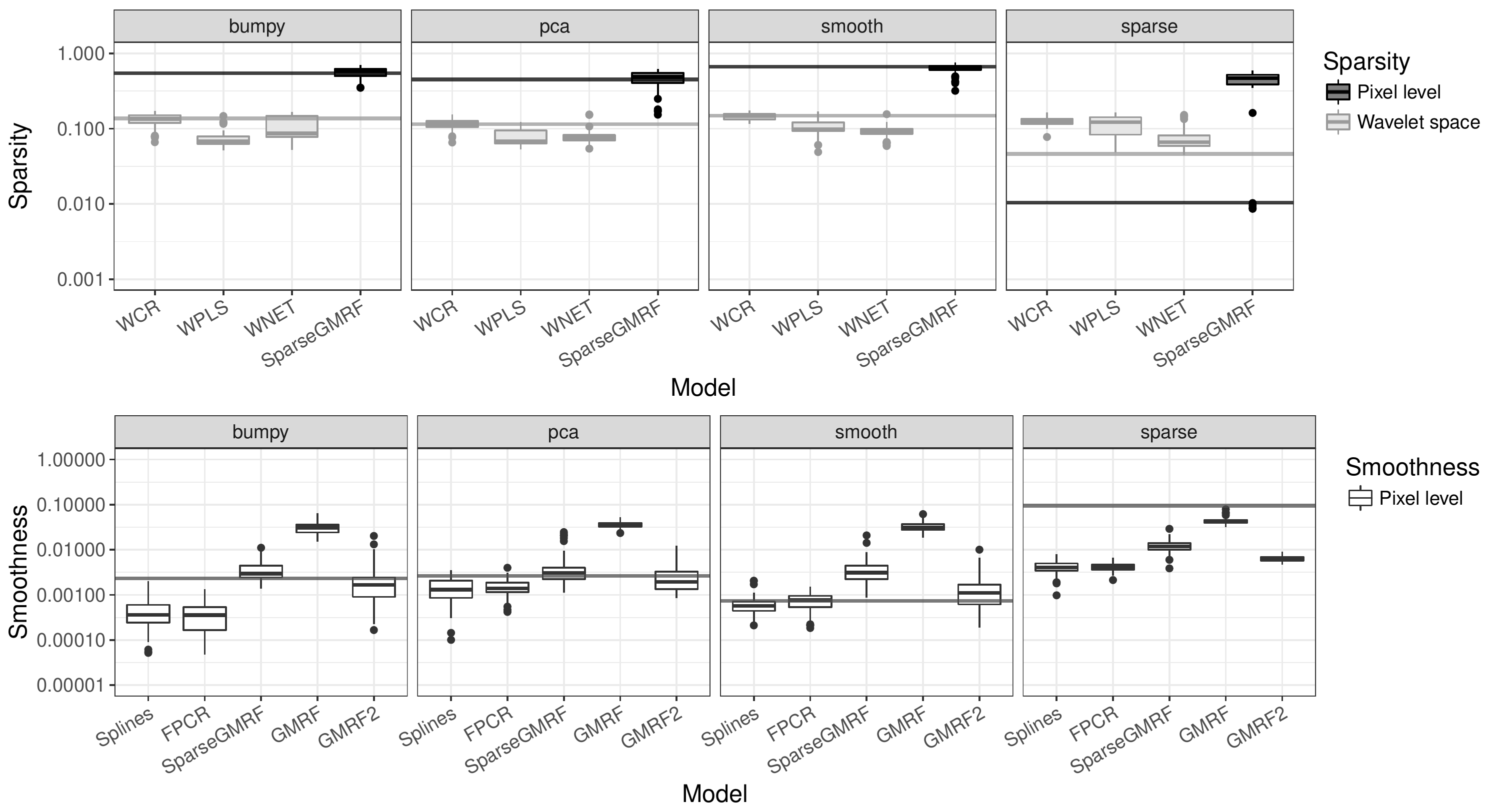}

\captionof{figure}{\protect Measures for underlying model assumptions in the simulation for $N = 250$  observations 
and $\SNR = 1$ over all $100$ simulation runs. Boxplots show the measures for the different models depending on the true coefficient image. All values on log-scale. Gray horizontal lines correspond to the values for the true coefficient images.}
\label{fig:simImplicitAss250_1}
\end{minipage}

\bigskip

\begin{minipage}{\textwidth}
\centering
\includegraphics[width = \textwidth]{\Path/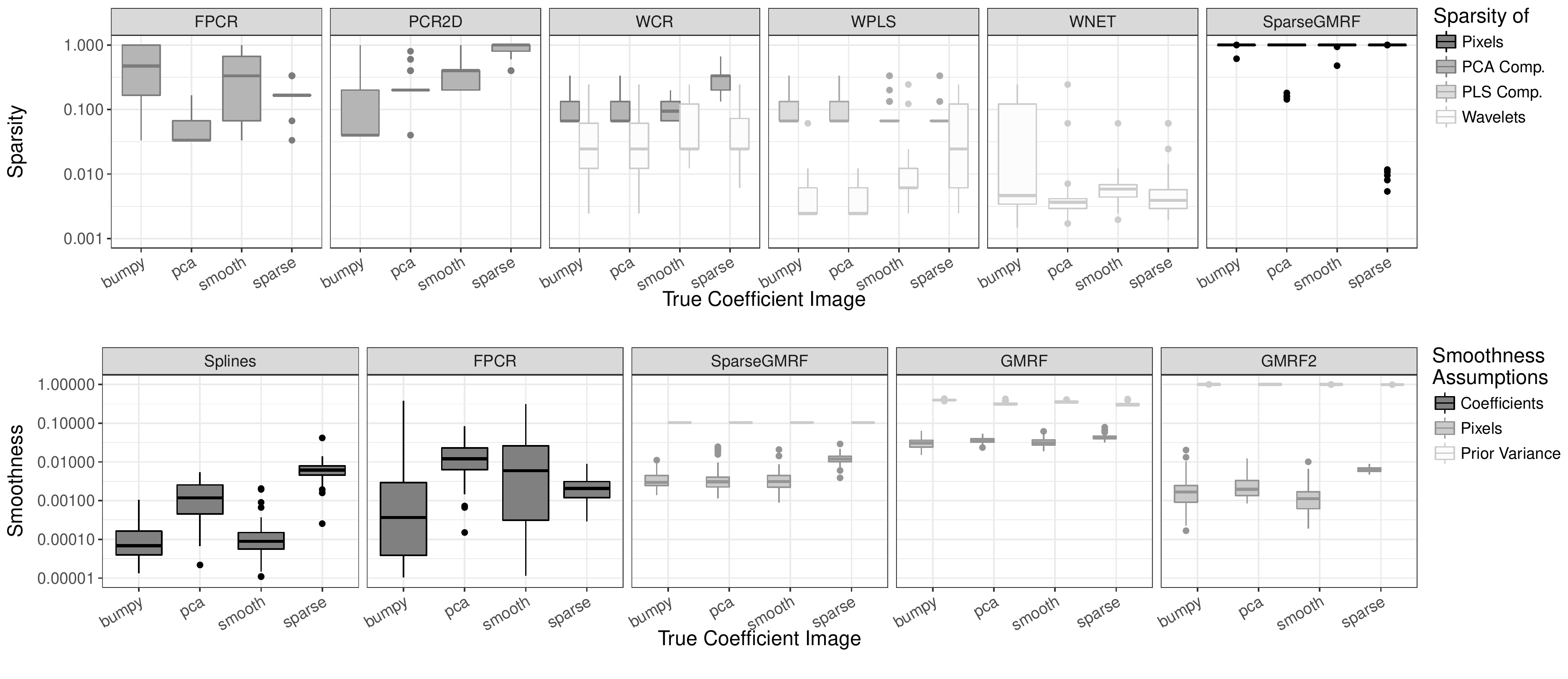}

\captionof{figure}{\protect Measures for parametric model assumptions in the simulation for $N = 250$ observations and $\SNR = 1$ over all $100$ simulation runs. Boxplots show the measures for the different coefficient images depending on the model. All values on log-scale.}
\label{fig:simExplicitAss250_1}
\end{minipage}

\bigskip

\begin{minipage}{\textwidth}
\centering
\includegraphics[width = 0.24\textwidth]{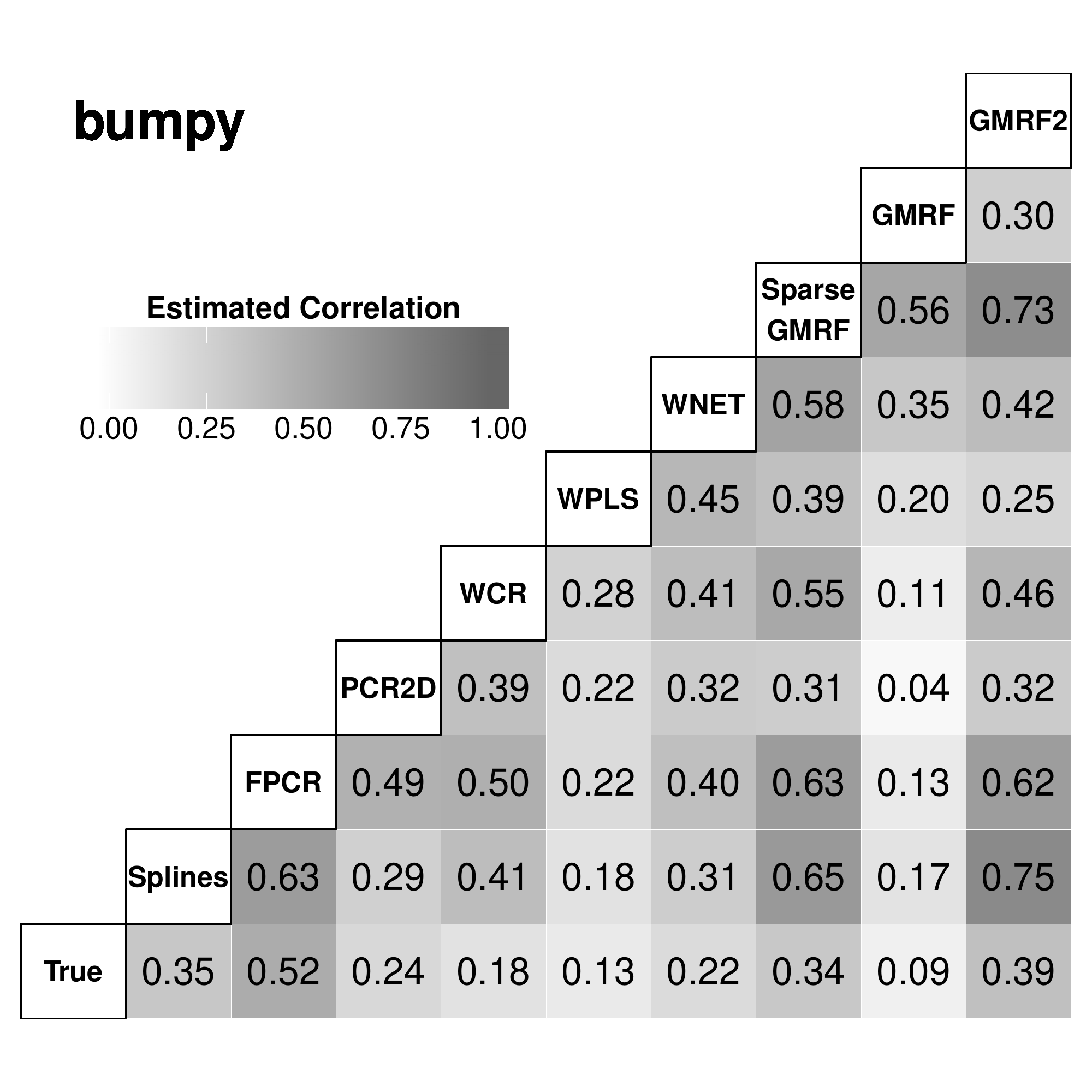}
\includegraphics[width = 0.24\textwidth]{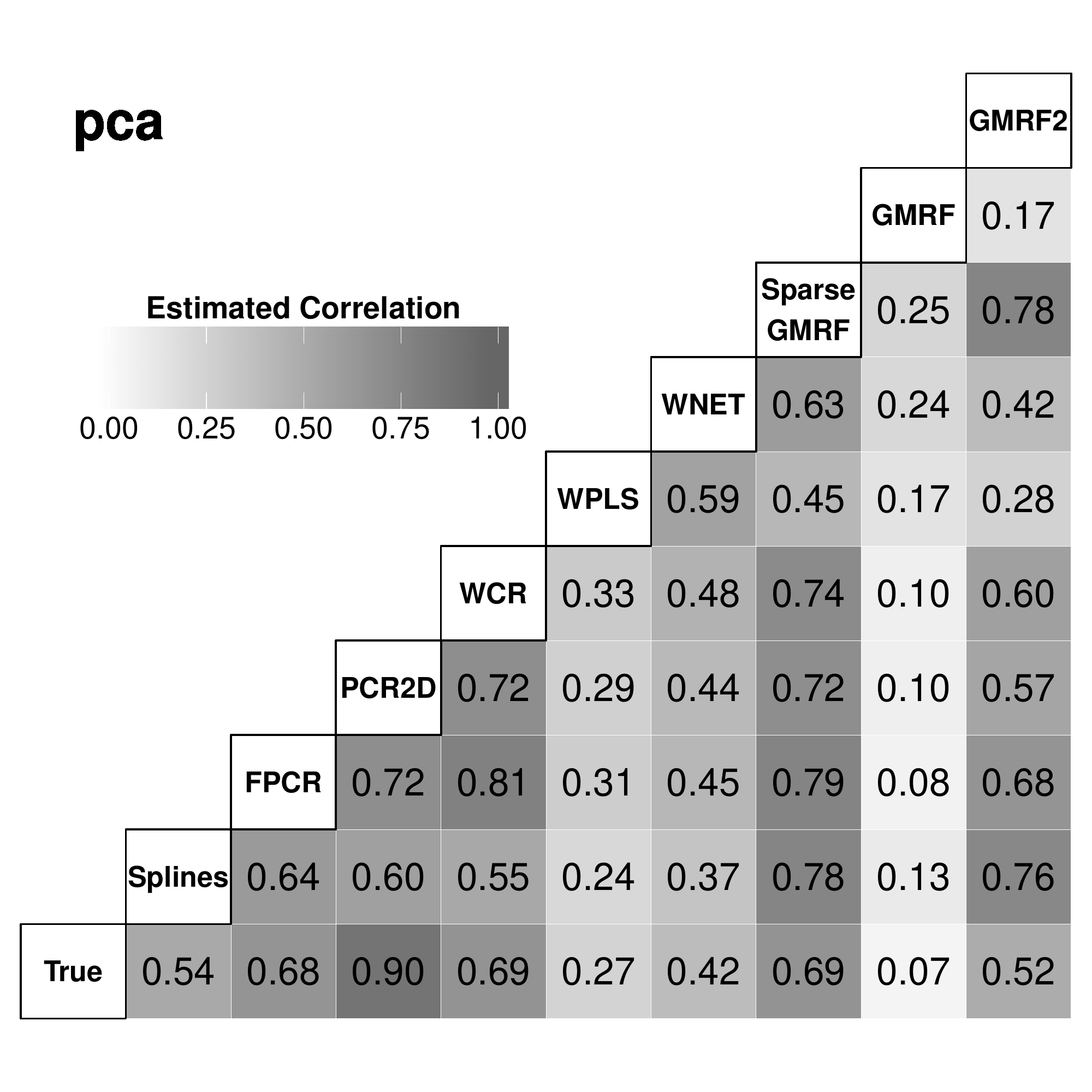}
\includegraphics[width = 0.24\textwidth]{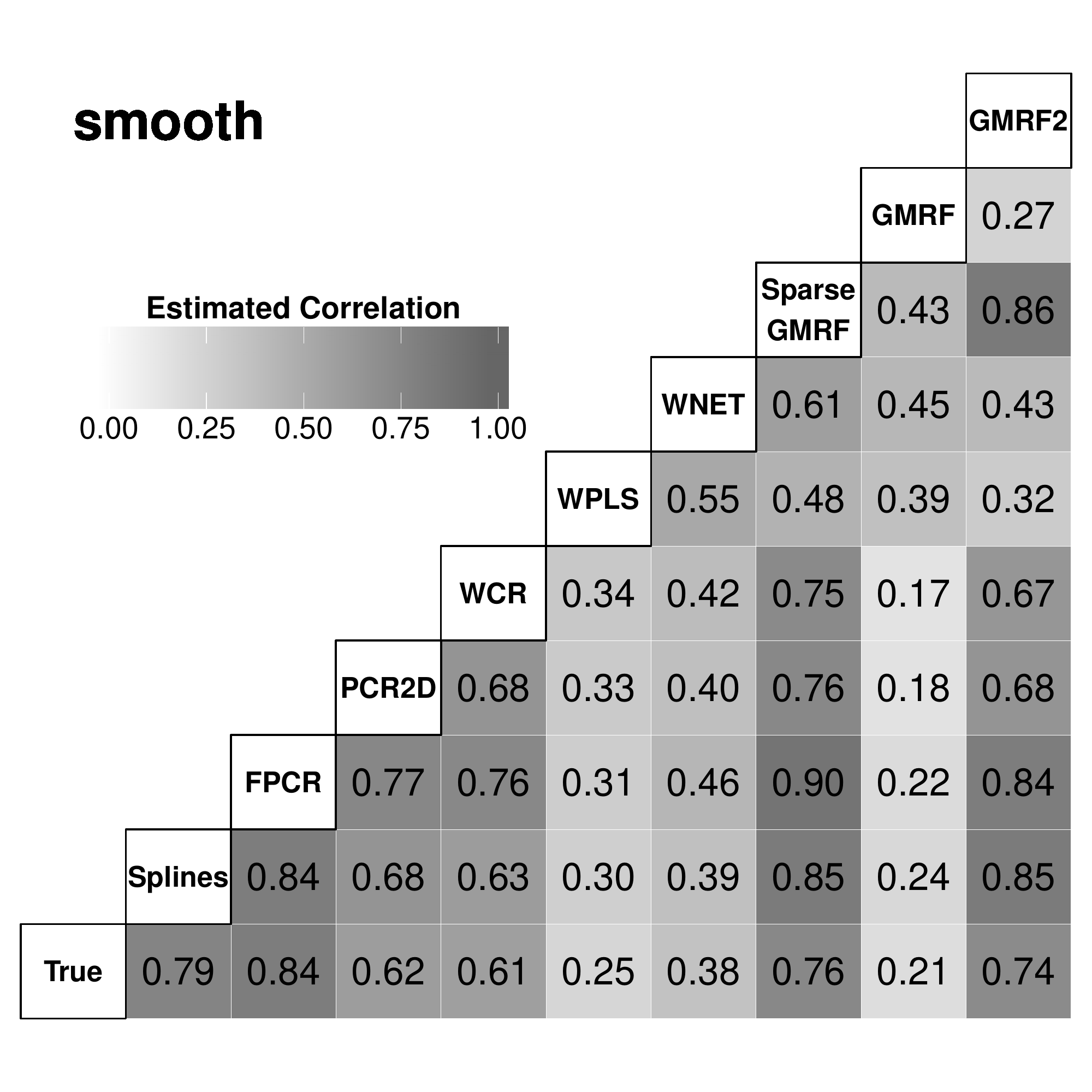}
\includegraphics[width = 0.24\textwidth]{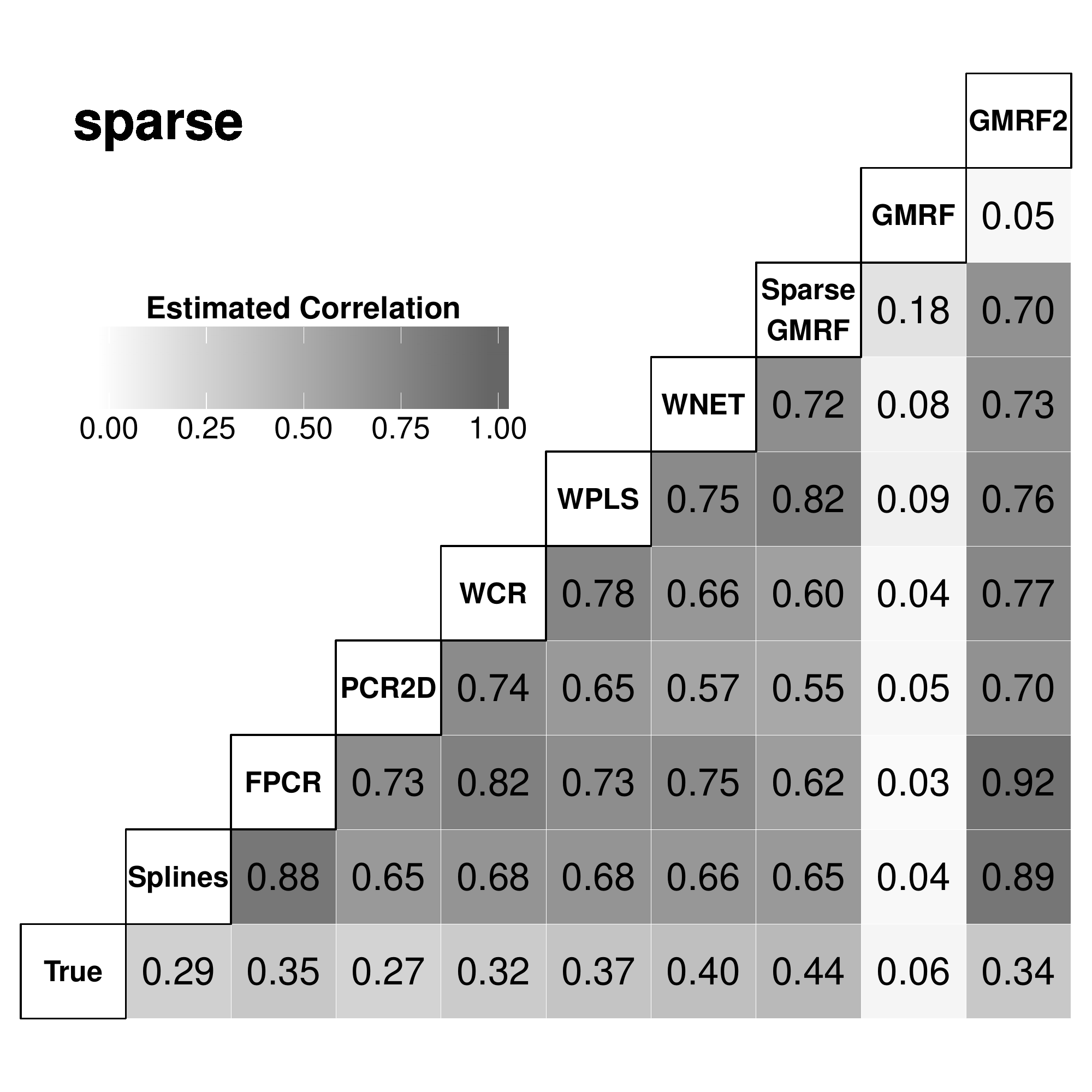}
\captionof{figure}{\protect Median correlation between the true coefficient images and the estimates for $N = 250$ observations and $\SNR = 1$ over all $100$ simulation runs. The figures show the median  correlation of the vectorized images depending on the true images and the models.}
\label{fig:simBetaCorr250_1}
\end{minipage}

\subsubsection{Results for N = 500 and SNR = 4}
\label{sec:res500_4}

\begin{minipage}{\textwidth}
\centering
\includegraphics[width = 0.85\textwidth]{\Path/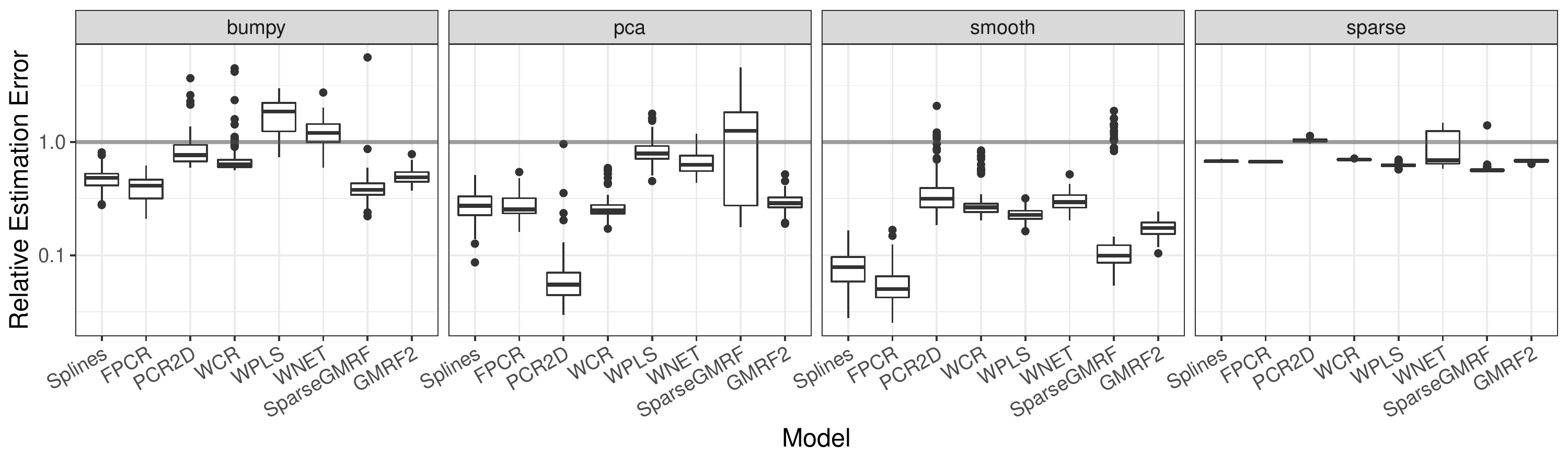}

\captionof{figure}{\protect Relative estimation errors for $N = 500$ observations and $\SNR = 4$ over all $100$ simulation runs.  
Boxplots show the errors for all models except \textit{GMRF} depending on the true coefficient image (\textit{GMRF}: median: $58.75$, sd: $54.75$).
Gray horizontal lines mark $1$, which corresponds to a constant coefficient image, having the average value of the true $\beta$ image.}
\label{fig:estErr_500_4}
\end{minipage}

\bigskip

\begin{minipage}{\textwidth}
\centering
\includegraphics[width = 0.85\textwidth]{\Path/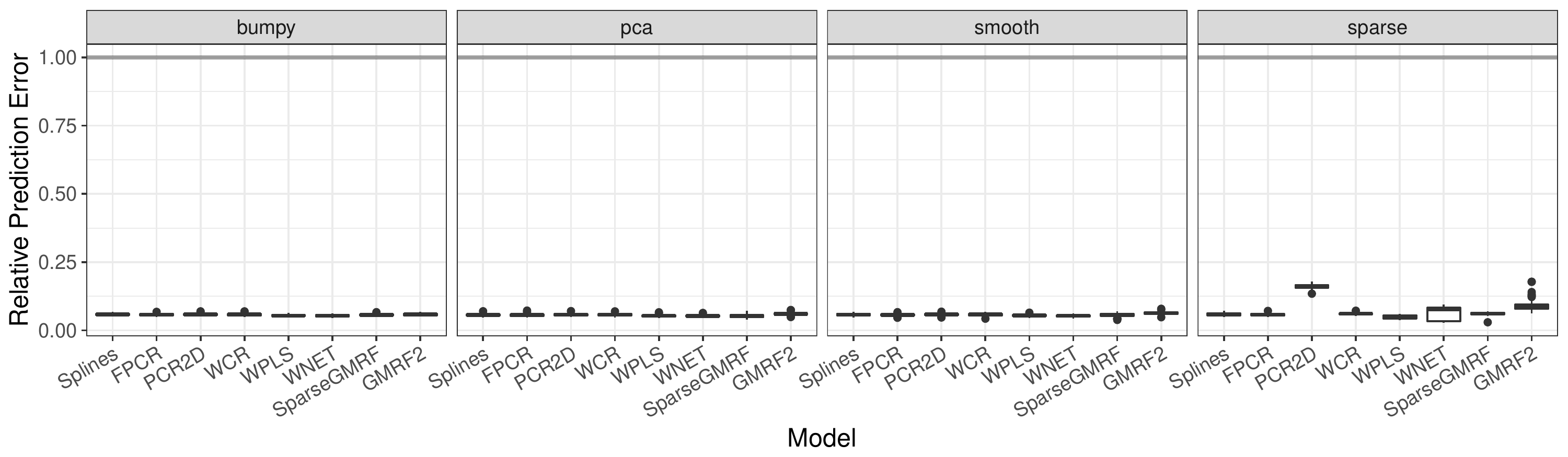}

\captionof{figure}{\protect Relative prediction errors for $N = 500$ observations and $\SNR = 4$ over all $100$ simulation runs.  
Boxplots show the errors for all models except \textit{GMRF} depending on the true coefficient image (\textit{GMRF}: median: $0.69$, sd: $62.22$).
Gray horizontal lines mark $1$, which corresponds to the simple intercept model.}
\label{fig:predErr_500_4}
\end{minipage}

\bigskip

\begin{minipage}{\textwidth}
\centering
\includegraphics[width = 0.85\textwidth]{\Path/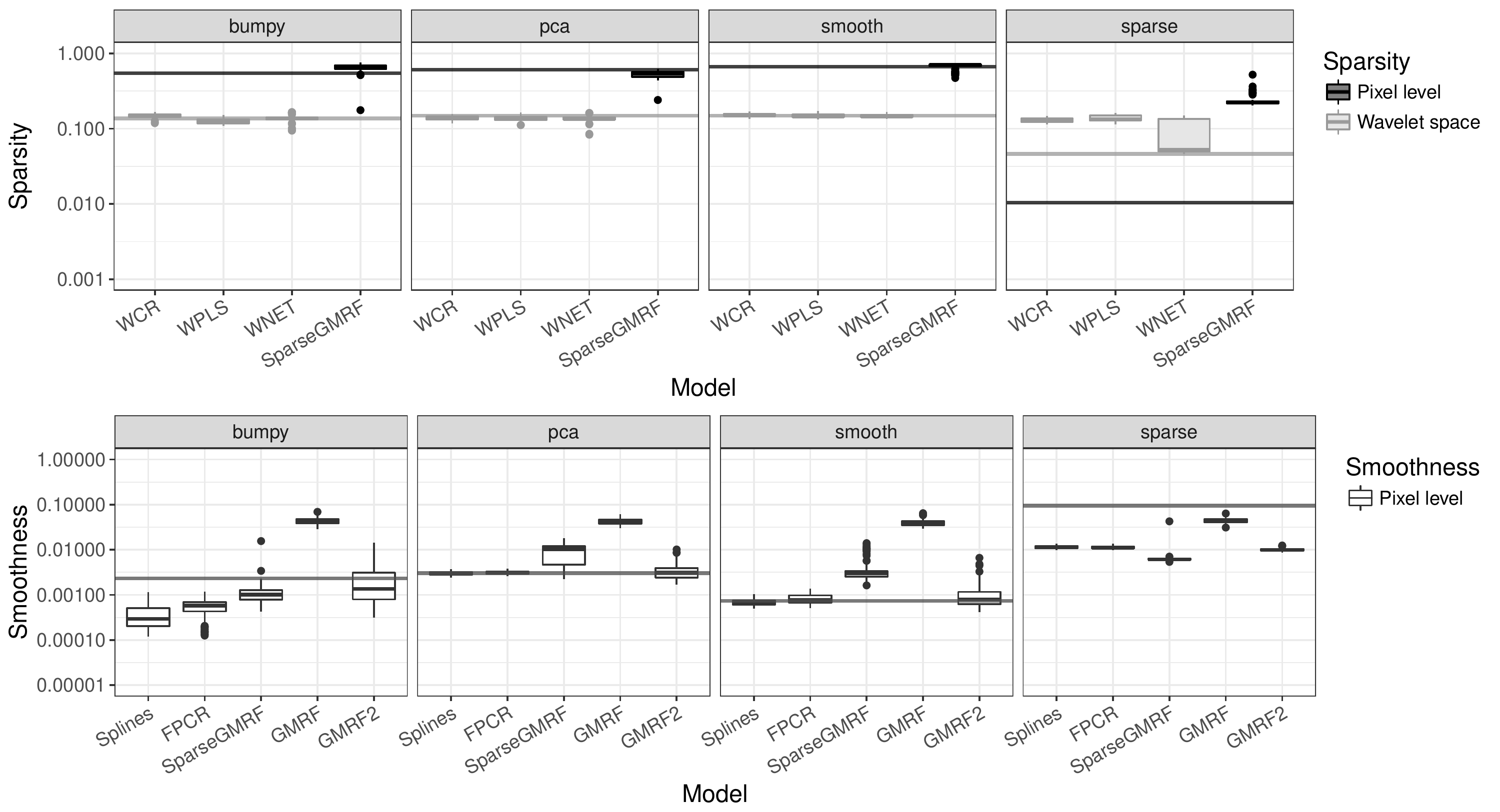}

\captionof{figure}{\protect Measures for underlying model assumptions in the simulation for $N = 500$  observations 
and $\SNR = 4$ over all $100$ simulation runs. Boxplots show the measures for the different models depending on the true coefficient image. All values on log-scale. Gray horizontal lines correspond to the values for the true coefficient images.}
\label{fig:simImplicitAss500_4}
\end{minipage}

\bigskip

\begin{minipage}{\textwidth}
\centering
\includegraphics[width = \textwidth]{\Path/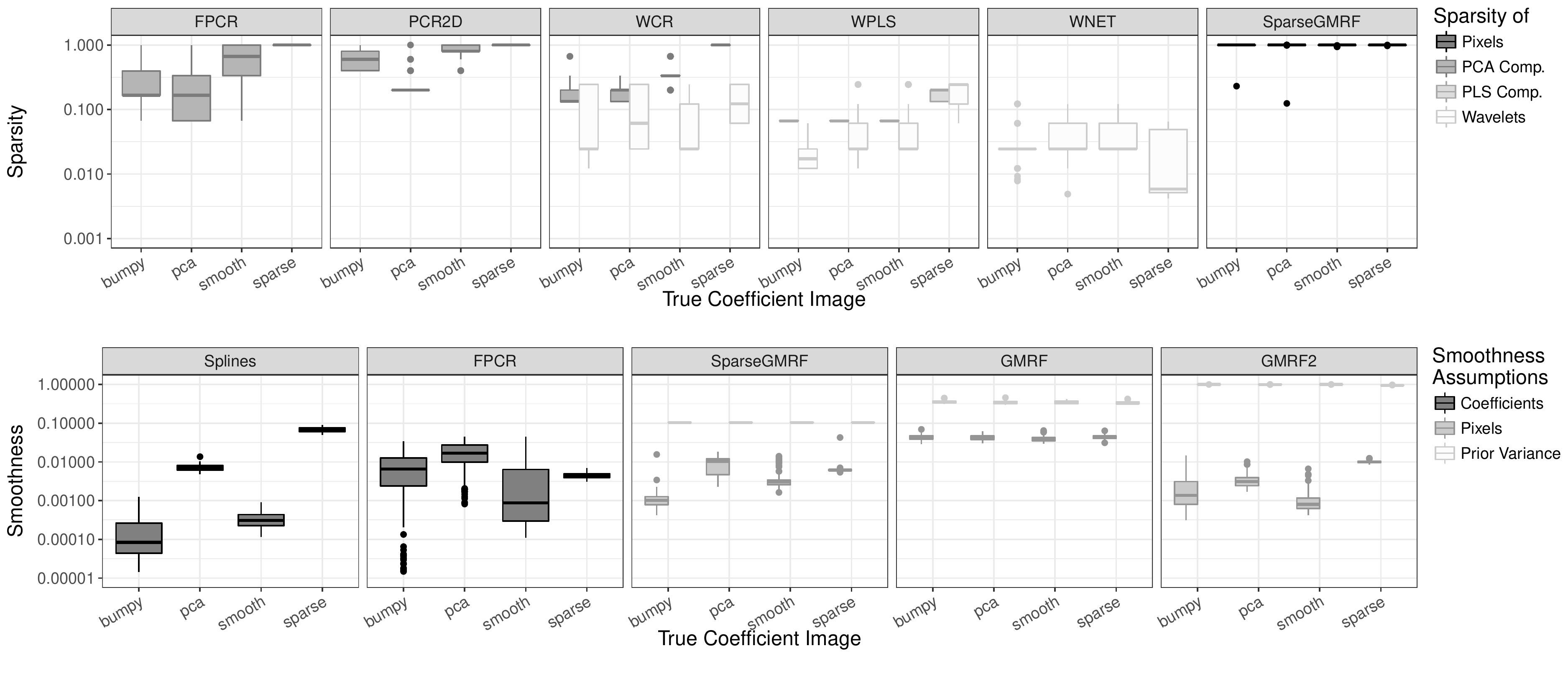}

\captionof{figure}{\protect Measures for parametric model assumptions in the simulation for $N = 500$ observations 
and $\SNR = 4$ over all $100$ simulation runs. Boxplots show the measures for the different coefficient images depending on the model. All values on log-scale.}
\label{fig:simExplicitAss500_4}
\end{minipage}

\bigskip

\begin{minipage}{\textwidth}
\centering
\includegraphics[width = 0.24\textwidth]{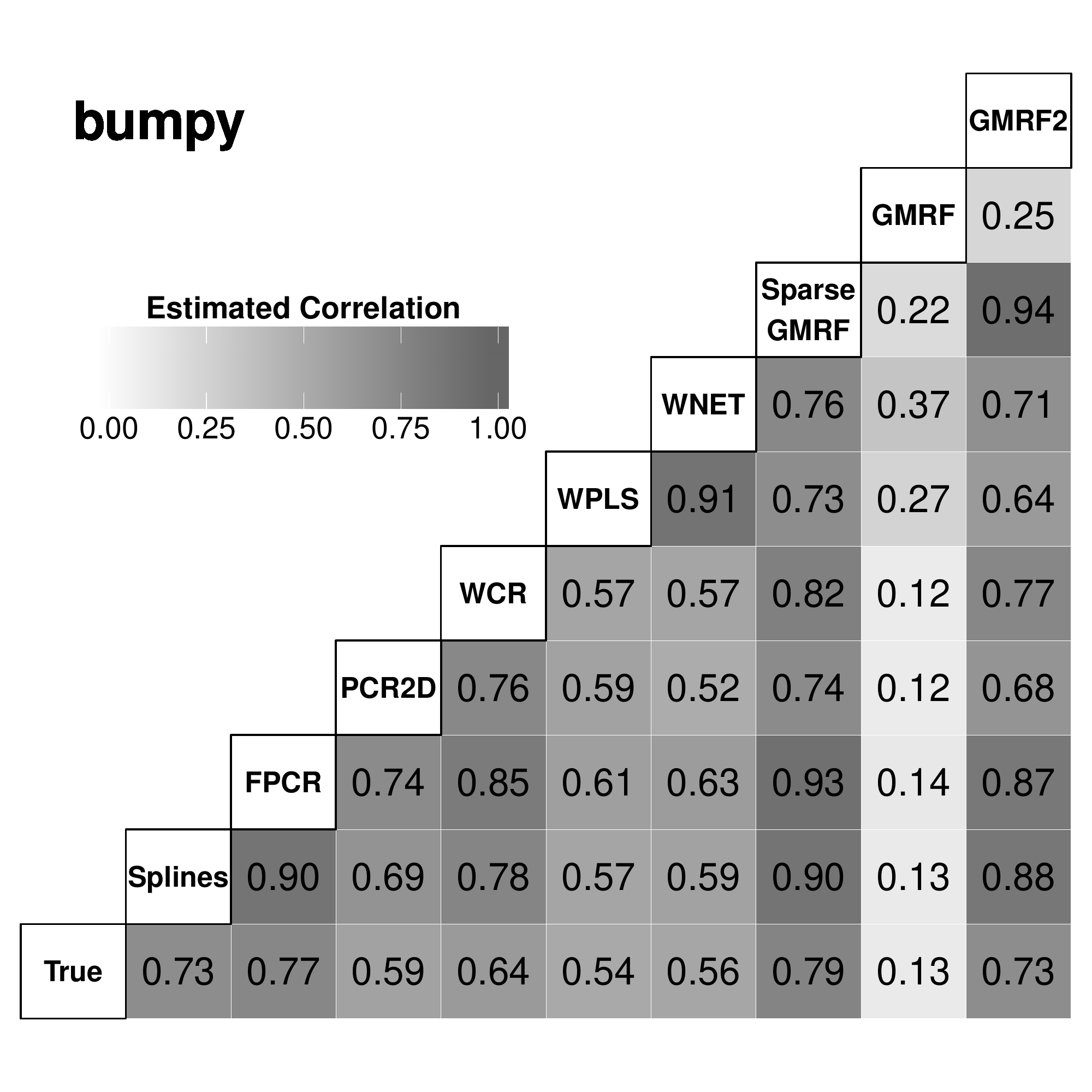}
\includegraphics[width = 0.24\textwidth]{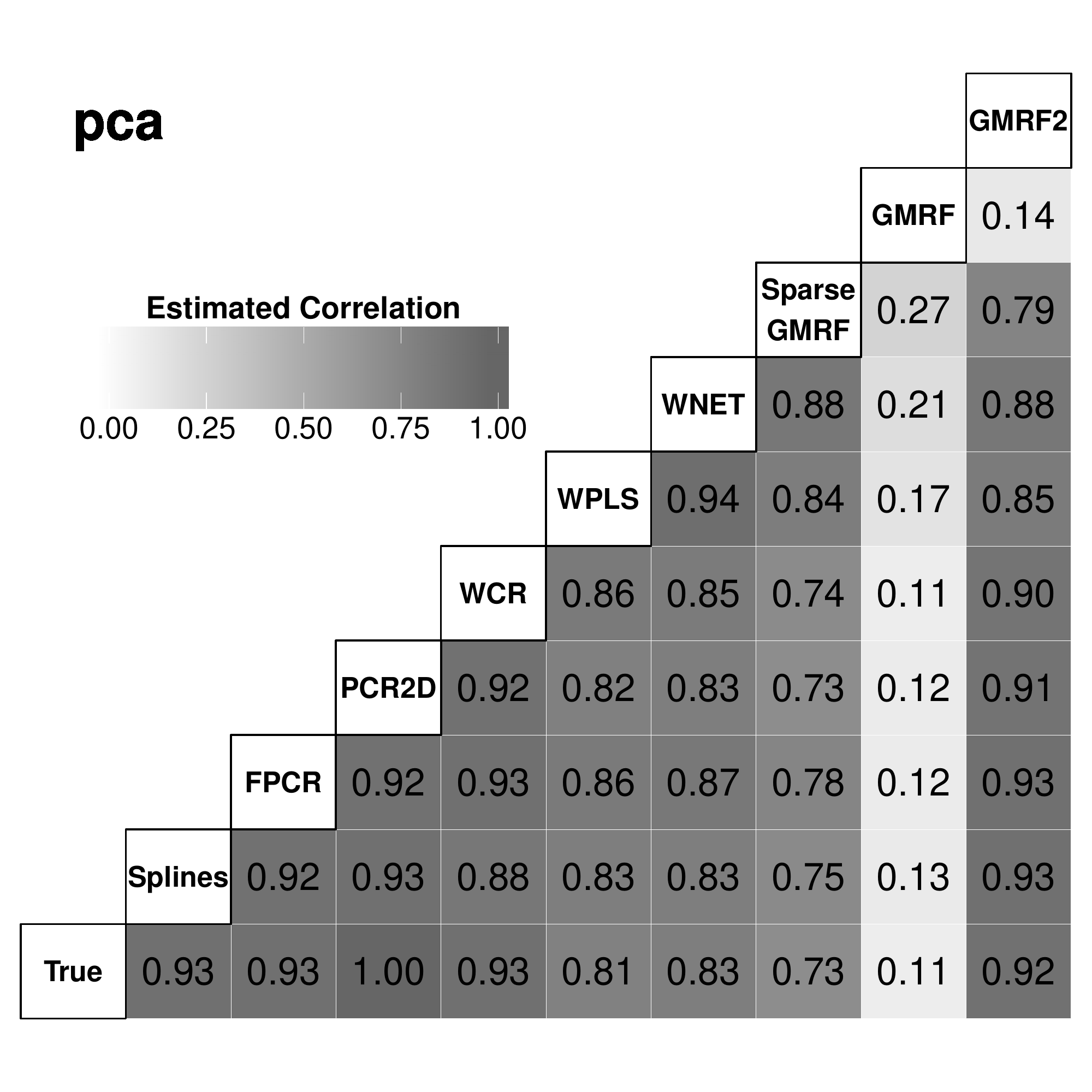}
\includegraphics[width = 0.24\textwidth]{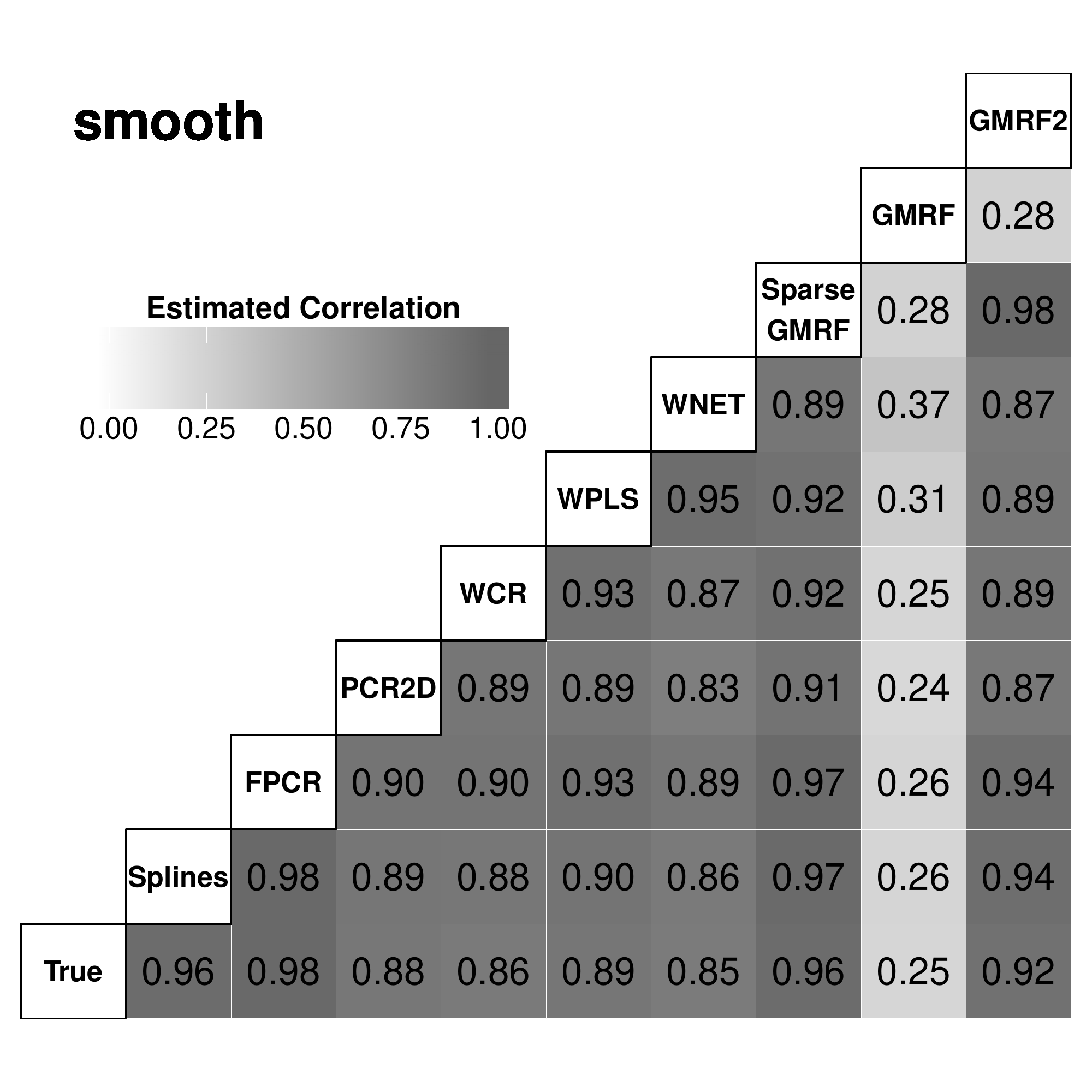}
\includegraphics[width = 0.24\textwidth]{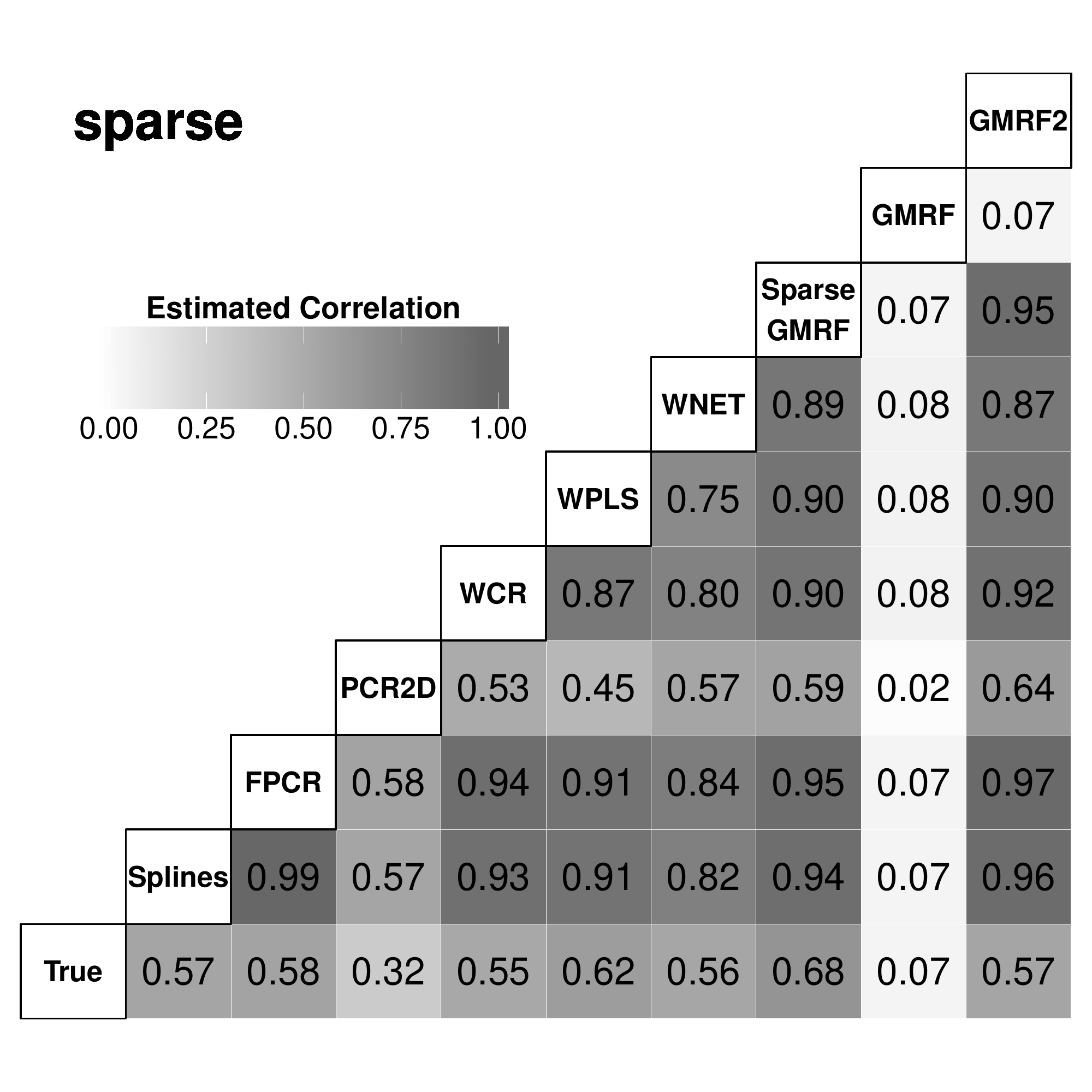}
\captionof{figure}{\protect Median correlation between the true coefficient images and the estimates for $N = 500$ observations and
 $\SNR = 4$ over all $100$ simulation runs. The figures show the median  correlation of the vectorized images depending on the true images and the models.}
\label{fig:simBetaCorr500_4}
\end{minipage}

\subsubsection{Results for N = 500 and SNR = 1}
\label{sec:res500_1}

\begin{minipage}{\textwidth}
\centering
\includegraphics[width = 0.85\textwidth]{\Path/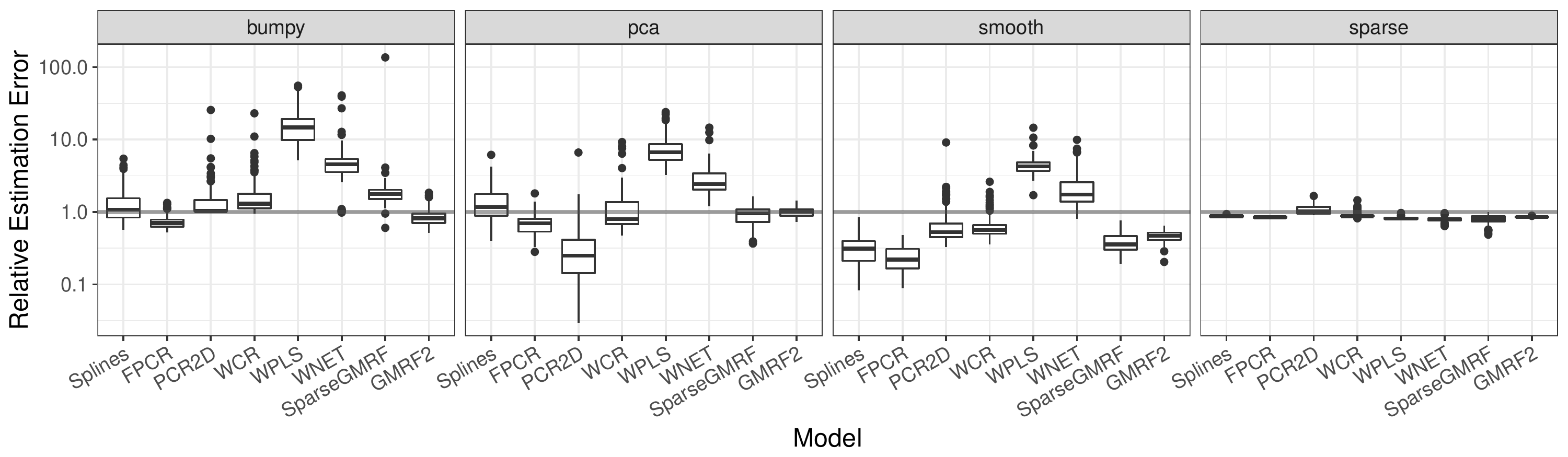}

\captionof{figure}{\protect Relative estimation errors for $N = 500$ observations and $\SNR = 1$ over all $100$ simulation runs.  
Boxplots show the errors for all models except \textit{GMRF} depending on the true coefficient image (\textit{GMRF}: median: $73.31$, sd: $78.83$).
Gray horizontal lines mark $1$, which corresponds to a constant coefficient image, having the average value of the true $\beta$ image.}
\label{fig:estErr_500_1}
\end{minipage}

\bigskip

\begin{minipage}{\textwidth}
\centering
\includegraphics[width = 0.85\textwidth]{\Path/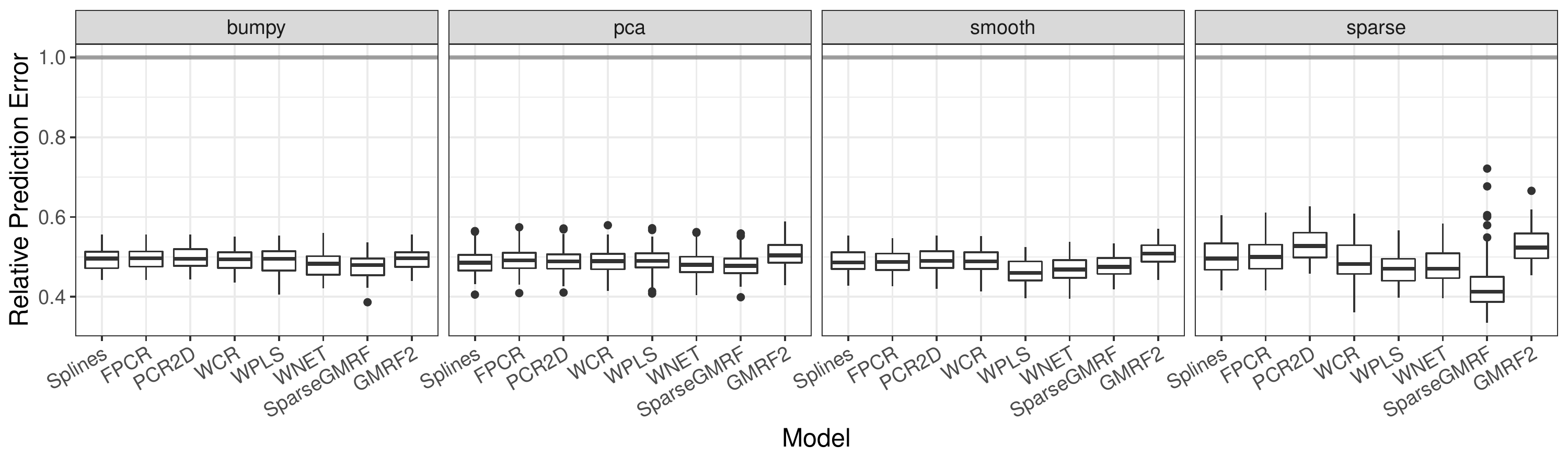}

\captionof{figure}{\protect Relative prediction errors for $N = 500$ observations and $\SNR = 1$ over all $100$ simulation runs.  
Boxplots show the errors for all models except \textit{GMRF} depending on the true coefficient image (\textit{GMRF}: median: $0.70$, sd: $37.90$).
Gray horizontal lines mark $1$, which corresponds to the simple intercept model.}
\label{fig:predErr_500_1}
\end{minipage}

\bigskip

\begin{minipage}{\textwidth}
\centering
\includegraphics[width = 0.85\textwidth]{\Path/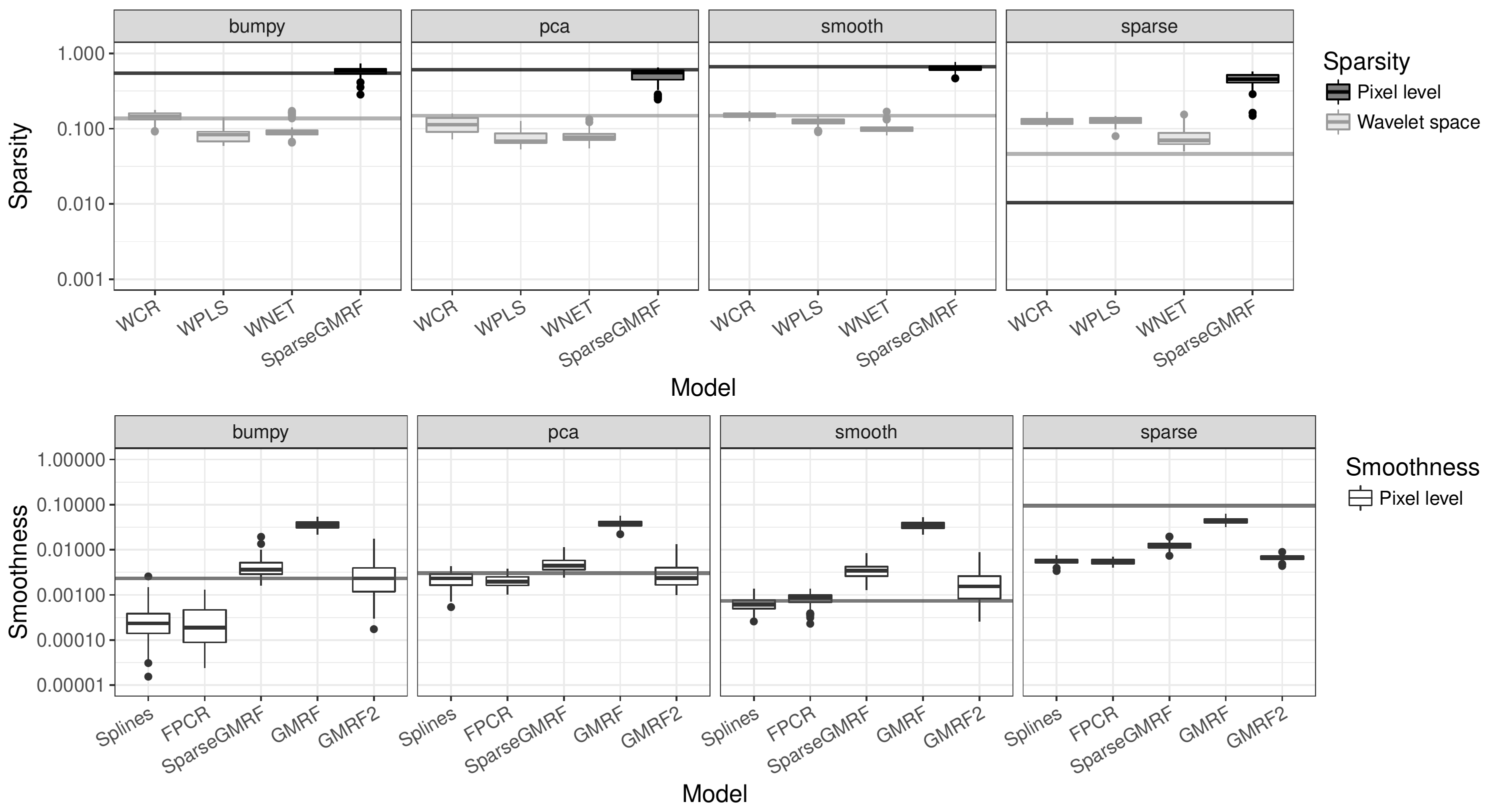}

\captionof{figure}{\protect Measures for underlying model assumptions in the simulation for $N = 500$  observations 
and $\SNR = 1$ over all $100$ simulation runs. Boxplots show the measures for the different models depending on the true coefficient image. All values on log-scale. Gray horizontal lines correspond to the values for the true coefficient images.}
\label{fig:simImplicitAss500_1}
\end{minipage}

\bigskip

\begin{minipage}{\textwidth}
\centering
\includegraphics[width = \textwidth]{\Path/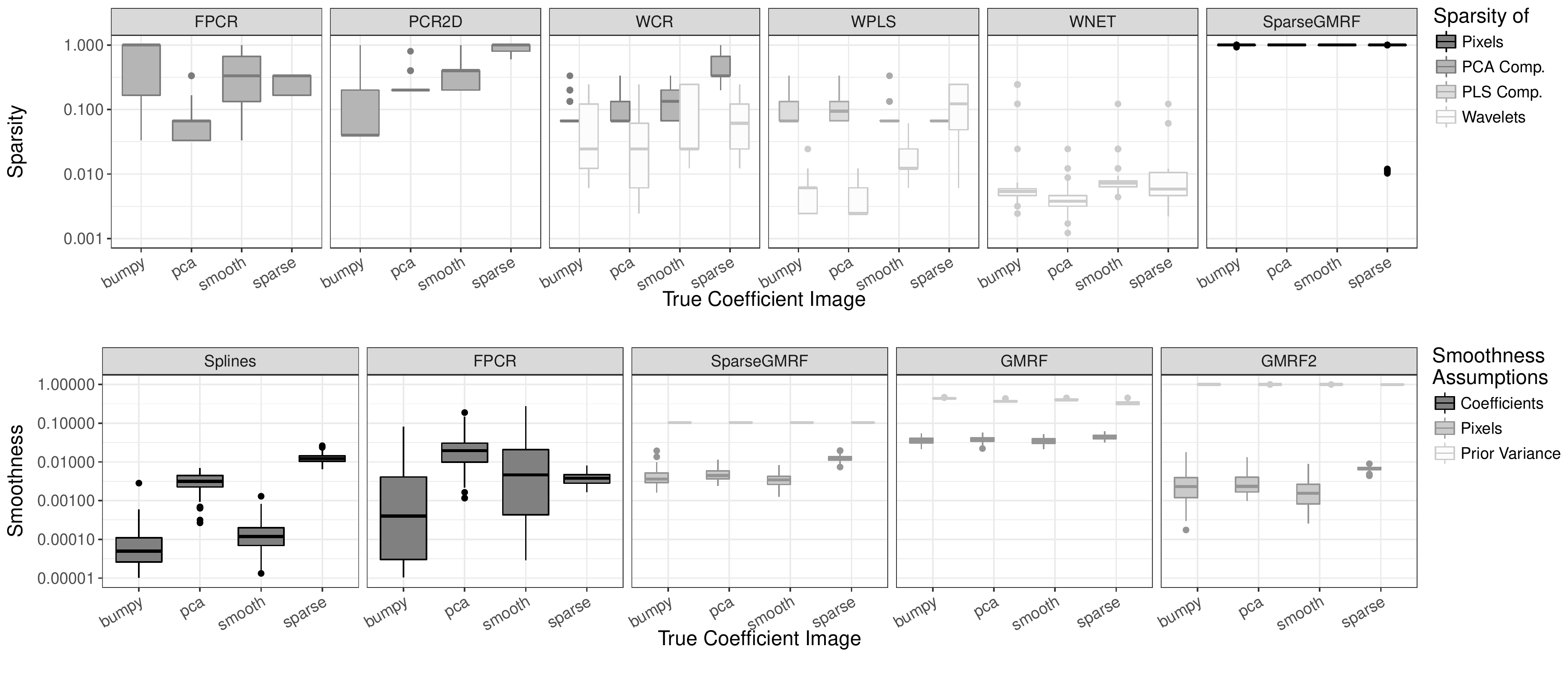}

\captionof{figure}{\protect Measures for parametric model assumptions in the simulation for $N = 500$ observations 
and $\SNR = 1$ over all $100$ simulation runs. Boxplots show the measures for the different coefficient images depending on the model. All values on log-scale.}
\label{fig:simExplicitAss500_1}
\end{minipage}

\bigskip

\begin{minipage}{\textwidth}
\centering
\includegraphics[width = 0.24\textwidth]{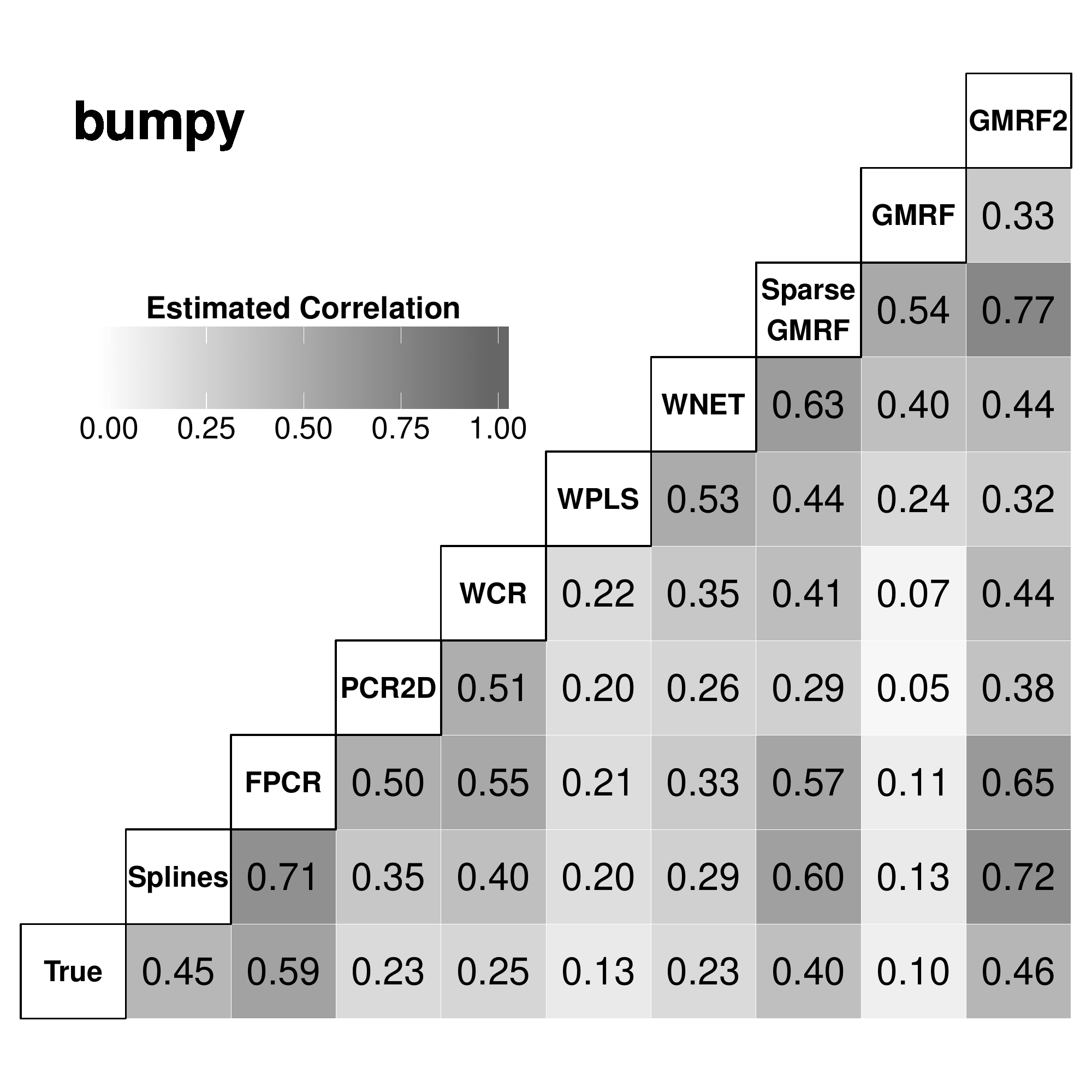}
\includegraphics[width = 0.24\textwidth]{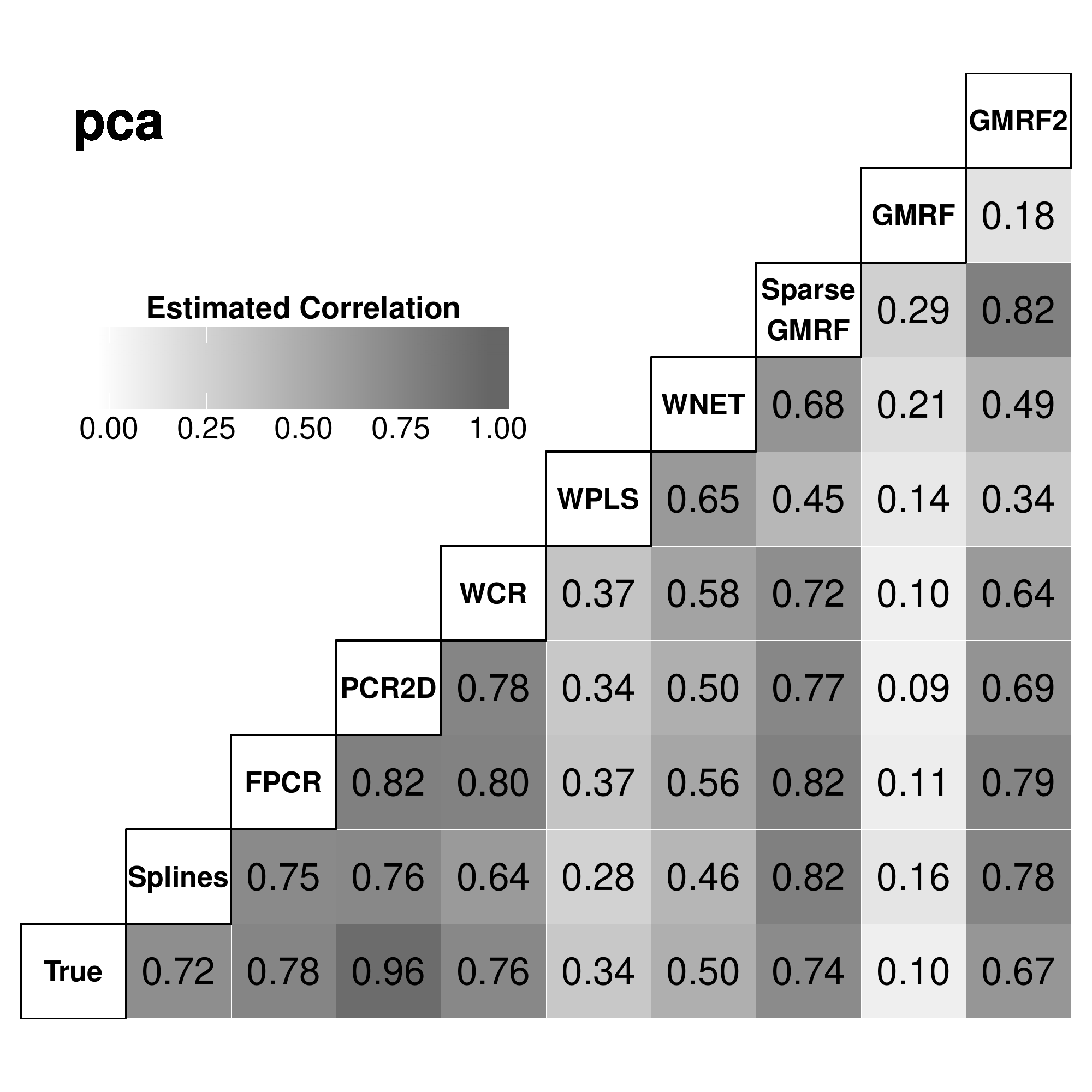}
\includegraphics[width = 0.24\textwidth]{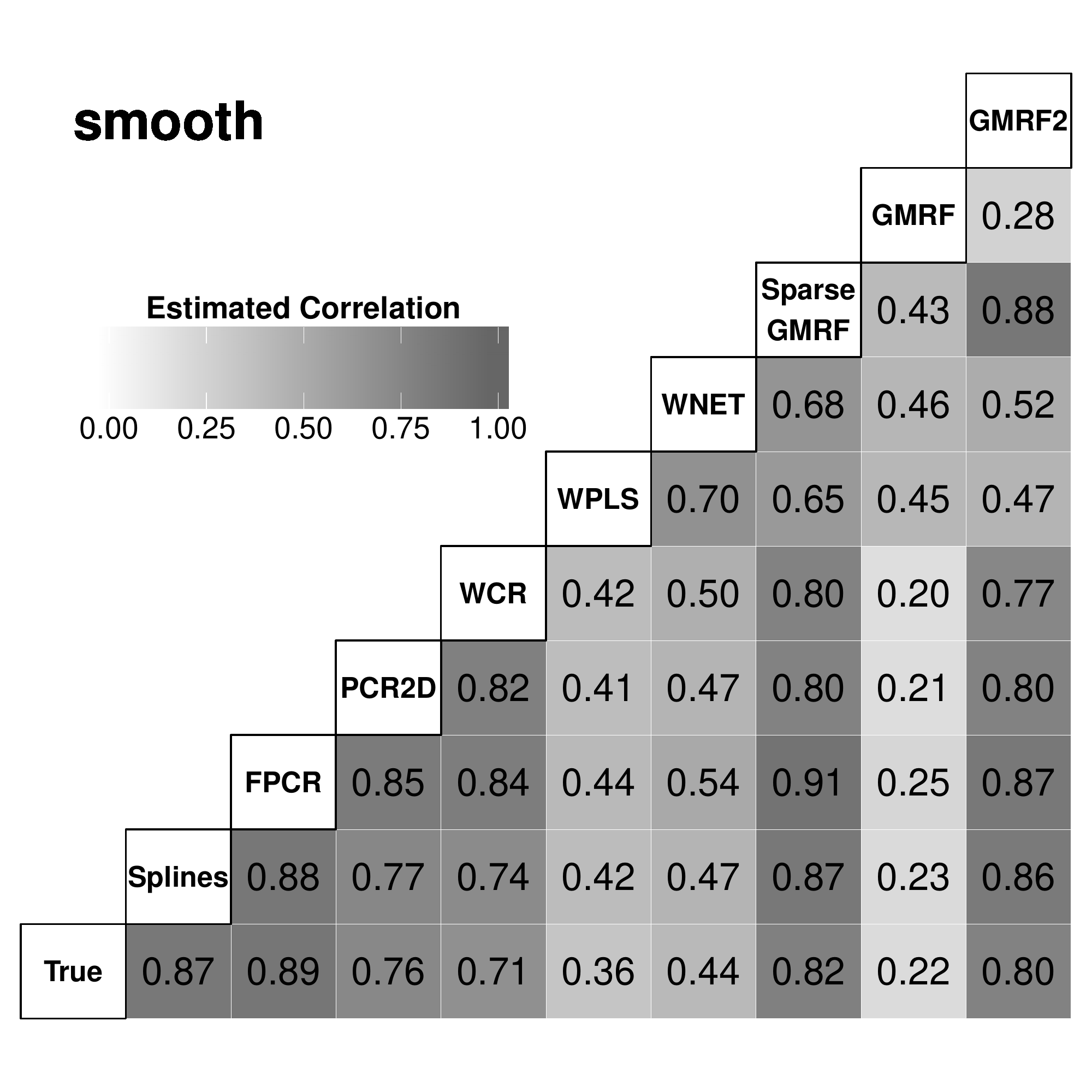}
\includegraphics[width = 0.24\textwidth]{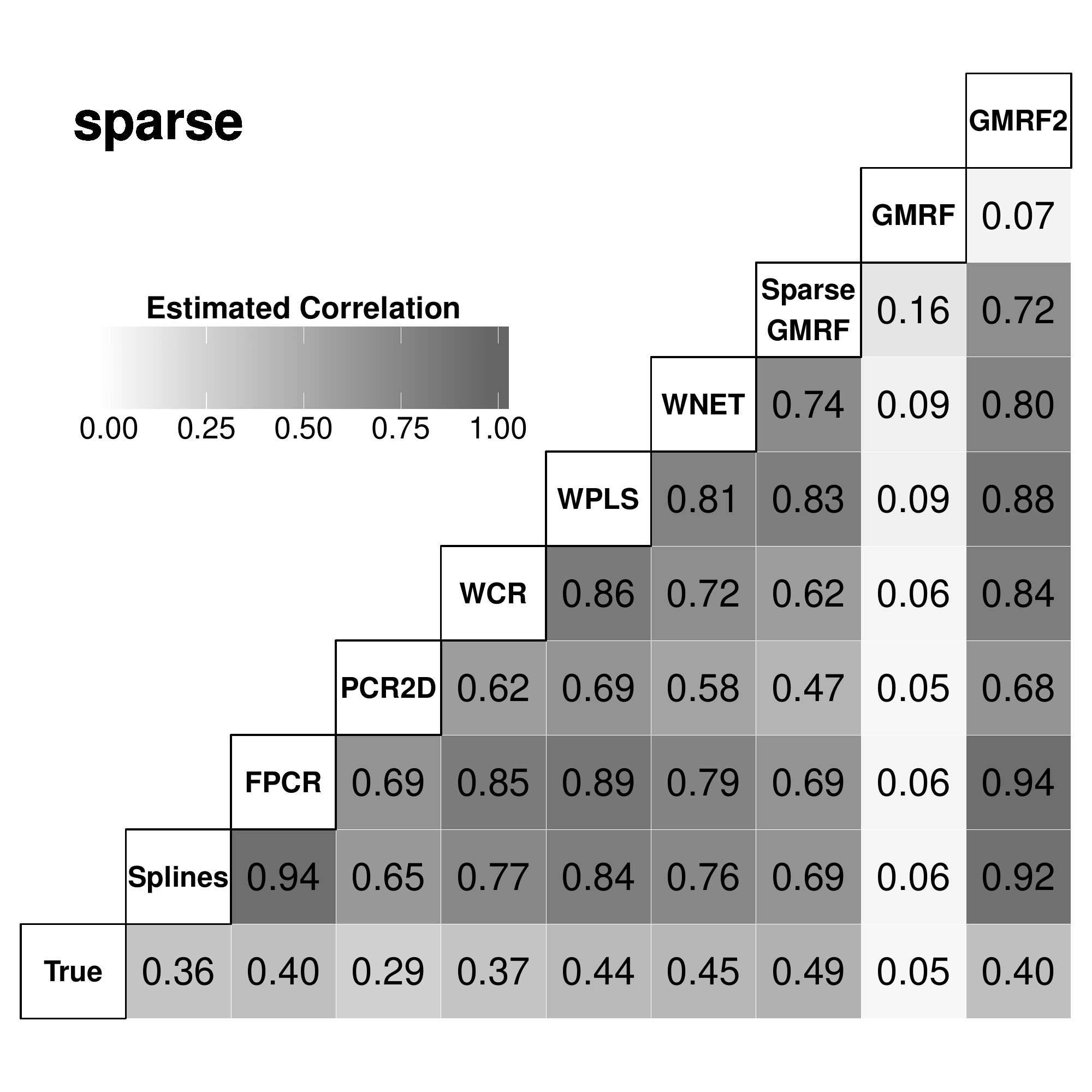}
\captionof{figure}{\protect Median correlation between the true coefficient images and the estimates for $N = 500$ observations and
 $\SNR = 1$ over all $100$ simulation runs. The figures show the median  correlation of the vectorized images depending on the true images and the models.}
\label{fig:simBetaCorr500_1}
\end{minipage}

\subsubsection{Computation Times}

\begin{minipage}{\textwidth}
\begin{center}
\includegraphics[width = 0.35\textwidth]{\Path/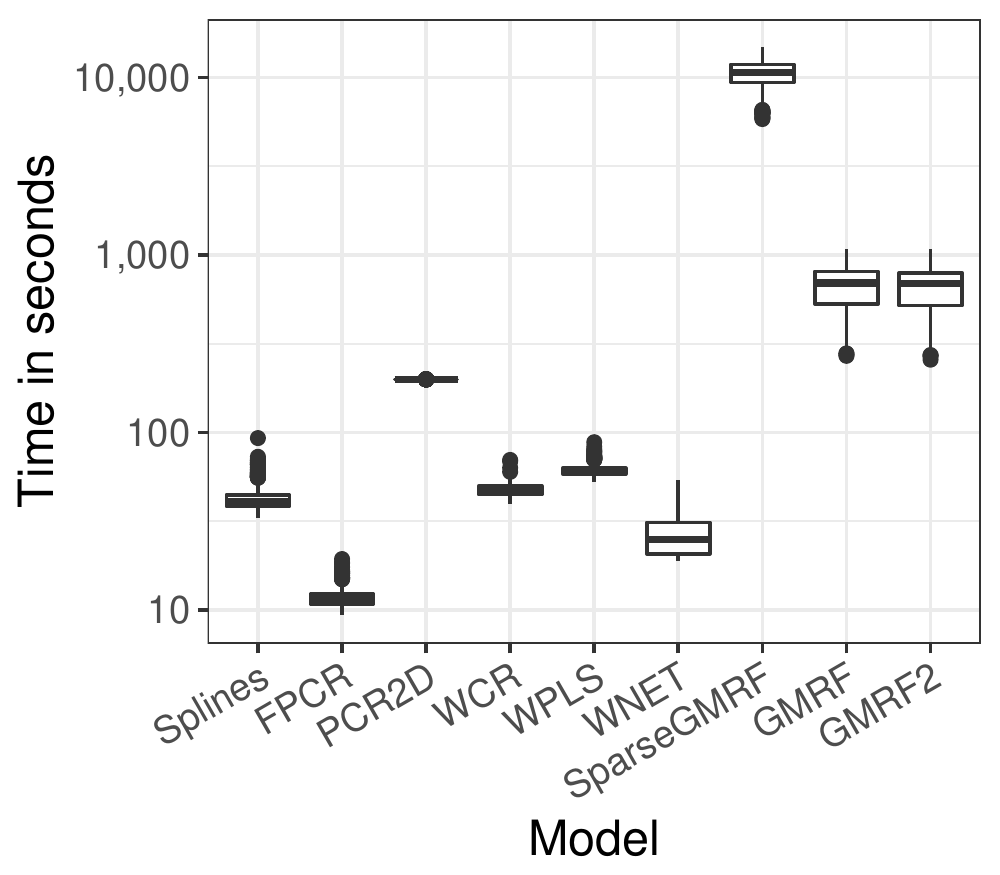}
\hspace*{0.5cm}
\includegraphics[width = 0.35\textwidth]{\Path/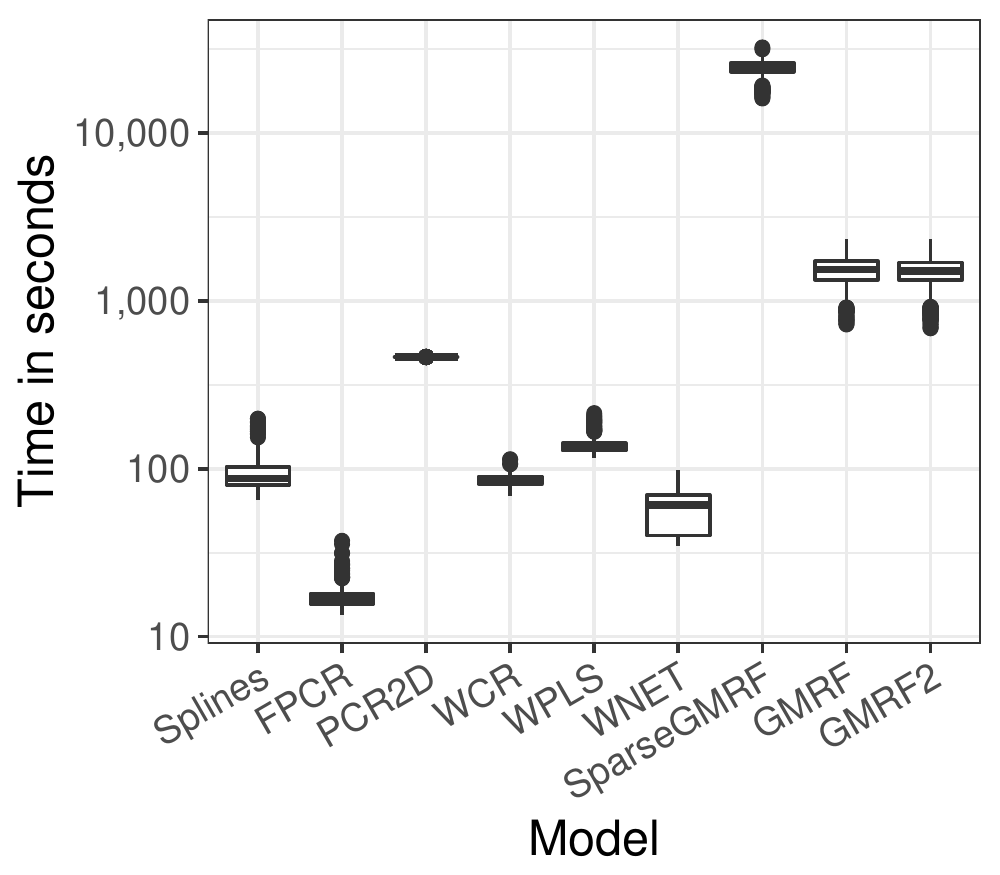}
\end{center}
\captionof{figure}{\protect Computation times for all nine models and $N = 250$ (left) / $N = 500$ (right) observations over all $100$ simulation runs. The boxplots contain the merged values for all coefficient images and signal-to-noise ratios.}
\label{fig:compTime250}
\label{fig:compTime500}
\end{minipage}

 \FloatBarrier
 \newpage

\subsection{Sensitivity Study}
\label{sec:sensStudy}

\begin{figure}[ht]
\includegraphics[width = \textwidth]{\Path/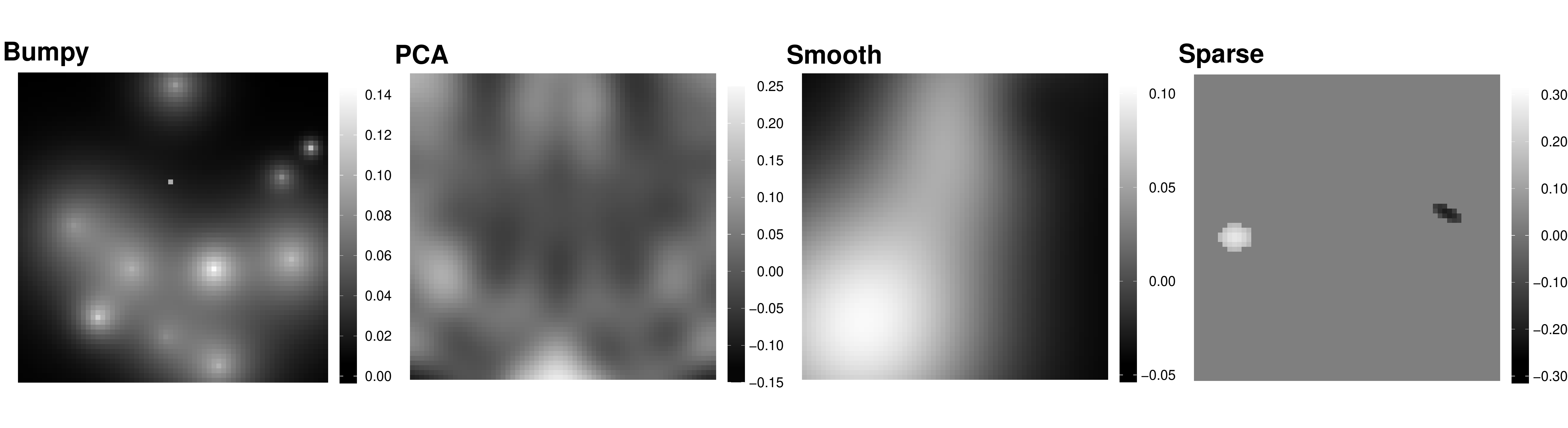}
\caption{\protect\input{\Path/captions/sensBeta}}
\label{fig:sensBeta}
\end{figure}

The results in Section~\ref{sec:sim} have been obtained for fixed coefficient images. As the covariate images $x_i$ do not have a constant variation over all pixels, some features of $\beta$ might be easier to find than others, notably if they are in areas with high variation and thus more information. In order to study the sensitivity of the results with respect to the spatial structure of $\beta$, a second study was conducted for $N = 250$ and $\SNR = 4$ with spatially varying coefficient images. Therefore, a new coefficient image was generated in each iteration of the simulation, sampling the locations of the features randomly (for \textit{bumpy, smooth} and \textit{sparse}) or with a random number of principal components and randomly chosen coefficients $b_k$ (for \textit{pca}). Examples for one iteration are shown in Fig.~\ref{fig:sensBeta}. 
In this study, we consider all models except for \textit{GMRF} due to extreme error rates in the first simulation study and \textit{SparseGMRF} due to long computations. The results are given in Fig.~\ref{fig:sensRes} (error rates) and~\ref{fig:sensBetaCorr} (correlations of the estimates with the true coefficient image and across models). Boxplots of the measures for underlying and parametric model assumptions are given in Figs.~\ref{fig:sensImplicitAss},~\ref{fig:sensExplicitAss} and~\ref{fig:sensProj}. 

Overall, the results are very similar to the ones from the previous simulation study with fixed image covariates. This shows that variations in the features of the true coefficient images $\beta$ have only marginal influence on the simulation results. Notable differences are found for the parametric model measures concerning principal components as well as in the results for the \textit{pca} coefficient image. This is plausible, as for varying coefficient images $\beta$, different numbers of principal components might be optimal in the   \textit{FPCR}, \textit{PCR2D} and \textit{WCR} models. For \textit{pca}, the higher variation can be explained by the fact that for this coefficient image, the number of eigenimages and their coefficients are resampled for generating new images $\beta$ and hence may lead to a higher variation.

\begin{minipage}{\textwidth}
\begin{center}
\includegraphics[width = 0.85\textwidth]{\Path/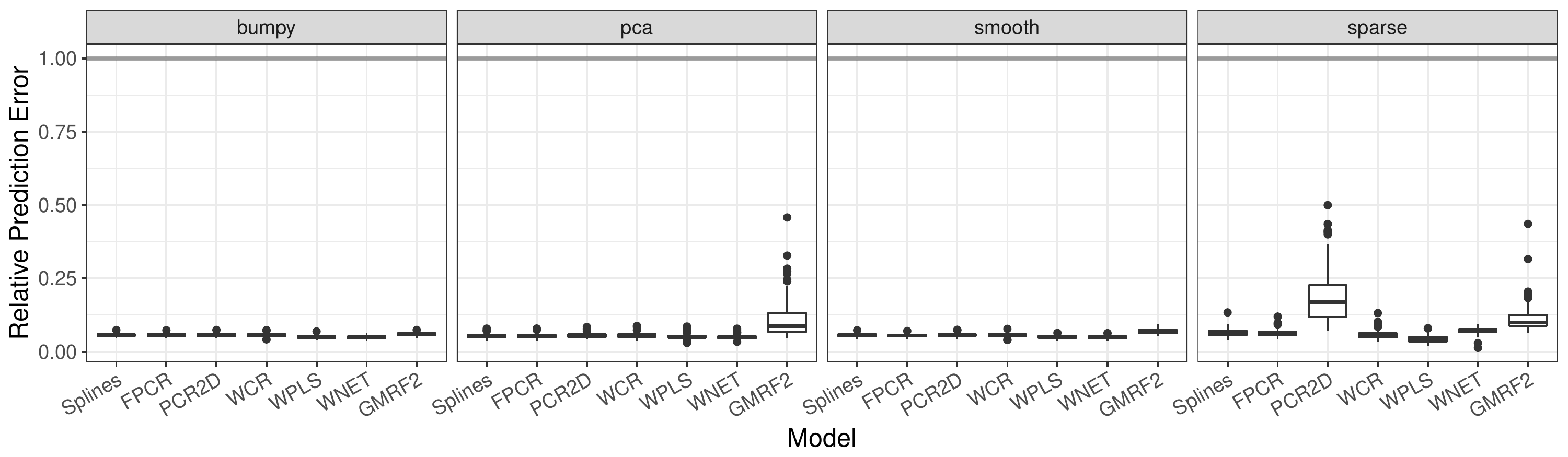}

\includegraphics[width = 0.85\textwidth]{\Path/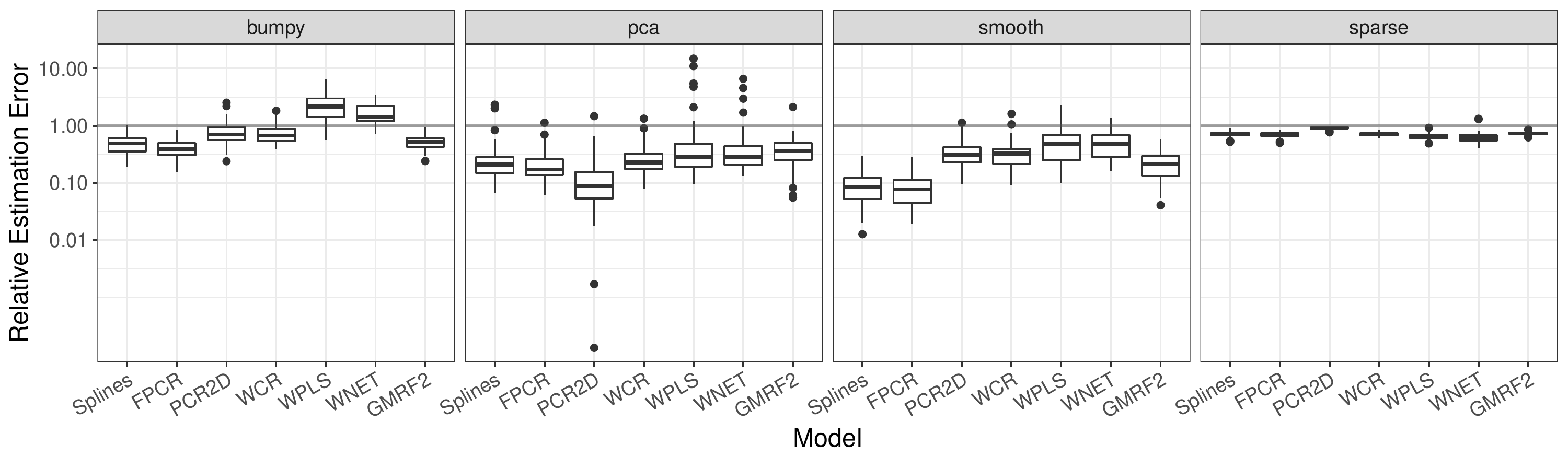}
\end{center}
\captionof{figure}{\protect Results of the sensitivity study. 
Boxplots show the relative prediction and estimation error for all seven models depending on the coefficient image over all $100$ simulation runs. 
Gray horizontal lines mark $1$, which corresponds to the simple intercept model (for prediction error) or to a constant coefficient image, having the average value of the true $\beta$ image (estimation error).}
\label{fig:sensRes}
\end{minipage}

\bigskip

\begin{minipage}{\textwidth}
\begin{center}
\includegraphics[width = 0.85\textwidth]{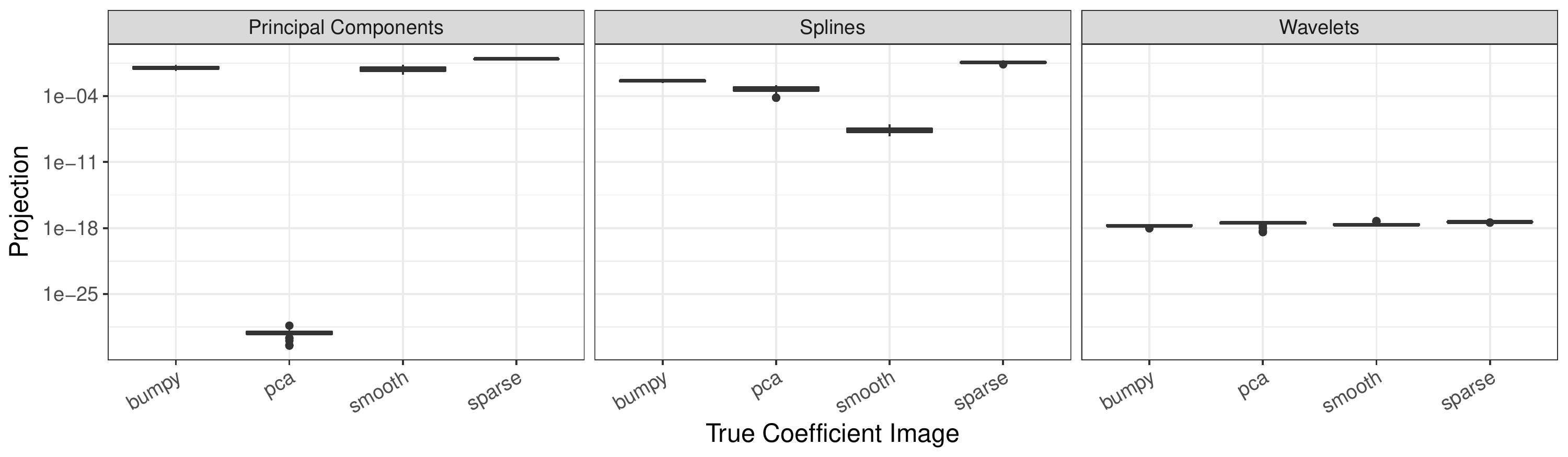}
\end{center}
\captionof{figure}{\protect Values of $m_\text{Projection}$ in the sensitivity study for the different coefficient images  depending on the basis functions used.}
\label{fig:sensProj}
\end{minipage}

\bigskip

\begin{minipage}{\textwidth}
\begin{center}
\includegraphics[width = 0.85\textwidth]{\Path/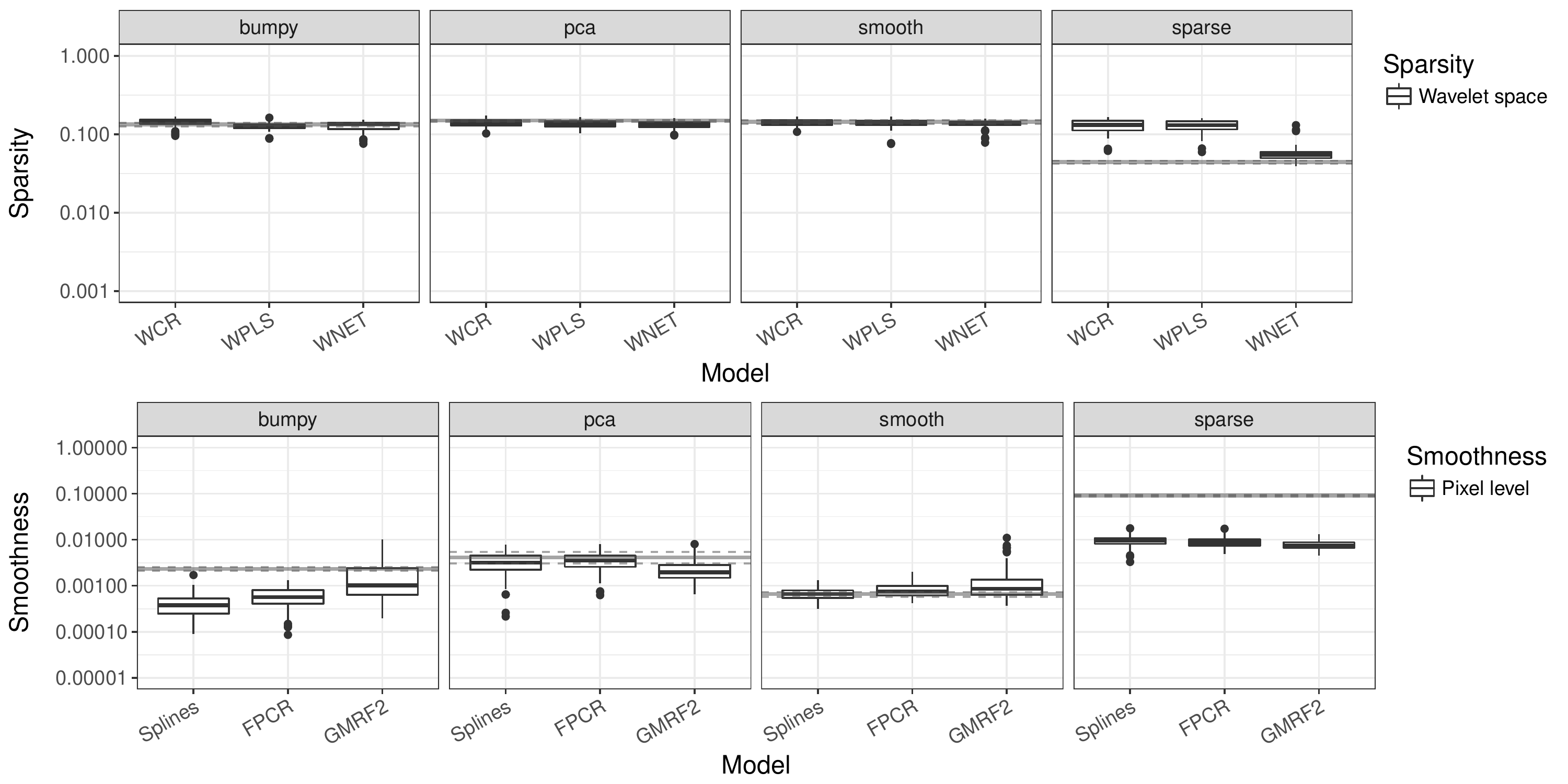}
\end{center}
\captionof{figure}{\protect Measures for underlying model assumptions in the sensitivity study. Boxplots show the measures for the different models depending on the true coefficient image over all $100$ simulation runs. All values on log-scale. Gray horizontal lines correspond to the median (solid line) and the $25\%$ and $75\%$ quantiles (dashed lines) for the true coefficient images.}
\label{fig:sensImplicitAss}
\end{minipage}

\bigskip

\begin{minipage}{\textwidth}
\begin{center}
\includegraphics[width = 0.85\textwidth]{\Path/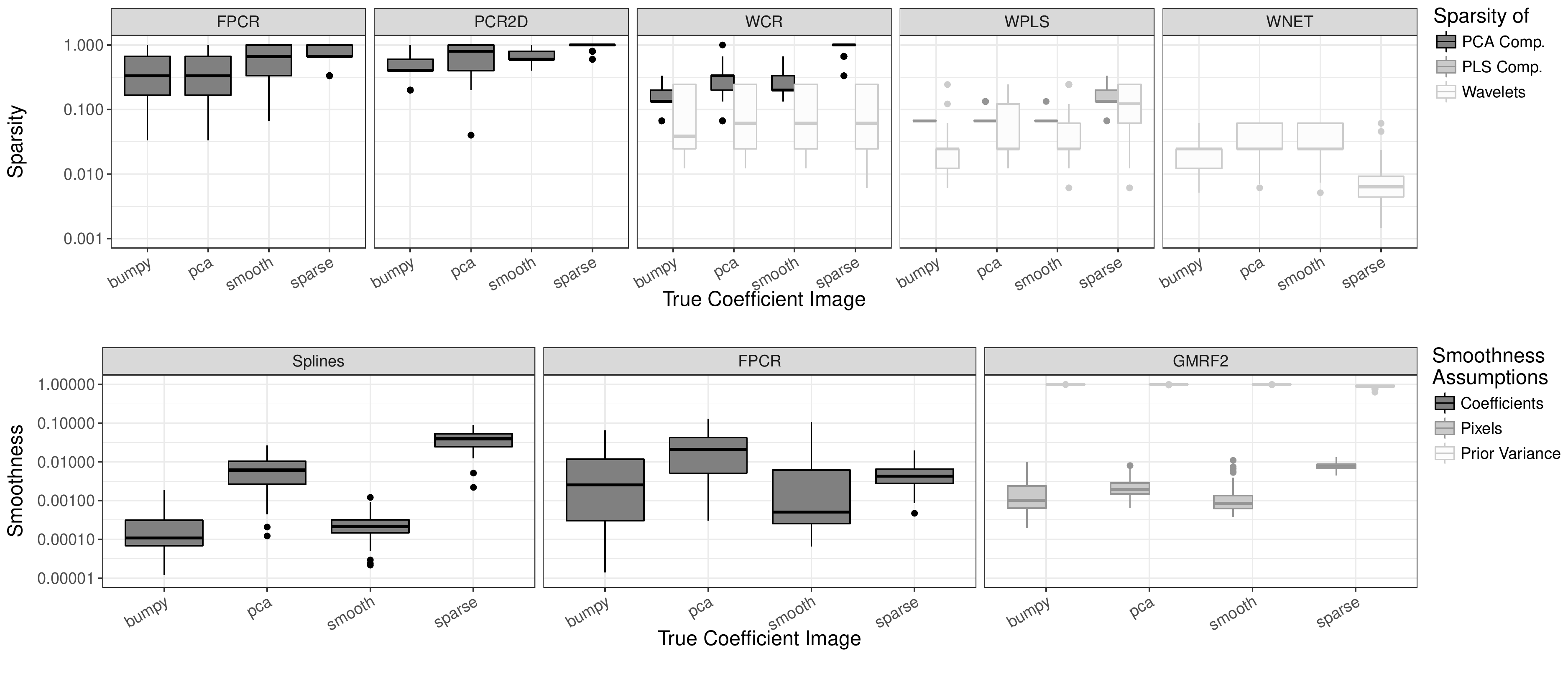}
\end{center}
\captionof{figure}{\protect Measures for parametric model assumptions in the sensitivity study. Boxplots show the measures for the different coefficient images depending on the model used over all $100$ simulation runs. All values on log-scale.}
\label{fig:sensExplicitAss}
\end{minipage}

\bigskip

\begin{minipage}{\textwidth}
\begin{center}
\centering
\includegraphics[width = 0.24\textwidth]{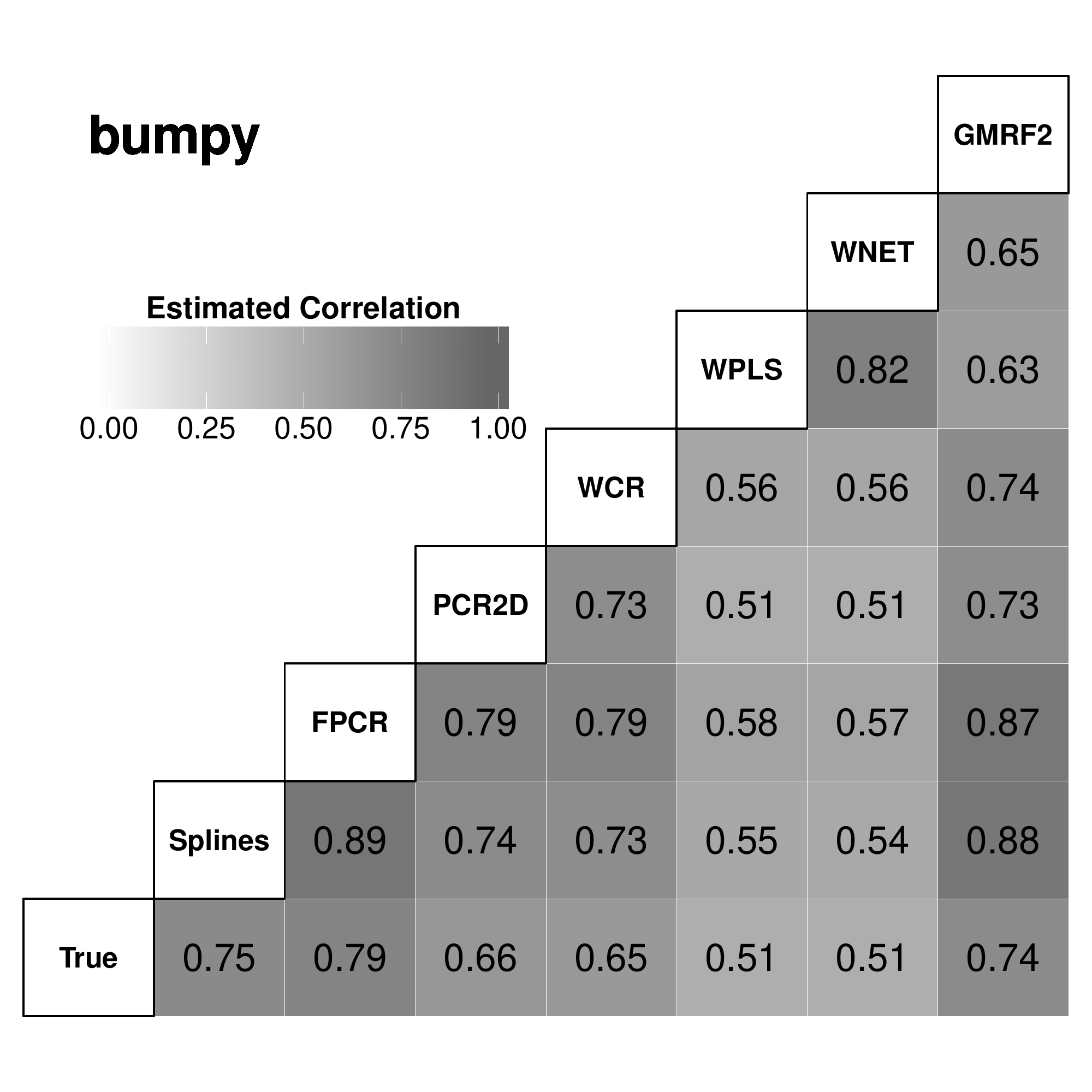}
\includegraphics[width = 0.24\textwidth]{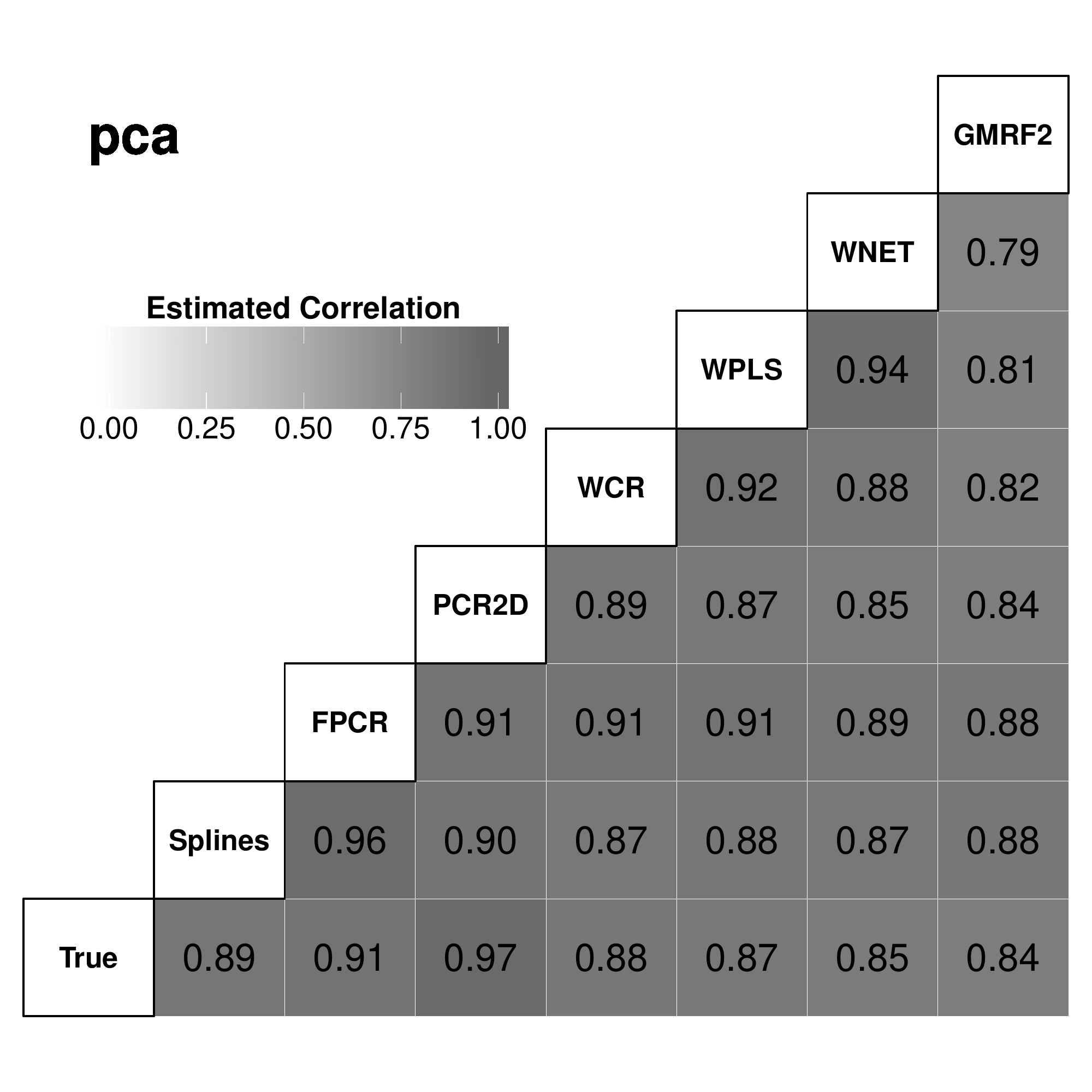}
\includegraphics[width = 0.24\textwidth]{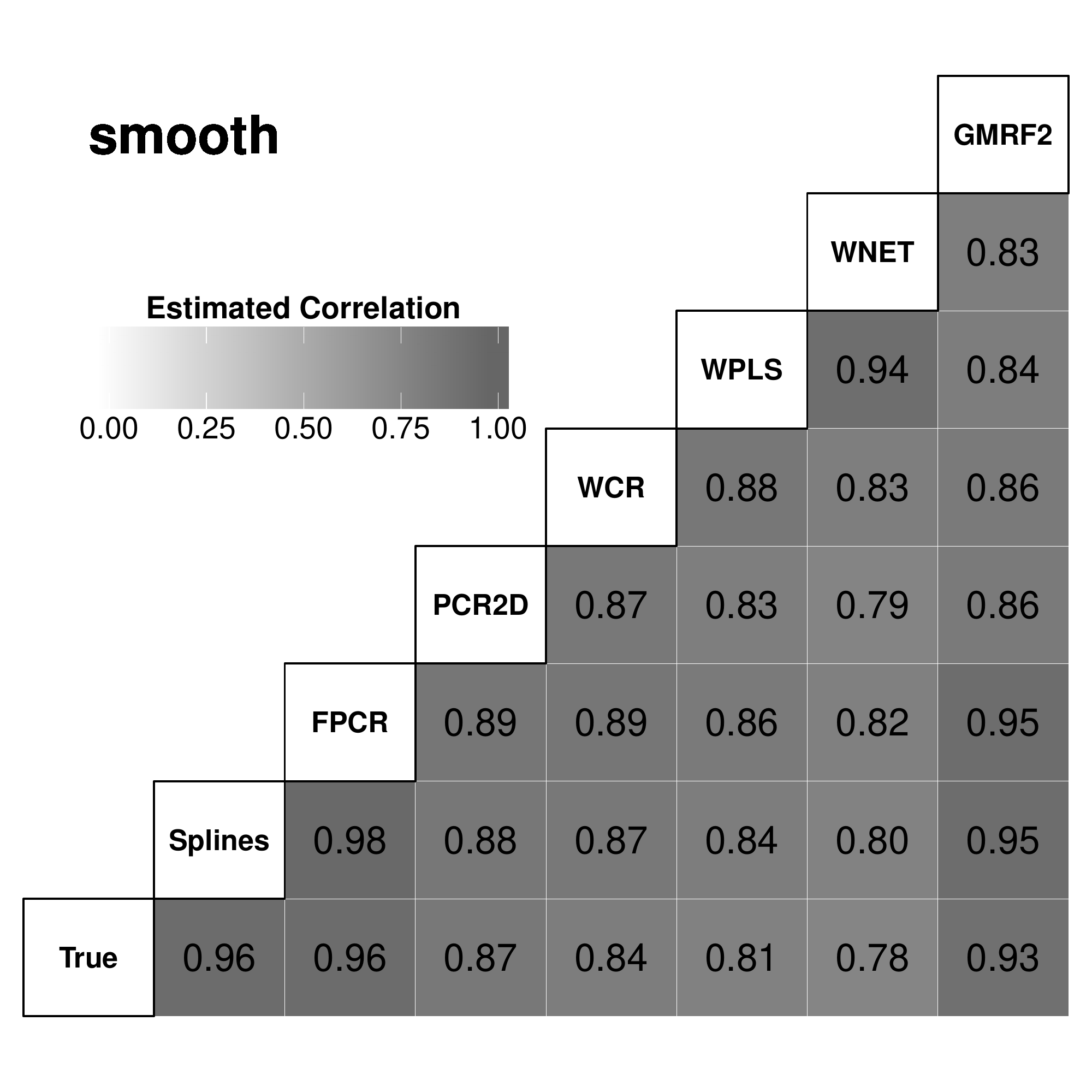}
\includegraphics[width = 0.24\textwidth]{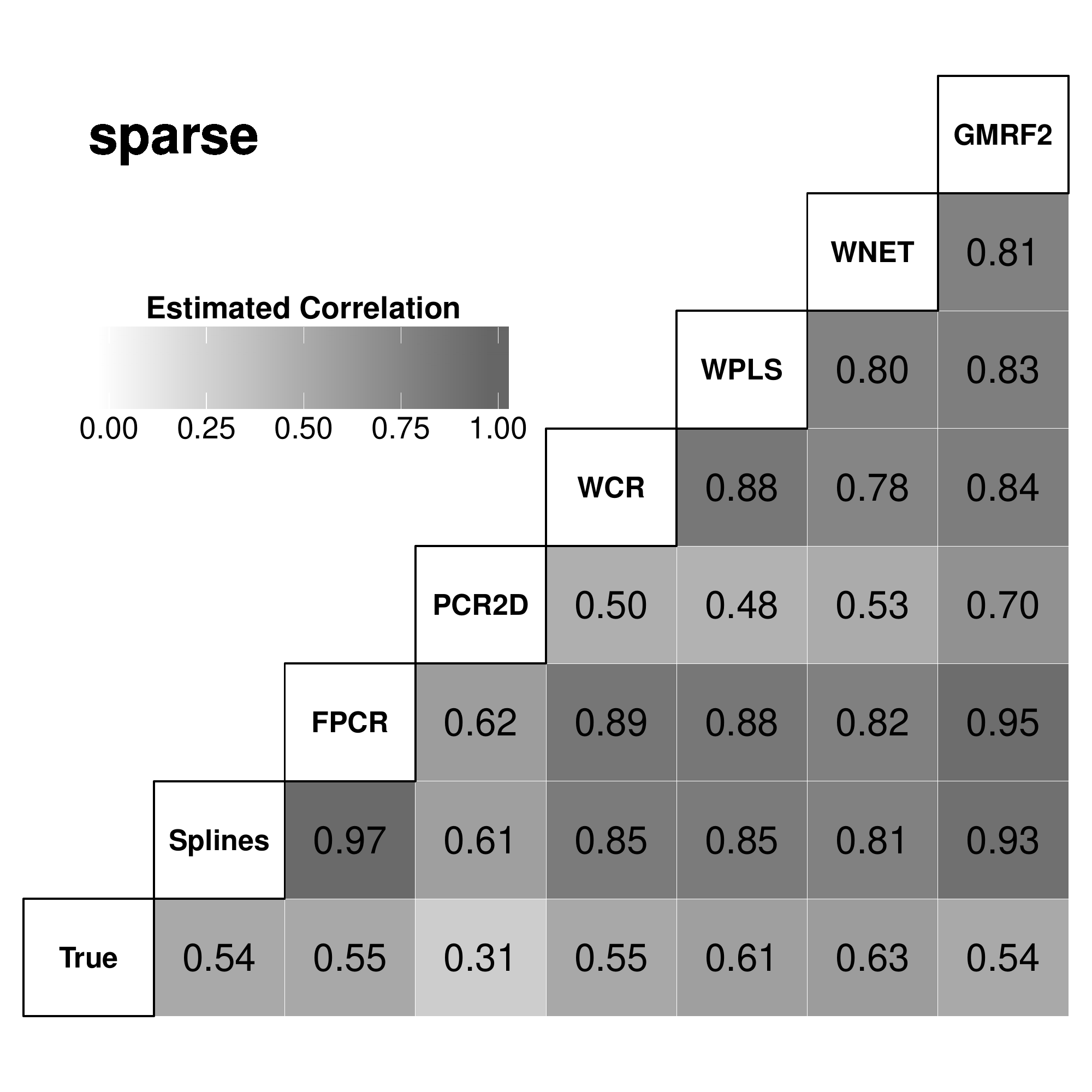}
\end{center}
\captionof{figure}{\protect Correlation between the true coefficient images and the estimates found by the different models in the sensitivity study. The figures show the median  correlation of the vectorized images over 100 simulation runs depending on the true images and the models used.}
\label{fig:sensBetaCorr}
\end{minipage}

\FloatBarrier

\section{Appendix -- Application}
\subsection{Calculation of Confidence/Credible Intervals}
\label{sec:appCI}

The confidence or credible intervals for $\hat \beta$ and $\hat \alpha$ in Figs.~\ref{fig:appBetaCI} and \ref{fig:appAlphaCI} 
have been obtained as follows:

\textit{Splines}: For $\hat \beta$, standard errors based on the Bayesian posterior covariance matrix of the model coefficients are calculated using the \texttt{predict.gam} function of the \texttt{mgcv} package \citep{mgcv}, giving pointwise standard errors conditional on the estimated smoothing parameters, while not including uncertainty of the intercept $\alpha$ (as this is considered separately). Using an approximate normality assumption, the pointwise confidence bands in a pixel $l = 1 \usw L$ are constructed as $95\%$ Wald confidence intervals
\begin{equation}
[\hat \beta_l + \Phi\inv(0.025) \cdot \widehat{\operatorname{se}}(\hat \beta_l),\hat \beta_l + \Phi\inv(0.975) \cdot \widehat{\operatorname{se}}(\hat \beta_l) ]
\label{eq:waldCIbeta}
\end{equation}
with $\Phi$ the distribution function of the standard normal distribution and $\widehat{\operatorname{se}}(\hat \beta_l)$ the standard error for $\hat \beta$ in pixel $l$.
For the $\hat \alpha$ coefficient, confidence intervals are constructed analogously, using the standard errors $\widehat{\operatorname{se}}(\hat \alpha_j)$ produced by the \texttt{summary.gam} function from the \texttt{mgcv} package:
\begin{equation}
[\hat \alpha_j + \Phi\inv(0.025) \cdot \widehat{\operatorname{se}}(\hat \alpha_j),\hat\alpha_j + \Phi\inv(0.975) \cdot \widehat{\operatorname{se}}(\hat \alpha_j) ].
\label{eq:waldCIalpha}
\end{equation}

\textit{FPCR}: Pointwise Bayesian standard errors for $\hat \beta$ are calculated using the \texttt{fpcr} function in \texttt{refund} \citep{refund}. In a next step, pointwise $95\%$ Wald confidence bands are obtained in full analogy to the \textit{Splines} model~\eqref{eq:waldCIbeta}. For $\hat \alpha$, we use again the \texttt{summary.gam} function from the \texttt{mgcv} package to obtain standard errors and calculate $95\%$ Wald confidence bands based on them as in~\eqref{eq:waldCIalpha}.

\textit{PCR2D}: The confidence intervals for $\hat \beta$ and $\hat \alpha$ are found based on a nonparametric bootstrap approach. To this end, the data was resampled $200$ times and the coefficients were re-estimated using the optimal number $K$ of eigenimages found for the original fit due to computational reasons. Pointwise confidence bands for $\hat \beta$ and for the $\hat \alpha$ coefficients are obtained as $95\%$ percentile bootstrap intervals.

\textit{WCR}/\textit{WPLS}/\textit{WNET}: For all three wavelet-based methods, the confidence bands for $\hat \beta$ and $\hat \alpha$ are also based on a nonparametric bootstrap with $200$ resampling iterations. For each bootstrap sample, the models are refit, using $M_0 = 3$ and the optimal parameters of the original fit ($K^\ast, K_0$ for \textit{WCR} and \textit{WPLS}; $K^\ast, \eta, \lambda$ for \textit{WNET}), similar to the case in \textit{PCR2D}. The confidence bands are calculated as  $95\%$ percentile bootstrap intervals for both $\hat \beta$ and $\hat \alpha$ on a pointwise basis.

\textit{SparseGMRF/GMRF/GMRF2}: For the Bayesian methods, we construct Bayesian $95\%$ credible intervals for each pixel in the coefficient image $\hat \beta_l$ and for each coefficient $\hat \alpha_j$ based on the posterior drawings produced by the Gibbs sampling algorithm. The credible intervals are obtained as $2.5\%$ and $97.5\%$ empirical quantiles of the samples after burnin and potential thinning.

\subsection{Supplementary Results for the Application}

\begin{minipage}{\textwidth}
\begin{center}
\includegraphics[width = 0.75\textwidth]{\Path/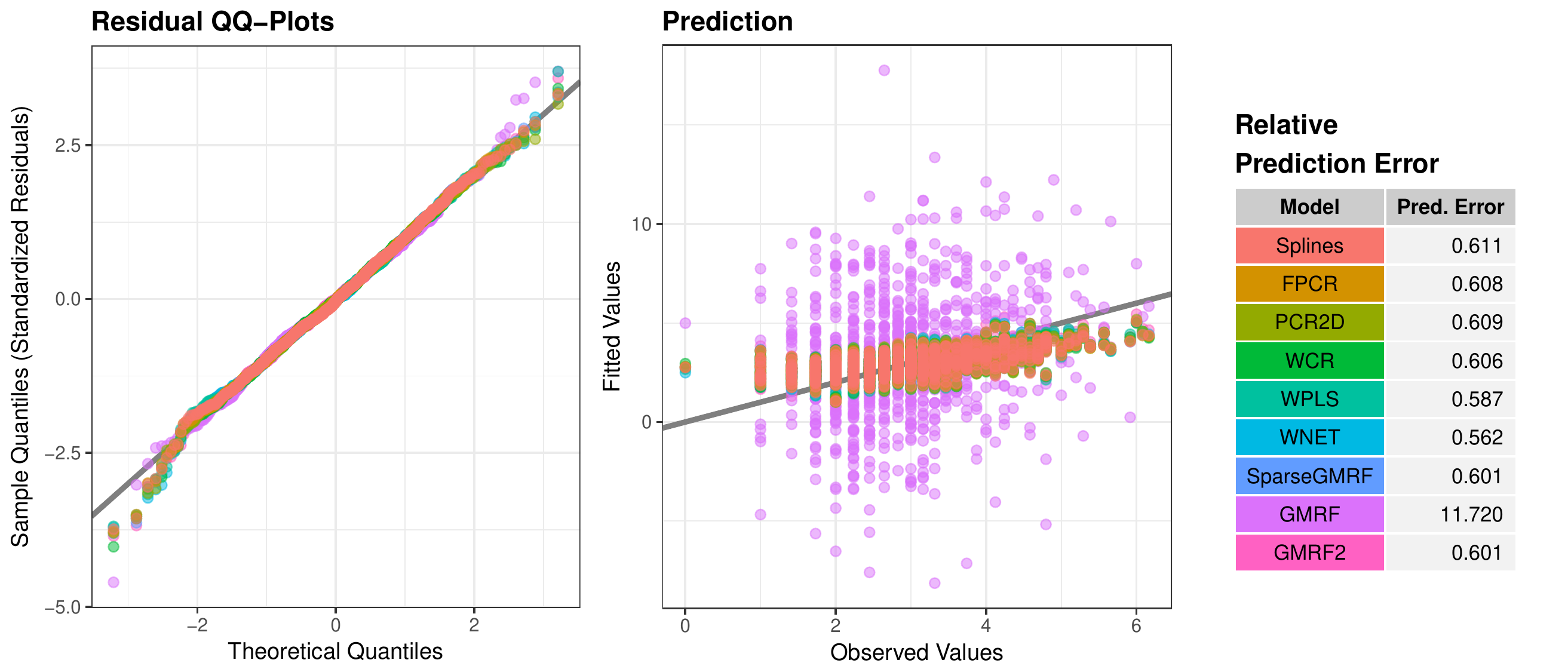}
\end{center}
\captionof{figure}{\protect Assessing the goodness of fit for the application. Left: Normal QQ-Plots for the standardized residuals in each model, showing that they are approximately normal.
Center: Observed response values $y_i$ vs. fitted values $\hat y_i$ found by the nine different models. The diagonal line corresponds to a perfect fit. Right: Relative prediction errors for each model.}
\label{fig:appGoF}
\end{minipage}

\bigskip

\begin{minipage}{\textwidth}
\begin{center}
\includegraphics[width = 0.75\textwidth]{\Path/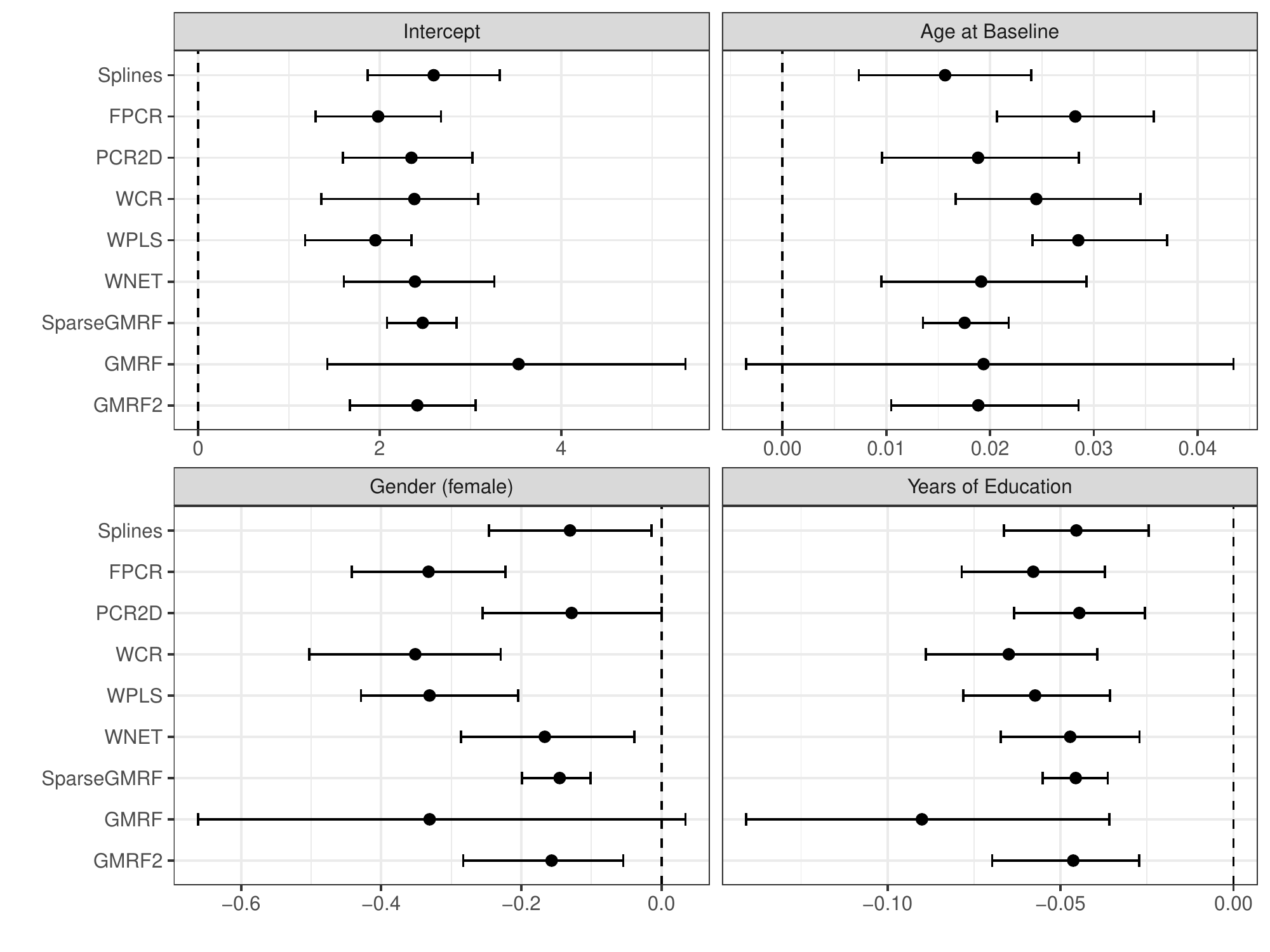}
\end{center}
\captionof{figure}{\protect Coefficient estimates for the scalar variables in the application with empirical $95\%$ confidence intervals. The solid point marks the coefficient estimate for each of the nine models and the horizontal lines correspond to the $95\%$ confidence intervals. The dashed vertical line marks zero.}
\label{fig:appAlphaCI}
\end{minipage}

\bigskip

\begin{minipage}{\textwidth}
\begin{center}
\includegraphics[width = 0.35\textwidth]{\Path/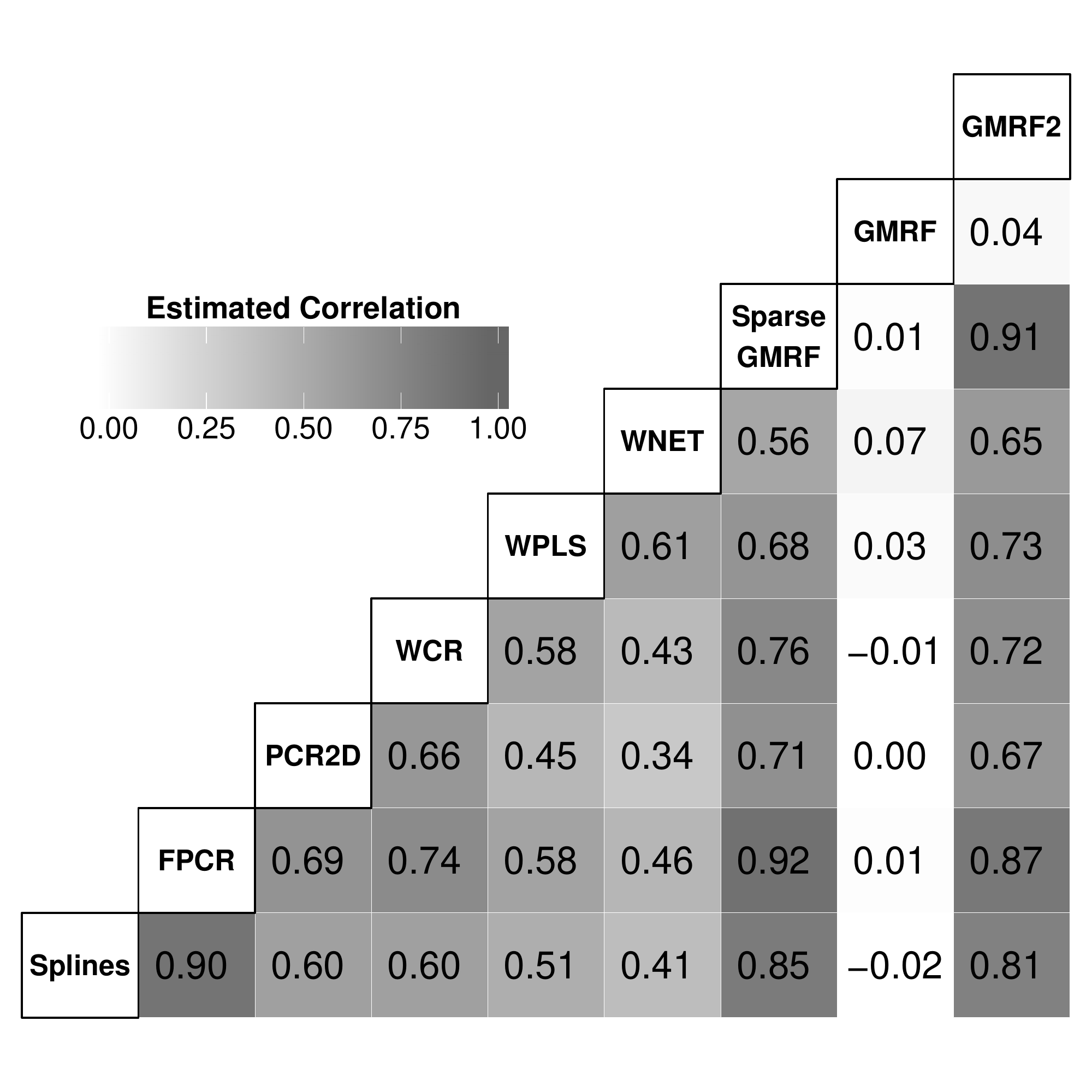}
\end{center}
\captionof{figure}{\protect Correlation between the vectorized estimated coefficient images $\hat \beta$  depending on the model used.}
\label{fig:appCorr}
\end{minipage}

\begin{table}[h]
\caption{\protect Measures for underlying and parametric model assumptions in the application.}
 \label{tab:appModelAss}
\footnotesize
\centering
\begin{filecontents*}{\Path/Code/ApplicationResults/allAss_reshaped.csv}
model,implicit_Smoothness_image,implicit_Sparsity_image,implicit_Sparsity_wavelets,explicit_Smoothness_coefs,explicit_Smoothness_pixels,explicit_Sparsity_pixels,explicit_Sparsity_PCs,explicit_Sparsity_PLSCs,explicit_Sparsity_wavelets,explicit_Prior_NA
{}\textit{Splines},0.002,-,-,0.001,-,-,-,-,-,-
{}\textit{FPCR},0.002,-,-,$< 10^{-3}$,-,-,0.667,-,-,-
{}\textit{PCR2D},-,-,-,-,-,-,0.800,-,-,-
{}\textit{WCR},-,-,0.146,-,-,-,0.200,-,0.244,-
{}\textit{WPLS},-,-,0.116,-,-,-,-,0.067,0.012,-
{}\textit{WNET},-,-,0.104,-,-,-,-,-,0.010,-
{}\textit{SparseGMRF},0.007,0.569,-,-,0.007,1.000,-,-,-,0.104
{}\textit{GMRF},0.043,-,-,-,0.043,-,-,-,-,0.300
{}\textit{GMRF2},0.010,-,-,-,0.010,-,-,-,-,0.998
\end{filecontents*}
\csvautobooktabular[table head= \toprule
& \multicolumn{3}{c}{\textbf{Underlying Assumptions}} & \multicolumn{7}{c}{\textbf{Parametric Assumptions}}  \\
  \cmidrule(lr){2-4} \cmidrule(lr){5-11} 
& Smoothness & \multicolumn{2}{c}{Sparsity} & \multicolumn{2}{c}{Smoothness} & \multicolumn{4}{c}{Sparsity} &Prior \\
\textbf{Model} &  Image & Image  & Wavelets & Coef. & Pixels&  Pixels & PCs & PLSCs & Wavelets & \multicolumn {1}{c}{$\sigma^2_\beta $} \\  \cmidrule(lr){1-1} \cmidrule(lr){2-2} \cmidrule(lr){3-4} \cmidrule(lr){5-6} \cmidrule(lr){7-10} \cmidrule(lr){11-11}]{\Path/Code/ApplicationResults/allAss_reshaped.csv}
\end{table}

\FloatBarrier

\end{document}